\documentclass[twocolumn,trackchanges]{aastex701}
\usepackage{siunitx}

\usepackage[dvipsnames]{xcolor}
\shorttitle{X-ray Spectral variability of PG 1553+113}
\shortauthors{Devanand 2026 et al.}
\usepackage{longtable}
\usepackage{amsmath}
\usepackage{xcolor}



\begin{document}

\title{X-Ray Spectral Variability of  the TeV HBL Blazar PG 1553+113 with XMM-Newton}

\author[orcid=0000-0003-3337-4861,gname='Devanand',sname='Pananchery Ullas']{P.\ U.\ Devanand}
\affiliation{Aryabhatta Research Institute of Observational Sciences (ARIES), Manora Peak, Nainital 263001, India}
\affiliation{Department of Applied Physics/Physics, Mahatma Jyotiba Phule Rohilkhand University, Bareilly 243006, India}
\email[show]{devanandullas@gmail.com}  
\correspondingauthor{P.\ U.\ Devanand}

\author[orcid=0000-0002-9331-4388]{Alok C.\ Gupta} 
\affiliation{Aryabhatta Research Institute of Observational Sciences (ARIES), Manora Peak, Nainital 263001, India}
\email{acgupta30@gmail.com}

\author[orcid=0000-0002-1029-3746]{Paul J.\ Wiita}
\affiliation{Department of Physics, The College of New Jersey, 2000 Pennington Rd., Ewing, NJ 08628-0718, USA}
\email{wiitap@tcnj.edu}

\author[orcid=0000-0002-6449-9643]{V.\ Jithesh}
\affiliation{Department of Physics and Electronics, Christ University, Hosur Main Road, Bengaluru  560029, India} 
\email{jitheshthejus@gmail.com}

\author{Archana Gupta}
\affiliation{Department of Applied Physics/Physics, Mahatma Jyotiba Phule Rohilkhand University, Bareilly 243006, India}
\email{archana.gupta@mjpru.ac.in }

\begin{abstract}
We present an extensive X-ray spectral variability study of the TeV photon-emitting high-energy-peaked BL Lacertae object PG 1553+113, using the data from \textit{EPIC-PN} camera of \textit{XMM-Newton}, which observed the source during its operational period from Sep 2001 to Nov 2024. X-ray spectra in this energy range, $0.6-7.0$ keV, were fitted with absorbed Power-law (PL) and absorbed Log-Parabola (LP) models. We found with 99$\%$ confidence that 14 of them were fit well by LP models having parameters in the range $\alpha\simeq2.13-2.80$, and $\beta\simeq0.04-0.18$, one spectrum favours a LP model with $\beta<0$, while simple PL models with $\Gamma\simeq2.53-2.69$ were sufficient to describe the X-ray spectra of the remaining 15. Two of these 30 observations showed strong signatures of an additional inverse Compton component, while one showed weaker indications. On fitting joint Optical Monitor and EPIC-PN data with LP models, we found synchrotron peaks in the energy range of $\nu_s\simeq4.59-48.61$ eV. This indicates that the spectral evolution is probably caused by variations in particle acceleration or cooling conditions within the jet.
\end{abstract}

\keywords{\uat{Blazars}{164} --- \uat{BL Lacertae objects}{158} --- \uat{X-ray active galactic nuclei}{2035}}

\section{Introduction} 
Blazars belong to a special class of radio-loud active galactic nuclei (AGNs) that exhibit intense flux and spectral variability throughout the entire electromagnetic spectrum on different timescales ranging from minutes to years. The main reason for this is their characteristic relativistic jets, oriented closely ($\leq 10^{\circ}$) to our line of sight \citep{Urr95}, which results in the blazars' non-thermal emission being strongly Doppler-boosted. Blazars comprise of BL Lac objects, which have spectra lacking prominent features or having very weak optical/UV emission lines, and Flat Spectrum Radio Quasars (FSRQs), which have prominent broad emission lines in their combined optical/UV spectra. 

Blazars' spectral energy distribution (SED) generally shows two distinct humps \citep[e.g.][]{Fos08}. The low-energy hump, peaking between the infrared and soft X-ray bands, is attributed to synchrotron emission from relativistic electrons spiraling in the magnetic fields of the relativistic jets. The high energy hump, peaking between the giga-electron (GeV) volts to tera-electron volts (TeV) is attributed to Inverse Compton (IC) processes in well-established leptonic models \citep[e.g.,][]{Bla95,Blo96}, or to processes such as proton synchrotron emission in hadronic models \citep[e.g.,][]{Man92}. \citet[][]{Abd10}, classified blazars into three subclasses: low energy peaked blazars (LBLs) / low synchrotron peaked blazars (LSPs; with the synchrotron peak frequency $\nu_s < 10^{14}$Hz);  intermediate energy peaked blazars (IBLs) / intermediate synchrotron peaked blazars (ISPs; $10^{14}$ Hz $<\nu_s < 10^{15}$Hz), and high-energy peaked blazars (HBLs) / high synchrotron peaked blazars (HSPs; $\nu_s > 10^{15}$ Hz). \\

TeV blazars are those that have been detected at such extremely high energies using ground-based Cherenkov telescopes, and are predominantly HBLs. Their X-ray spectra have long been modeled with simple and broken power-law (PL) models, ascribed to the synchrotron emission component in leptonic models; however, later works have shown that many HBLs exhibit significant curvature and are better depicted by log-parabolic (LP) models \citep[e.g.,][]{Gio02,Tra07,Mas08}. The LP model was initially proposed by \cite{Lan86} and \cite{Jon86}, with further modifications made by \cite{Kre99}. \cite{Mas04} interpreted this model in terms of statistical particle acceleration. They argued that the radiation spectrum is curved because the underlying particle distribution is also curved, and that this curved particle distribution arises when the probability of a particle gaining energy decreases with increasing particle energy.\\

In our previous work \citep[][hereafter Paper I]{Dev25}, we performed X-ray spectral analyses of 13 HBLs observed by XMM-Newton in the energy range of $0.6-10$ keV. When we fitted the X-ray spectra with different models, such as absorbed power law (PL), absorbed log parabola (LP), and absorbed broken power law (BPL), we found that most spectra fit best with LP models. For the LP best-fitted X-ray spectra, we estimated their synchrotron peak frequency and isotropic peak luminosity, and performed correlation studies among different parameters. \\

This study is the continuation of that work, focused on PG 1553+113 (also known as 1ES 1553+113, $\alpha_{2000}=15^h55^m43^s.04$, $\delta_{2000} = +11^\circ 11' 24\farcs36$),  which has been observed by {\it XMM-Newton}  42 times since its launch till the end of November 2025. It was initially discovered as a blue stellar object in the Palomar-Green (PG) survey of ultraviolet excess stellar sources \citep{Gre86}. Afterward, it was categorized as a BL Lac object since its optical spectrum was featureless. The redshift of the source remains uncertain, although several values have been reported in recent years, with the most commonly used value being $z=0.433$ \citep[e.g.,][]{Nic18,Joh19,Dor21}. The source is particularly interesting, as a nearly simultaneous quasi-periodic oscillation (QPO) in $\gamma$-rays, optical and radio bands with a time period of roughly 2.2 years has been claimed for it \citep{Ack15,Abd24}. Such a QPO could arise from intrinsic and/or geometric scenarios. \citet{Abd24} list five possible physical interpretations for the observed QPO: (a) pulsational accretion instabilities in magnetically dominated/arrested accretion flows (MDAF/MAAF) that modulate the outflow efficiency \citep[e.g.,][]{Fra09,Kar12,Pin14}; (b) geometric scenarios involving jet precession \citep[e.g.,][and references therein]{Rom00,Sti03}, rotation/nutation \citep[e.g.,][and references therein]{Cam92,Vla98}, or intrinsic helical jet structures \citep[e.g.,][and references therein]{Con93,Rai15}, all of which can cyclically change the viewing angle and Doppler factor; (c) accretion-outflow coupling mechanisms analogous to low-frequency QPOs in high-mass X-ray binaries, involving Lense-Thirring type precession of the jet-launching region \citep{Wil72,Fen04,Ing09}; (d) periodic accretion modulations driven by eccentric/inclined orbiting massive objects (e.g., massive stars or an intermediate-mass black hole) around a single SMBH; and (e) a gravitationally bound binary SMBH system \citep[and references therein]{Beg80} with total mass $\gtrsim10^{8}\,M_{\odot}$ and milliparsec-scale separation in an early inspiral, GW-driven regime \citep[e.g.,][and references therein]{Pet64,Ses15}. \\

PG 1553+113 is also bright in X-rays and has been extensively observed by \textit{Einstein Observatory (HEAO-2), Rossi XTE, BeppoSAX, Swift, NuSTAR, Suzaku, Chandra, XMM-Newton, and {\it IXPE}}. Many studies based on the data from these space satellites show data well described by different LP models \citep[e.g., see][]{Mas08,Rei08,Ale12,Ale15,Rai15,Mid23,Abd24} with galactic absorption fixed at different values for each study, but in the range of 3.60--3.79$\times 10^{20} {\rm cm}^{-2}$. There have also been instances where a simple PL with fixed galactic absorption \citep[e.g., see][]{Mas08,Cor12,Hua21,Mag24} and sometimes free galactic absorption \citep[e.g., see][]{Rai17} was sufficient to describe X-ray spectra of the source. However, to date, no extensive study using all the observations made by \textit{XMM-Newton} has been conducted, providing the motivation for this work.\\

Here we present a detailed analysis of all publicly available {\it XMM-Newton EPIC-PN} spectra of PG 1553+113. Multiple models, including PL, LP, and BPL, were fitted to the X-ray spectra over the energy range $0.6-7.0$ keV. Additionally, we performed a joint spectral fitting of X-ray data (from EPIC-PN) and optical/UV data (from the optical monitor, OM) for all these observations. The \autoref{sec2} gives a brief overview of the {\it XMM-Newton} instruments we used, the data selection criteria, and the reduction processes. The spectral analysis methodologies we followed in fitting X-ray spectra, as well as joint X-ray - Optical/UV spectra, are described in \autoref{sec3}. Our results and a discussion are presented in \autoref{sec4}, followed by our conclusions in \autoref{sec5}.

\section{Instrumentation, Data Selection, and  Data Reduction}\label{sec2}
\subsection{Instrumentation}\label{sec2.1}
Three science instruments on board \textit{XMM-Newton} are the European Photon Imaging Camera (EPIC), Optical Monitor (OM), and Reflection Grating Spectrometer (RGS). In the present work, we will primarily focus on EPIC-PN and OM. We discussed the EPIC instrument in some detail in \cite{Dev22}. The OM \citep{Mas01} onboard \textit{XMM-Newton} allows us to obtain simultaneous Ultraviolet(UV)/optical data. It has a total field of view of 17 arcmin and its three UV filters (UVW1, UVM2, UVW2) and three Optical filters (V, U, B) cover a wavelength range from 180 nm to 580 nm. The availability of OM data provides us with the opportunity to perform joint spectral fitting of X-ray and optical/UV data.

\begin{deluxetable*}{cccccccc}
\tabletypesize{\scriptsize}
\tablewidth{400pt}
\tablecaption{{\it XMM-Newton EPIC PN+OM} Observation Log for  HBL PG 1553+113 \label{tab1}}
\tablehead{
\colhead{Obs-Date} & \colhead{Obs-ID}& \colhead{Window Mode$^a$} & \colhead{Pile-up} & \colhead{Exposure Time}& \colhead{Good Exposure Time $^b$}&\colhead{Mean Count rate $^c$}&\colhead{OM Filters$^d$}\\
\nocolhead{} & \nocolhead{} & \nocolhead{} &\nocolhead{} &\colhead{(ks)} &\colhead{(ks)} &\colhead{Counts s$^{-1}$}&\nocolhead{}
} 
\colnumbers
\startdata 
2001 Sep 06 & 0094380801 & PSW & No & 6.41 & 5.10 & 32.44$\pm$0.99 &1\\
2010 Aug 06 & 0656990101 & PSW & No & 21.91 & 21.50 & 9.73$\pm$0.40 & 1, 2, 3, 4, 5, 6 \\
2013 Jul 24 & 0727780101 & PSW & No & 33.42 & 33.00 & 22.17$\pm$0.60& 1, 2, 3, 4, 5, 6 \\
2014 Jul 28 & 0727780201 & PSW & No & 35.22 & 34.80 & 12.97$\pm$0.46 & 1, 2, 3, 4, 5, 6 \\
2015 Jul 29 & 0761100101 & PFW & Yes & 137.32 & 122.40 & 6.45$\pm$0.51& 1, 2, 3, 4, 5, 6 \\
2015 Aug 02 & 0761100201 & PFW & Yes & 137.81 & 121.50 & 5.80$\pm$0.50 & 1, 2, 3, 4, 5, 6 \\
2015 Aug 04 & 0761100301 & PFW & Yes & 137.82 & 132.90 & 5.91$\pm$0.50 & 1, 2, 3, 4, 5, 6 \\
2015 Aug 08 & 0761100401 & PFW & Yes & 137.81 & 120.30 & 5.78$\pm$0.50& 1, 2, 3, 4, 5, 6 \\
2015 Aug 16 & 0761100701 & PSW & No & 88.92 & 88.50 & 7.12$\pm$0.34 & 1, 2, 3, 4, 5, 6 \\
2015 Aug 30 & 0761101001 & PFW & Yes & 137.91 & 114.50 & 7.60$\pm$0.56 & 1, 2, 3, 4, 5, 6 \\
2015 Sep 04 & 0727780301 & PSW & No & 28.92 & 28.50 & 7.20$\pm$0.34& 1, 2, 3, 4, 5, 6 \\
2016 Aug 17 & 0727780401 & PSW & No & 28.91 & 28.50 & 8.23$\pm$0.36 & 1, 2, 3, 4, 5, 6 \\
2017 Feb 01 & 0790380501S$^{*}$ & PFW & Yes & 142.08 & 26.60 & 7.66$\pm$0.56 & 1, 2, 3, 4, 5, 6 \\
2017 Feb 01 & 0790380501U$^{*}$ & PFW & Yes & 142.08 & 42.30 & 7.01$\pm$0.51 & 1, 2, 3, 4, 5, 6 \\
2017 Feb 05 & 0790380601 & PFW & Yes & 142.12 & 78.90 & 7.79$\pm$0.58 & 1, 2, 3, 4, 5, 6 \\
2017 Feb 07 & 0790380801 & PFW & Yes & 142.12 & 95.80 & 8.71$\pm$0.61 & 1, 2, 3, 4, 5, 6 \\
2017 Feb 11 & 0790380901 & PFW & Yes & 142.12 & 107.60 & 5.81$\pm$0.50 & 1, 2, 3, 4, 5, 6 \\
2017 Feb 13 & 0790381401 & PFW & Yes & 144.62 & 98.90 & 4.95$\pm$0.41 & 1, 2, 3, 4, 5, 6 \\
2017 Feb 15 & 0790381501 & PFW & Yes & 144.62 & 130.70 & 7.81$\pm$0.57 & 1, 2, 3, 4, 5, 6 \\
2017 Feb 21 & 0790381001 & PSW & No & 95.92 & 95.20 & 7.47$\pm$0.35 &  2, 3, 4, 5, 6 \\ 
2017 Aug 22 & 0727780501 & PSW & Yes & 28.93 & 28.60 & 34.68$\pm$0.88 & 1, 2, 3, 4, 5, 6 \\
2018 Aug 25 & 0810830101 & PSW & No & 34.02 & 33.10 & 12.75$\pm$0.46 & 1, 2, 3, 4, 5, 6 \\
2019 Aug 27 & 0810830201 & PSW & No & 31.87 & 31.10 & 15.37$\pm$0.50 & 1, 2, 3, 4, 5, 6 \\
2020 Aug 26 & 0810830801 & PSW & No & 41.87 & 35.90 & 12.18$\pm$0.45 & 1, 2, 3, 4, 5, 6 \\
2022 Aug 26 & 0810831201 & PSW & No & 29.82 & 29.10 & 31.76$\pm$0.71 & 1, 2, 3, 4, 5, 6 \\
2023 Jan 31 & 0902112101 & PSW & No & 14.02 & 13.30 & 30.38$\pm$0.70 &5 \\
2023 Feb 28 & 0923770201 & PSW & Yes & 124.45 & 113.80 &46.03$\pm$1.08 & 1, 2, 3, 4, 5, 6 \\
2023 Aug 19 & 0810831401 & PSW & No & 32.91 & 32.20 & 21.12$\pm$0.58& 1, 2, 3, 4, 5, 6 \\
2024 Aug 12 & 0810831601 & PSW & No & 44.55 & 34.90 & 7.78$\pm$0.36& 1, 2, 3, 4, 5, 6 \\
2025 Aug 29&0810831801& PSW& No& 28.91 &27.70 &12.53$\pm$0.45& 1, 2, 3, 4, 5, 6\\
\enddata
\tablecomments{
 Cols 3-7 are for EPIC PN data. a: PSW - Prime Small Window mode, PFW - Prime Full Window mode.\\
 b: We define good exposure time as the sum of all good time intervals (GTIs) and can be retrieved from the GTI file.\\
 c: Mean count rate  in the energy range of 0.6-7 keV.\\
d: OM filters 1-UVW2, 2-UVM2, 3-UVW1, 4-U, 5-V, 6-B.\\
*: The observation ID 0790380501 consists of two PN event  files; we are treating them as two independent events named 0790380501S (Scheduled), and 0790380501U (Unsheduled).
}
\end{deluxetable*}

\begin{figure}
% \figurenum{1}
 \centering
    \includegraphics[width=6.5cm, height=6.3cm]{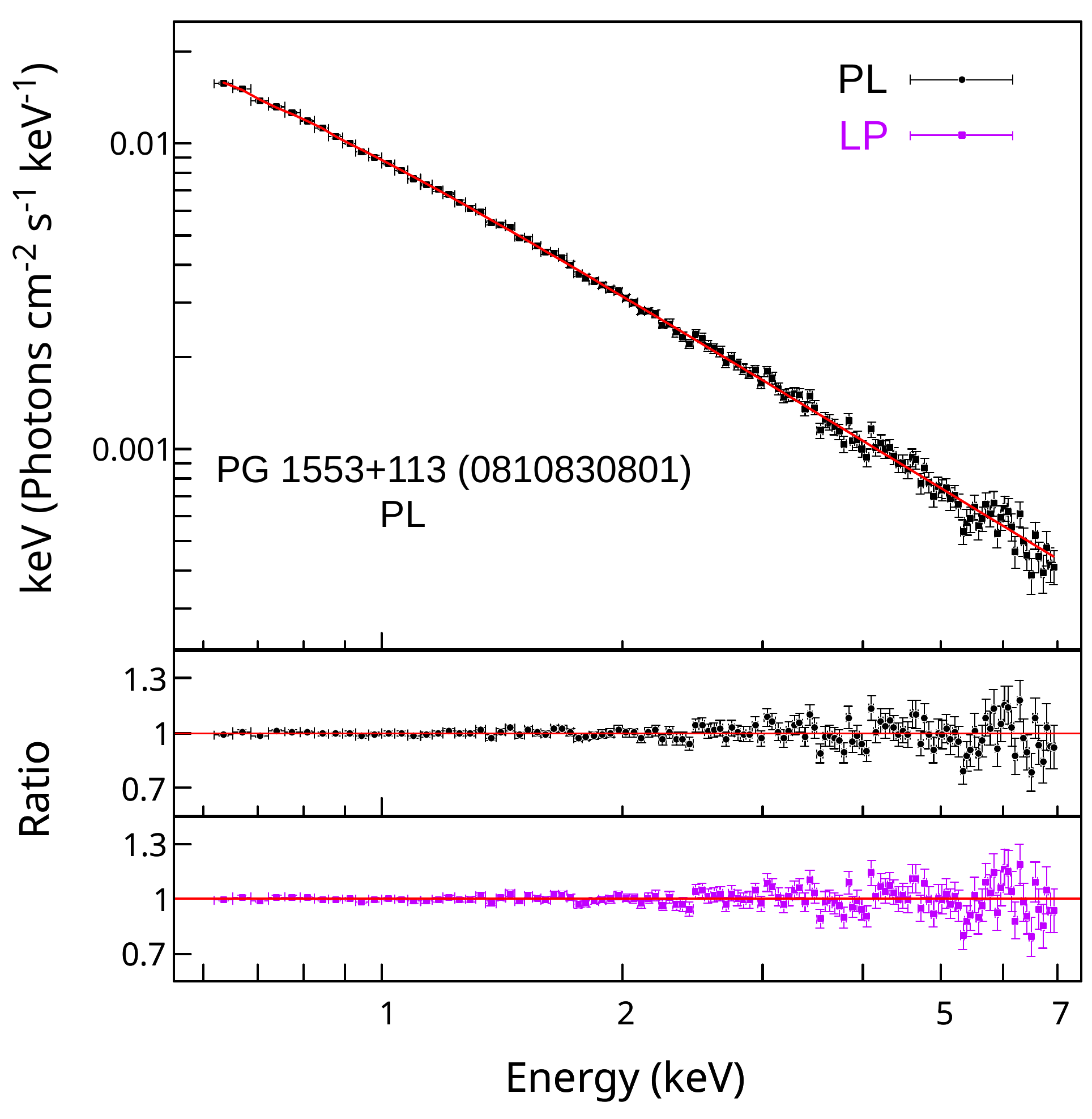}
    \includegraphics[width=6.5cm, height=6.3cm]{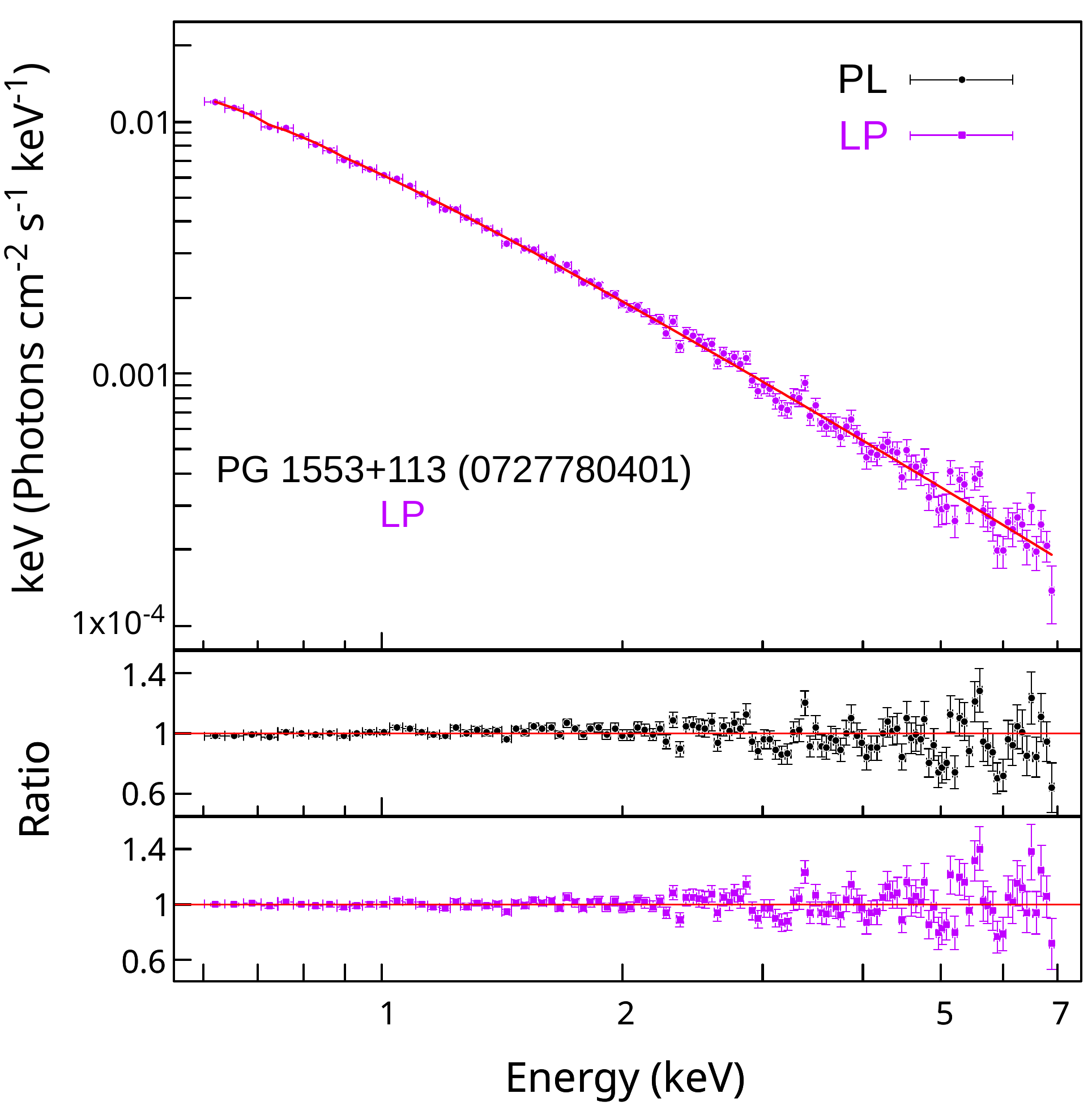}
    \caption{Sample of best-fit models to EPIC PN X-ray spectra of PG 1553+113. From top: (a) Power Law (PL); (b) Log Parabolic (LP);  In the upper portion of each panel, the photon flux of the best-fit model is given for the labeled observation ID, while in the lower portion of the panels, ratios of the data to different models are shown.   PL and LP spectral fits are represented by black-filled circles and dark-magenta-filled squares, respectively. Plots are re-binned for better pictorial representation. Similar spectral plots for all 30 X-ray spectra are available in Appendix \autoref{A1}.\label{fig1}}
\end{figure}

\subsection{Data Selection}\label{sec2.2}
\textit{XMM-newton} observed PG 1153+113 a total of 42 times during its operation. We have downloaded these observational data from NASA's HEASARC archive\footnote{\url{https://heasarc.gsfc.nasa.gov/cgi-bin/W3Browse/w3browse.pl}}. Alternatively, data can also be downloaded from the XMM-Newton Science Archive\footnote {\url{https://nxsa.esac.esa.int/nxsa-web/}}. As of November 2025, data from five of these observations have not yet been made public, and thus, we were left with 37 observations to study. Eight of these observations do not contain EPIC science products, so we were limited to 29 pointed observations. 
Furthermore, the \textit{EPIC PN} observation of February 1, 2017 (Obs ID:0790380501), includes both scheduled (S) and unscheduled (U) observations, which we treat as two different observations. Thus, we have 30 \textit{EPIC PN} X-ray spectra to be studied, and 27 of them have simultaneous OM data in all six bands. Details of \textit{PN} and \textit{OM} data for each observation are in \autoref{tab1}

\subsection{Data Reduction}\label{sec2.3}
We have used the Science Analysis System (SAS) version 22.1.0 to reduce and analyze the data collected by \textit{XMM-Newton}. Below, we discuss the data reduction for each instrument we used in this study.

\subsubsection{EPIC-PN}
The \textit{EPIC-PN} data reduction has been explained in detail in Paper I. Here, we focus on updates to our earlier methodology. From raw event files, we produced cleaned and calibrated event files using various SAS tasks including \textit{epproc}, \textit{tabgtigen}, and \textit{evselect}. Among the 30 \textit{EPIC PN} observations, 17 were in Prime Small Window (PSW), while the remaining  13 were in Prime Full Window (PFW) Imaging mode. PN observations in PFW modes are more susceptible than PSW modes to Out-Of-Time (OOT) events, which distort the original spectrum. Thus, the PFW PN observations were subjected to OOT event corrections as described in the data analysis thread\footnote{\url{https://www.cosmos.esa.int/web/xmm-newton/sas-thread-epic-oot}}. \\

The SAS task \textit{epatplot} can be used to investigate the presence of pile-ups in {\it EPIC} observations (see SAS data analysis thread\footnote{\url{https://www.cosmos.esa.int/web/xmm-newton/sas-thread-epatplot}}). Mode-dependent pile-up thresholds for \textit{EPIC PN} and \textit{EPIC MOS} are discussed in \citet{Jet15}. Since a significant portion of our observations are in PFW, which has a lower pile-up threshold, pile-ups are expected in these observations. For those observations which are free of pile-up, we extract source events from circular regions of radius $\SI{37.5}{\arcsecond}$ to  $\SI{47.5}{\arcsecond}$. We selected these regions to include most of the source emission, while excluding the outer CCD areas to minimize the edge effects. For the 14 PN observations where pile-up was a problem, the outer radius of the annular region is in the same range as the circular region described above, but an inner region of radius varying between $\SI{4.1}{\arcsecond}$ and  $\SI{12}{\arcsecond}$ is excluded. For all PN observations, background regions are also selected with the same shape and size as that of the source region, but away from the source on the same CCD. Source spectrum, background spectrum, redistribution matrix, and ancillary files were generated using  SAS tasks such as \textit{evselect}, \textit{rmfgen}, \textit{arfgen}. The SAS task \textit{specgroup} was used to create group spectra that have all four files linked together. The \textit{specgroup} task also performs the job of rebinning the grouped spectra such that each spectral channel has a minimum of 25 counts and that the energy resolution was not over-sampled by more than a factor of 3.\\

In our previous work, we adopted a lower energy limit of 0.6 keV for spectral analysis due to significant instrumental uncertainties below this threshold, especially in timing mode data, where observed and expected event patterns showed substantial disparities. In the present study, as we are working with imaging mode data alone, we initially extended the lower energy limit to 0.3 keV. However, both PL and LP model fits in this range resulted in relatively large reduced chi-square ($\chi^2_r$) values. The mean reduced chi-square of the best-fit PL and LP models decreased from 2.96 and 1.51 to 1.52 and 1.17, respectively, when the lower energy limit was changed from 0.3 to 0.6 keV. Therefore, we chose to restrict the lower energy limit of spectral analysis to 0.6 keV. Before fitting spectra with any model, we inspected source and background spectra in count space using XSPEC \citep{Arn96}. XSPEC commands \textit{setplot back} and \textit{setplot ldata} were used to do so. This inspection showed an increasing relative background contribution towards high energies. In 17 out of 30 observations, the background spectrum becomes comparable to the source spectrum at the high-energy end and in some cases overlaps it. In five additional observations, the background approaches the source closely above $\sim$7 keV. Inclusion of these energy ranges leads to large fractional uncertainties in the highest-energy bins. Only eight observations show the source spectrum clearly dominant over the background across the full 0.6--10 keV range. We therefore adopt 7 keV as a uniform and conservative upper energy limit, ensuring that the spectral fits are not influenced by background-dominated bins. Thus, the spectral analysis was performed over the 0.6--7.0 keV range.

\subsection{EPIC MOS}
The \textit{EPIC-MOS} data were reduced following the standard procedure outlined in Paper I. We analyzed MOS data only in select cases where the PN spectrum, when fitted with a LP model, exhibited significant negative curvature. The SAS task \textit{emproc} was used to produce an uncalibrated event list, and we applied a good time interval filtering by excluding time periods where soft proton background count rate exceeded $0.35\hspace{0.1cm}ct \hspace{0.1cm}s^{-1}$. As we see in the results section, we have reported spectral results from MOS data for three observations only, and  \textit{MOS1} and \textit{MOS2} data of these observations were in prime partial Window2 (PW2) imaging mode. Observations in PW2 mode do not have a background region present in the same CCD, and thus we had to depend on background sky files \citep{Car07} for background extraction. Furthermore, we selected only events with a PATTERN in the range of 0–12 and a FLAG equal to 0 for source and background spectral extraction. Beyond these MOS-specific steps, the rest of the reduction procedure follows the same as that for \textit{EPIC-PN}.

\subsubsection{OM}
We used standard OM imaging mode data for our studies. These imaging mode data are fully processed by a single SAS task \textit{omichain}, which is a Perl script that executes several tasks in the correct order using appropriate parameters. This SAS task generates a substantial number of outputs, including a combined source list that contains corrected coordinates, count rates, magnitudes, and additional information for every source detected by OM during the observation period. We obtain corrected coordinates for our source of interest by comparing the cleaned science-ready images corresponding to each filter with the corresponding source list for that filter. Using these corrected coordinates, we can get information about source count rates in different bands. Next, we used the SAS task \textit{om2pha} to convert OM photometric data into an OGIP II format file, which contains the count rate for each filter and can be directly read into \textit{XSPEC}.

\section{Spectral analysis}\label{sec3}
We used XSPEC \citep{Arn96} version 12.15.0 to perform spectral analysis of the X-ray spectra and later the combined  X-ray and optical/UV data. We adopted the same three models as described in our earlier work (Paper I): a power-law (PL) model, a log-parabolic (LP) model, and a broken power-law (BPL) model. The definitions of these models and their associated parameters are explained in detail in Paper I and are briefly described here.
\begin{enumerate}
  \item Power Law model, defined as 
  \begin{equation}\label{eq1}
    F(E) = N E^{-\Gamma},
  \end{equation}
  with normalization $N$ and photon index $\Gamma$.

  \item Log-parabolic model, defined as
  \begin{equation}\label{eq2}
    F(E) = N \left(\frac{E}{E_1}\right)^{-(\alpha + \beta \log(E/E_1))},
  \end{equation}
  with normalization $N$; photon index at pivot energy $E_1 = 1\,\text{keV}$, $\alpha$; and curvature parameter $\beta$.

  \item Broken Power Law model, defined as
  \begin{equation}\label{eq3}
    F(E) =
    \begin{cases}
      N E^{-\Gamma_1}, & \text{for } E \le E_{\text{break}}, \\
      N E_{\text{break}}^{\Gamma_2-\Gamma_1} \left(\dfrac{E}{1\,\text{keV}}\right)^{-\Gamma_2}, & \text{for } E > E_{\text{break}},
    \end{cases}
  \end{equation}
  with break energy $E_{\text{break}}$; photon indices $\Gamma_1$, and $\Gamma_2$; and normalization $N$. This model is used only for spectra with significant concave curvature ($|\beta| \ge  \beta_{\text{err}}$).
\end{enumerate}
To account for galactic absorption, each of these models is convolved with the {\it tbabs} model (Tuebingen-Boulder ISM absorption model) \citep{Wil00}. Thus, any references to these models throughout this paper implicitly include pre-multiplication by the {\it tbabs} model. The hydrogen column density ($n_H$) / Galactic absorption of the source is taken from the nH calculator\footnote{\url{https://heasarc.gsfc.nasa.gov/cgi-bin/Tools/w3nh/w3nh.pl}} tool using the HI 4 Pi Survey map \citep{HI416} and has a value of 3.61$\times10^{20} cm^{-2}$. Initially, we fit the  X-ray spectrum of an observation with both a PL model and a LP model.

 We use the F-test to examine whether the LP model offers a better fit than the PL model. We treat the PL as the null hypothesis (NH). Since the LP model reduces to a PL for $\beta=0$ (pivot energy E$_1$=1 keV fixed), the models are nested and differ by only a single additional parameter $\beta$. If the F-test between the two models results in $p_{null}$ $\leq$0.01, we consider LP to be the best fit. Otherwise, the PL model is considered to adequately describe the X-ray spectrum. In the cases where, on fitting a LP model, a significant negative curvature ($|\beta|\geq\beta_{err}$) is obtained, we fit the X-ray spectrum with a BLP model as well.

The X-ray spectra of HBLs typically show curvature associated with synchrotron emission and lack distinct features over wide energy ranges \citep[e.g.][]{Gio05,Per07,Mas08}. However, there are occasions when, due to instrumental energy constraints, brief exposures, or spectral peaks lying outside the observed energy band, it is difficult to detect possible curvature. In such instances, the PL model serves as a sufficiently accurate representation of the X-ray spectra. In some instances, there is intermixing of the high-energy end of synchrotron emission and the low-energy end of inverse Compton (IC) emission, causing the flattening of the former component by the latter, which results in concave curvature of X-ray spectra. In such cases, BPL provides a more accurate description of X-ray spectra. It is to be noted that concave curvature can also arise when there is excess absorption of soft X-rays by local gas within the blazar.

\subsection{Finding the Synchrotron Peaks}\label{sec3.1}
 We follow an approach similar to that of Paper I. While the parameters of the best-fitting LP models can be employed to estimate the synchrotron peak energy E$_{p,LP}$ and peak height S$_{p,LP}$ using \citep{Mas04} 
\begin{equation}
     E_{p,LP} = E_1 10^{(2-\alpha)/2\beta} \hspace{0.4cm} \rm{(keV)}\label{eq4}
 \end{equation}
 and
 \begin{equation}\label{eq5}
     S_{p,LP} =(1.60 \times 10^{-9})NE_1^2 10^{(2-\alpha)^2/4\beta} \hspace{0.2cm} \rm(erg \hspace{0.1cm}cm^{-2} \hspace{0.1cm}s^{-1}),
 \end{equation}
Making error estimations from these equations is a tiresome task. Thus, we use a different version of the LP model in the form of the eplogpar (ELP) model \citep{Tra07,Tra09,Wan19}, which in turn has the peak energy as a parameter. The model is defined as 
\begin{equation}\label{eq6}
    F(E) = K_{ELP}\hspace{0.1cm}10^{-\beta_{ELP}(log(E/E_{p,ELP}))^2}/E^2,
\end{equation}
in units of ph cm$^{-2}$ s$^{-1}$, where the parameters of the model are: $E_{ELP}$, the peak energy; $K_{ELP}$, the  flux at the peak energy; and $\beta_{ELP}$, the curvature parameter. We give the $E_{p,LP}$ value computed using \autoref{eq4} as the initial value of E$_{p,ELP}$. Again, throughout the paper, it is implied that when we mention ELP, it is convolved with \textit{tbabs}.
The SED peak value S$_{p,ELP}$ can be easily computed using
\begin{equation}\label{eq7}
    S_{p,ELP} = 1.60 \times 10^{-9} K_{ELP} \hspace{0.3cm} \rm(erg \ cm^{-2} \ s^{-1}).
\end{equation}
As in Paper I, and following \cite{Tra09} and \cite{Wan19}, we checked the reliability of both versions of the log parabola fits, LP and ELP, by asserting the conditions that $E_{p,ELP}$ exceeds its  $1\sigma$ uncertainty ($E_{ELP}$/$\sigma_{E_{ELP}}$ $ > $ 1), as well as that both peak values, $E_{ELP}$ and $E_{LP}$, agree within a $1\sigma$ uncertainty. Taking cosmological corrections into consideration, the peak energy in the rest frame, $E_p$, is calculated as,
\begin{equation}
    E_p = (1+z)E_{p,ELP}  \hspace{0.3cm} \rm{(keV)}.\label{eq8}
\end{equation}
We take the redshift, $z$ to be 0.433, which is the most widely accepted value. The isotropic peak luminosity \citep{Mas08}, which is the measure of power in the rest frame, is calculated as
\begin{equation}\label{eq9}
    L_p \simeq 4 \pi D_L^2 S_{p,ELP} \hspace{0.3cm} \rm(erg \hspace{0.1cm}s^{-1}).
\end{equation}
The luminosity distance, $D_L$, of PG 1553+113  is \citep[][]{Pee93}, 
\begin{equation}
    D_L = \frac{c}{H_0}(1+z) \int_{0}^{z}\frac{dz}{\sqrt{\Omega_M(1+z)^3+\Omega_{\Lambda}}},\label{eq10}
\end{equation} 
where we make use of the following flat cosmological parameters \citep{Pla16}: $\Omega_{\Lambda}=$ 0.692; $\Omega_M=$ 0.308; and $H_{0}=$ 67.8 km s$^{-1} Mpc^{-1}$.

\subsection{Joint PN+OM fitting}\label{sec3.2}
As we attempted to estimate $E_p$, we found that in every case, the peak lies significantly below 0.2 keV, making them poorly constrained using X-ray data alone. We therefore attempted to expand our studies into the Optical/UV regime using OM data. The output of {\it omichain} task contains a combined source list with photometric information such as count rates, magnitudes, and fluxes for each source corresponding to each filter. The SAS task {\it om2pha} is then used on this combined source list along with information of source coordinates to obtain an OGIP type II file, which, along with response files, can be fed to  XSPEC. The canned response files for each filter are publicly available\footnote{\url{https://sasdev-xmm.esac.esa.int/pub/ccf/constituents/extras/responses/OM/}}. We can then use XSPEC for jointly fitting {\it EPIC PN} and {\it OM} data as described in the SAS XSPEC session thread\footnote{\url{https://www.cosmos.esa.int/web/xmm-newton/sas-thread-xspec}}. A similar approach has been followed elsewhere \citep[e.g.,][]{Rai17,Jag18}.

Our initial course of action was fitting \textit{tbabs*redden*logpar} model on combined PN+OM data. This absorbed LP model with reddening correction will henceforth be referred to as LP$_R$. We used a reddening value E(B-V) = 0.0446 \citep[][]{Sch11} for PG 1553+113, which is used to account for the effects of galactic reddening in optical/UV wavelengths. E(B-V) as well as $n_H$ values are fixed to their respective values during the fitting process. On initial fitting, most observations yielded large reduced chi-squares in the range 2--15. So we applied a systematic error of 3$\%$ on OM data alone. This systematic error was introduced by raising the uncertainties on the OM data point by $3\%$ of their respective flux values without changing the flux value itself. This was done using \textit{ftcal} tool available in HEASoft, applied to the OM spectral file in OGIP II format. Other authors \citep[e.g.,][]{Lob16,Jag18,Xu21}, have previously applied similar systematic errors to OM data. These systematic errors mainly account for any cross-normalization issues between \textit{EPIC PN }and \textit{OM}. The inclusion of systematic errors helped keep $\chi_r \leq 2$ for most observations.

\begin{figure}
% \figurenum{1}
 \centering
    \includegraphics[width=6.5cm, height=6.3cm]{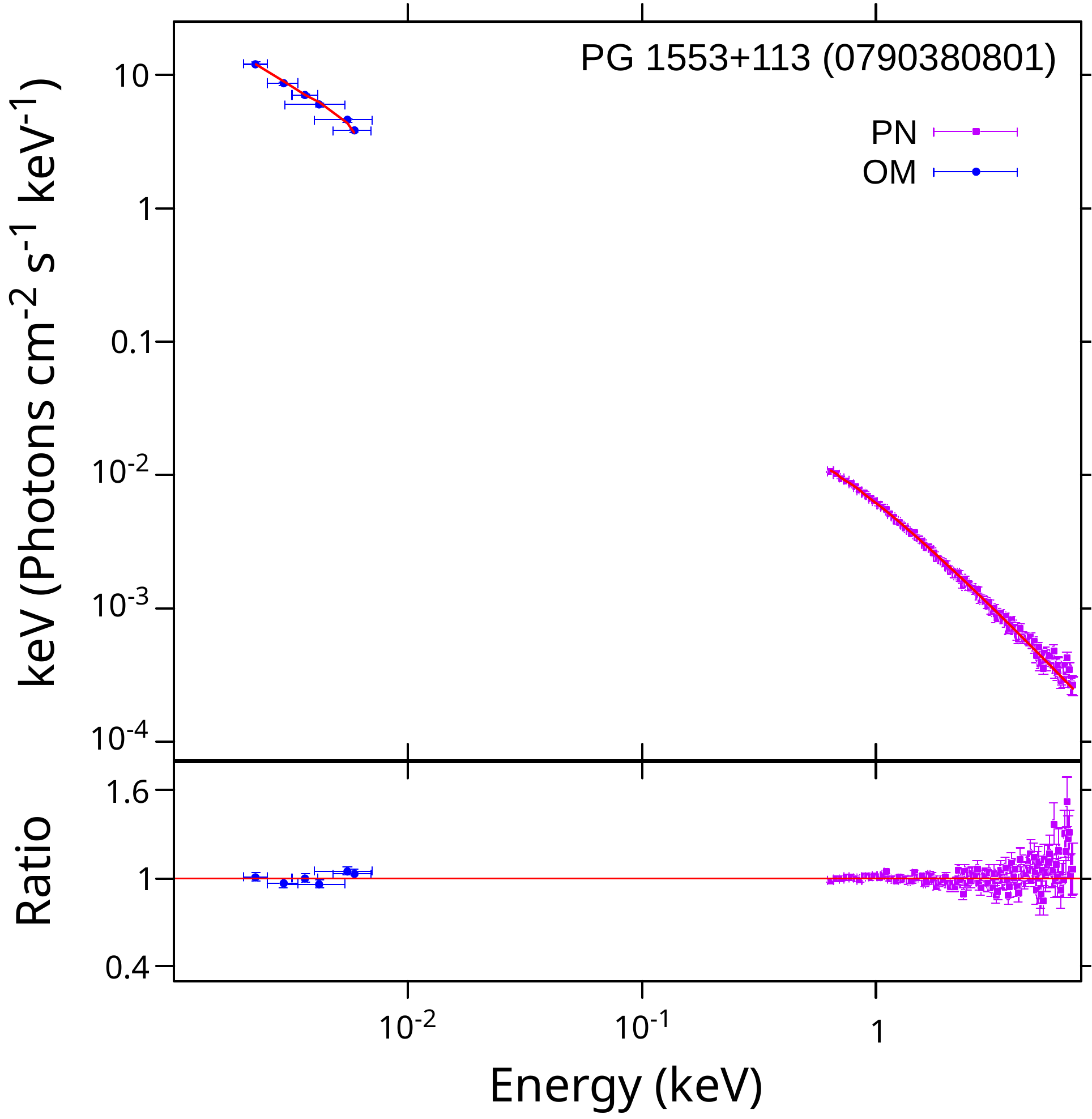}
    \caption{Sample of Log Parabolic (\textit{tbabs*redden*logpar}) fit  to joint \textit{PN+OM} spectra of PG 1553+113. In the upper panel, the plot of photon flux of the model is given for the labeled observation ID, while in the lower panel, the ratio of the data to the model is shown.   PN and OM data points are represented by dark-magenta-filled squares and blue-filled circles, respectively. Plots are re-binned for better pictorial representation. Similar spectral plots for all 27 joint Optical/UV+X-ray spectra are available in Appendix \autoref{A2}.\label{fig2}}
\end{figure}

\subsection{Search for Signatures of IC component}\label{sec3.3}
Following \cite{Zha08} and Paper I, we searched for signatures of the IC emission component in the spectra of PG 1553+113. This emission process in soft X-rays is expected to contribute to FSRQs and LBL/IBL, assuming leptonic emission models. The search is conducted mainly for those \textit{PN} observations that yield a significant negative curvature, $|\beta|\geq \beta_{err}$, when fitting the LP model to PN spectra in the energy range of $0.6-7$ keV. For these observations, a BPL model with spectral hardening ($\Gamma_1 > \Gamma_2$) tends to describe the X-ray spectrum better. We first check if the source spectra counts are higher than the background spectral counts at higher energies. If so, we can use the energy range of $0.6-10$ keV for analysis. If background counts are comparable to source counts at higher energies, we stick to a lower energy range (e.g $0.6-7$ or $0.6-8$ keV range). Typically, the EPIC \textit{PN} spectra are generated by taking both single and double events (PATTERN $\leq4$). We generate a modified \textit{PN} spectra (hereafter mod \textit{PN}) by using only single events (PATTERN = 0), removing the central core of at least $\SI{5}{\arcsecond}$ (if the spectrum is piled-up, the radius of the excised core would increase), and choosing a different background region. We then check if the modified \textit{PN} spectra also exhibit negative curvature when fitting the LP model, and if the spectra are well-fitted by a BPL model with spectral hardening. We also analyze \textit{EPIC-MOS} data from these observations and see if the \textit{EPIC-MOS} spectrum can be fitted with the LP model, which features significant negative curvature. For all these data, the condition of significance for $\beta$ is $ |\beta|\geq 1\beta_{err}$. If it does display significant negative curvature, we then fit BPL to the \textit{MOS} spectra. Additionally, we tried joint spectral fitting of \textit{PN}, \textit{MOS1}, and \textit{MOS2} spectra with PL, LP, and BPL models using \textit{XSPEC}. A cross-calibration constant was added for each instrument, which was kept constant for \textit{PN} and allowed to vary for \textit{MOS}. For this joint spectral fit, we generated contour plots between the BPL fit parameters to see if soft and hard photon indices are completely independent at 99$\%$ confidence level. \\

We consider an observation to have a significant claim for signatures of IC X-ray component when the following criteria are met:
\begin{enumerate}
\item The X-ray spectra generated from {\it EPIC PN}, \textit{mod-PN}, \textit{MOS1}, \textit{MOS2}, and the PN+MOS combined data show significant negative curvatures, $|\beta|>\beta_{err}$, on fitting LP models.
\item  The X-ray spectra generated from {\it EPIC PN},\textit{ mod-PN}, \textit{MOS1}, \textit{MOS2}, and the PN+MOS combined data are well described by stable BPL models with well-constrained values for both soft and hard photon indices.
\item The soft photon index and hard photon index of the BPL models are independently constrained at 99$\%$ confidence levels.

\end{enumerate}

\begin{deluxetable*}{cccccccc}
\tabletypesize{\scriptsize}
\tablewidth{400pt}
\renewcommand{\arraystretch}{0.9}
\tablecaption{Spectral Properties of PG 1553+113 obtained by fitting X-ray spectra with PL, and LP models\label{tab2}}
\tablehead{
\colhead{Obs ID} & \colhead{Model} &
\colhead{$\Gamma/\alpha$} &
\colhead{$\beta$} &
\colhead{$N$} &
\colhead{$\chi^2_r$ (dof)} &
\colhead{$F_{0.6-7}$} &
\colhead{F-test (p-value)}
}
\colnumbers
\startdata
0094380801 & PL & $2.41\pm0.01$ &\dots & $24.47\pm0.14$ & 1.29(117) & $74.79_{-0.45}^{+0.46}$ & NH \\
 & \textbf{LP} & $2.37\pm0.02$ &$0.09\pm0.04$ & $24.58\pm0.15$ & 1.15(116) & $74.38_{-0.48}^{+0.49}$ & 15.0($1.8\times10^{-04}$) \\
0656990101 & PL & $2.84\pm0.01$ & \dots   & $7.81\pm0.04$ & 1.19(113) & $20.0_{-0.10}^{+0.10}$ & NH \\
 & \textbf{LP} & $2.80\pm0.02$ & $0.12\pm0.04$ & $7.87\pm0.04$ & 0.93(112) & $19.88_{-0.11}^{+0.11}$ & 31.8($\ll$ 0.0001) \\
0727780101 & PL & 2.37 &  \dots &16.83 & 3.81(125) & 52.53 & NH \\
 & \textbf{LP} & $2.30\pm0.01$ &$0.18\pm0.02$ & $16.99\pm0.05$ & 1.35(124) & $52.02_{-0.16}^{+0.16}$ & 227.9($\ll$ 0.0001) \\
0727780201 & PL & $2.39\pm0.01$ &\dots  & $9.90\pm0.03$ & 1.77(123) & $30.60_{-0.12}^{+0.11}$ & NH \\
 & \textbf{LP} & $2.34\pm0.01$ &$0.11\pm0.02$ & $9.95_{-0.04}^{+0.03}$ & 1.22(122) & $30.40_{-0.12}^{+0.12}$ & 56.7($\ll$ 0.0001) \\
0761100101 & \textbf{PL} & $2.54\pm0.01$ & \dots &$4.85\pm0.02$ & 1.15(120) & $13.93_{-0.06}^{+0.06}$ & NH \\
 & LP & $2.53\pm0.01$ & $0.03\pm0.03$ & $4.86\pm0.02$ & 1.14(119) & $13.90_{-0.06}^{+0.07}$ & 2.2($>0.05$) \\
0761100201 & \textbf{PL} & $2.53\pm0.01$ &\dots & $4.22\pm0.02$ & 0.99(120) & $12.18_{-0.06}^{+0.06}$ & NH \\
 & LP & $2.52\pm0.01$ & $0.01\pm0.03$ & $4.22\pm0.02$ & 0.99(119) & $12.17_{-0.06}^{+0.06}$ & 0.4($>0.05$) \\
0761100301 & \textbf{PL} & $2.53\pm0.01$ & \dots & $4.36\pm0.02$ & 0.97(121) & $12.61_{-0.06}^{+0.05}$ & NH \\
 & LP & $2.53\pm0.01$ &  $-0.001\pm0.029$ & $4.36\pm0.02$ & 0.98(120) & $12.61_{-0.07}^{+0.06}$ & 0.01($>0.05$) \\
0761100401 & \textbf{PL} & $2.53\pm0.01$ &  \dots&$4.32\pm0.02$ & 1.08(120) & $12.49_{-0.06}^{+0.06}$ & NH \\
 & LP & $2.54\pm0.01$ & $-0.04\pm0.03$ & $4.31\pm0.02$ & 1.05(119) & $12.52_{-0.06}^{+0.07}$ & 4.1($4.4\times10^{-02}$) \\
0761100701 & PL & $2.58\pm0.01$ &\dots & $5.53\pm0.01$ & 1.29 (124) & $15.58_{-0.05}^{+0.04}$ & NH \\
 & \textbf{LP} & $2.56\pm0.01$   & $0.06\pm0.02$ & $5.55\pm0.02$ & 1.07(123) & $15.52_{-0.05}^{+0.05}$ & 26.7($\ll$ 0.0001) \\
0761101001 & \textbf{PL} & $2.55\pm0.01$ &\dots & $5.72\pm0.02$ & 1.07 (121) & $16.35_{-0.07}^{+0.07}$ & NH \\
 & LP & $2.54\pm0.01$ &$0.04\pm0.03$ & $5.73\pm0.02$ & 1.04(120) & $16.31_{-0.07}^{+0.08}$ & 5.2($2.5\times10^{-02}$) \\
0727780301 & \textbf{PL} & $2.56\pm0.01$ & \dots & $5.68\pm0.03$ & 1.20(118) & $16.15_{-0.09}^{+0.09}$ & NH \\
 & LP & $2.55_{-0.01}^{+0.02}$& $0.03\pm0.03$ & $5.69\pm0.03$ & 1.19(117) & $16.12_{-0.09}^{+0.09}$ & 1.9($>0.05$) \\
0727780401 & PL & $2.78\pm0.01$ &\dots & $6.70\pm0.03$ & 1.63(115) & $17.46_{-0.08}^{+0.09}$ & NH \\
 & \textbf{LP} & $2.74\pm0.01$ & $0.13\pm0.03$ & $6.76\pm0.03$ & 1.30(114) & $17.35_{-0.09}^{+0.09}$ & 30.2($\ll$ 0.0001) \\
0790380501S & \textbf{PL} & $2.68\pm0.02$ & \dots& $5.79\pm0.05$ & 0.87(106) & $15.66_{-0.14}^{+0.13}$ & NH \\
 & LP & $2.62\pm0.01$ & $-0.01\pm0.06$ & $5.78\pm0.05$ & 0.88(105) & $15.67_{-0.15}^{+0.15}$ & 0.1($>0.05$) \\
0790380501U & \textbf{PL} & $2.68\pm0.01$ & \dots&$5.37\pm0.03$ & 0.98 (109) & $14.55_{-0.10}^{+0.11}$ & NH \\
 & LP & $2.66\pm0.02$ &$0.06\pm0.05$ & $5.39\pm0.04$ & 0.95(108) & $14.50_{-0.12}^{+0.11}$ & 4.1($4.4\times10^{-02}$) \\
0790380601 & PL & $2.69\pm0.01$ &\dots  & $5.93\pm0.03$ & 1.42(116) & $16.01_{-0.08}^{+0.08}$ & NH \\
 & \textbf{LP} & $2.66\pm0.01$ & $0.10\pm0.04$ & $5.97\pm0.03$ & 1.23(115) & $15.91_{-0.09}^{+0.08}$ & 18.8($\ll$ 0.0001) \\
0790380801 & PL & $2.64\pm0.01$ &\dots& $6.61\pm0.03$ & 1.50(119) & $18.2_{-0.08}^{+0.08}$ & NH \\
 & \textbf{LP} & $2.60\pm0.01$&$0.10\pm0.03$ & $6.65\pm0.03$ & 1.25(118) & $18.08_{-0.08}^{+0.09}$ & 24.2($\ll$ 0.0001) \\
0790380901 & \textbf{PL} & $2.65\pm0.01$ &\dots&$4.44\pm0.02$ & 0.90(118) & $12.17_{-0.07}^{+0.06}$ & NH \\
 & LP & $2.66\pm0.02$ &$-0.03\pm0.03$ & $4.43\pm0.02$ & 0.89(117) & $12.19_{-0.07}^{+0.07}$ & 2.3($>0.05$) \\
0790381401$^*$ & PL & $2.69\pm0.01$ & \dots& $3.80\pm0.02$ & 1.38(117) & $10.25_{-0.05}^{+0.06}$ & NH \\
 & \textbf{LP} & $2.71\pm0.02$ & $-0.08\pm0.04$ & $3.78\pm0.02$ & 1.27(116) & $10.32_{-0.06}^{+0.05}$ & 10.9($1.3\times10^{-03}$) \\
0790381501 & \textbf{PL} & $2.55\pm0.01$ & \dots & $5.89\pm0.02$ & 1.14(121) & $16.85_{-0.07}^{+0.06}$ & NH \\
 & LP & $2.56\pm0.01$ & $-0.03\pm0.03$ & $5.88\pm0.02$ & 1.12(120) & $16.88_{-0.07}^{+0.08}$ & 3.3($>0.05$) \\
0790381001 & \textbf{PL} & $2.68\pm0.01$ & \dots  & $5.89\pm0.02$ & 1.36 (124) & $15.93_{-0.04}^{+0.05}$ & NH \\
 & LP & $2.69\pm0.01$ & $-0.03\pm0.02$ & $5.88\pm0.02$ & 1.32(123) & $15.97_{-0.05}^{+0.05}$ & 5.5($2.0\times10^{-2}$) \\
0727780501 & PL & 2.43 & \dots  & 26.72 & 2.33(124) & 80.78 & NH \\
 & \textbf{LP} & $2.38\pm0.01$ & $0.12\pm0.02$ & $26.87\pm0.07$ & 1.31(123) & $80.20_{-0.25}^{+0.24}$ & 98.0($\ll$ 0.0001) \\
0810830101 & PL & $2.63\pm0.01$ & \dots&$10.19\pm0.03$ & 1.50 (122) & $28.10_{-0.10}^{+0.11}$ & NH \\
 & \textbf{LP} & $2.61\pm0.01$ & $0.07\pm0.02$ & $10.24\pm0.04$ & 1.30(121) & $27.99_{-0.11}^{+0.11}$ & 19.4($\ll$ 0.0001) \\
0810830201 & \textbf{PL} & $2.51\pm0.01$ & \dots& $12.02\pm0.04$ & 0.95 (123) & $35.03_{-0.13}^{+0.12}$ & NH \\
 & LP & $2.51\pm0.01$ &$0.004\pm0.022$ & $12.03\pm0.04$ & 0.96(122) & $35.02_{-0.13}^{+0.13}$ & 0.1($>0.05$)\\
0810830801 & \textbf{PL} & $2.59\pm0.01$ &\dots & $9.66\pm0.03$ & 0.96 (123) & $27.13_{-0.10}^{+0.10}$ & NH \\
 & LP & $2.59\pm0.01$ & $0.01\pm0.02$ & $9.67\pm0.03$ & 0.96(122) & $27.11_{-0.11}^{+0.11}$ & 0.6($>0.05$) \\
0810831201 & PL & $2.15\pm0.01$ & \dots& $22.84\pm0.05$ & 1.52 (126) & $81.24_{-0.21}^{+0.21}$ & NH \\
 & \textbf{LP} & $2.13\pm0.01$ & $0.04\pm0.01$ & $22.88\pm0.06$ & 1.31(125) & $81.02_{-0.22}^{+0.22}$ & 21.7($\ll$ 0.0001) \\
0902112101 & PL & $2.45\pm0.01$ & \dots& $23.07\pm0.08$ & 1.44 (122) & $69.07_{-0.27}^{+0.26}$ & NH \\
 & \textbf{LP} & $2.42\pm0.01$ & $0.09\pm0.02$ & $23.17\pm0.09$ & 1.16(121) & $68.71_{-0.29}^{+0.28}$ & 30.8($\ll$ 0.0001) \\
0923770201 & PL & 2.43 & \dots& 34.51 & 4.57 (126) & 104.39 & NH \\
 & \textbf{LP} & $2.40\pm0.0$ & $0.09\pm0.01$ & $34.7\pm0.04$ & 1.99(125) & $103.84_{-0.15}^{+0.15}$ & 164.7($\ll$ 0.0001) \\
0810831401 & PL & $2.50\pm0.01$ & \dots& $16.45\pm0.04$ & 1.14(124) & $48.20_{-0.15}^{+0.14}$ & NH \\
 & \textbf{LP} & $2.48\pm0.01$ & $0.04\pm0.02$ & $16.48\pm0.05$ & 1.07(123) & $48.08_{-0.15}^{+0.16}$ & 9.6($2.4\times10^{-3}$) \\
0810831601 & \textbf{PL} & $2.69\pm0.01$ &\dots  & $6.25\pm0.03$ & 0.96(119) & $16.88_{-0.08}^{+0.08}$ & NH \\
 & LP & $2.68\pm0.01$ & $0.03\pm0.03$ & $6.26\pm0.03$ & 0.95(118) & $16.85_{-0.09}^{+0.08}$ & 2.3($>0.05$) \\
 0810831801 & \textbf{PL} & $2.65\pm0.01$ &\dots  & $10.04\pm0.04$ & 0.95(120) & $27.50_{-0.11}^{+0.11}$ & NH \\
 & LP & $2.64\pm0.01$ & $0.03\pm0.03$ & $10.05\pm0.04$ & 0.93(119) & $27.45_{-0.13}^{+0.12}$ & 3.7($>0.05$) \\
\enddata
\tablecomments{(1) : Observation IDs. (2) : PL--Absorbed power-law, LP--Absorbed Log Parabolic models. (3) : $\Gamma$--PL photon index, $\alpha$--LP local photon index (at 1 keV). (4): $\beta$--LP curvature parameter. (5) : $N$--Normalization constant for models in units of $10^{-3}$ photons keV cm$^{-2}$. (6) : Reduced chi-square and degrees of freedom of each fit. (7) : Spectral flux in units of  $10^{-12}$ erg cm$^{-2}$ s$^{-1}$  in the energy range 0.6--7.0 keV. (8) : Obtained F-test values and null hypothesis (NH) probability value;  PL adequately describing an X-ray spectra is the NH.\\
Note that errors are estimated only when $\chi_r^2\leq2$  and are at 90$\%$ confidence; errors $\leq$ 0.01 are listed as 0.01. Best-fitting models are highlighted in Col.(2).
}
\end{deluxetable*}

\section{Results}\label{sec4}
 Several previous papers have utilized some of the \textit{XMM-Newton} data to analyze the X-ray spectra or to fit X-ray and Optical/UV data together. For instance, \cite{Mas08} analyzed \textit{MOS} spectra of the very first observation of PG 1553+113 (Obs ID: 0094380801) and described its X-ray spectra in the energy range $0.5-10$ keV with a PL model having $\Gamma\simeq2.09$, while \cite{Rai15} described combined \textit{PN+MOS} spectra of Obs ID: 0722780101 in the energy range $0.3-12$ keV with a LP model having $\alpha\simeq2.283$ and $\beta\simeq0.166$. \cite{Rai17} analyzed combined \textit{PN+MOS} spectra of six observations between July 29$-$Sep 1, 2015, in the energy range of $0.3-12$ keV and found that LP models with parameters $\alpha\simeq2.476-2.521$ and $\beta\simeq0.057-0.151$, and PL models having free absorption, with parameters in the range n$_H\simeq4.57-5.65 \times 10^{20} {\rm cm}^{-2}$ and $\Gamma\simeq2.551-2.594$, described the spectra best. Lastly, \cite{Mid23}, performed a spectral analysis of a single observation, Obs ID: 0902112101, in the energy range of $0.3-10$ keV using PN data alone, and found that a LP model with parameters $\alpha\simeq2.50, \beta\simeq0.13$ best described that spectrum. The other spectra of this source are less explored. So here we present spectral analyses of 21 of them for the first time. We now present the results of our analyses, as described in \autoref{sec3}.

\subsection{Spectral Fitting results}
We fitted PL and LP models to the X-ray spectra in the energy range of $0.6-7.0$ keV. In cases where a significant negative curvature was obtained upon fitting the LP model, we attempted to fit the BPL model and conducted further studies on it, as discussed in \autoref{sec3.3}. Sample X-ray spectral plots for the best-fit PL and LP models are shown in \autoref{fig1}, while the remaining X-ray spectral plots are provided in the appendix. The results of spectral fitting on 30 X-ray spectra are reported in \autoref{tab2}. Out of the  30 X-ray spectra, we found  14 of them to be best fitted with LP models with local photon index (at 1 keV), $\alpha$, in the range $2.13-2.80$, and curvature parameter $\beta$ in the range $0.04-0.18$. One observation (Obs ID 0790381401) favored LP with negative curvature ($\beta\simeq-0.08\pm0.04$). The PL  model was sufficient to describe the X-ray spectra of 15 of the other spectra. The photon energy index $\Gamma$ of the PLs took values in the range of $2.53-2.69$.

 For the LP best-fit spectra, we attempted to search for correlations between different parameters using linear Pearson's correlations. We would say that the correlation is significant if the associated p-value, $p_{r,lin} \leq 0.05$. However, we found no correlation between $\alpha$ and $\beta$, or $\beta$ and spectral flux in the energy range $0.6-7.0$ keV ($F_{0.6-7}$).  The lack of correlation between $\beta$ and $F_{0.6-7}$ signifies that the spectral curvature does not show any systematic dependence on the source flux. This is possibly due to competing effects: stronger cooling comes in at high flux states and tends to increase $\beta$; however, efficient acceleration at high flux states tends to decrease it \citep{Kap16a}. These offsetting effects can yield no significant correlation in either direction.

A linear correlation between $\alpha$ and $\beta$ is expected under a simple statistical acceleration scenario in which the probability of a particle getting accelerated is a decreasing function of energy, leading to a log-parabolic electron energy distribution \citep[e.g.,][]{Mas04}. However, for the 14 spectra best-fit by LP (with $\beta>0$) we find no significant $\alpha$–$\beta$ correlation ($r_{\rm lin}=0.146$, $p=0.618$). The absence of any such correlation suggests that the spectral curvature is not governed solely by a simple energy-dependent acceleration scenario. Log parabolic spectra can also arise in the framework of stochastic (second order Fermi) acceleration through momentum diffusion \citep{Tra09}. The combined operation of statistical and stochastic acceleration can cause additional scatter in the $\alpha-\beta$ relation, thereby diluting any intrinsic linear correlation between them \citep[e.g.,][]{Kap16a,Kap20,Kir23}. Further, when acceleration and radiative cooling timescales become comparable, the spectral slope and curvature evolve under competing influences, which can further reduce the strength of a simple linear relation between $\alpha$ and $\beta$. \cite{Kir20} attribute a weak, insignificant correlation they obtained between the $\alpha$ and $\beta$ parameters for the HBL, H 2356$-$309, to additional acceleration mechanisms and cooling processes.

We did find a strong negative correlation between $\alpha$  and  $F_{0.6-7}$ with correlation coefficient, $r_{lin}=-0.744$ and associated $p_{r,lin}=0.002$. This suggests that for PG 1553+113, the spectrum becomes harder as the flux increases. This behavior is common among HBLs. It implies either more rapid variability in hard X-rays than in soft X-rays, or the injection of fresh electrons with a harder energy distribution than the previously cooled population \citep{Mas02}. 

For the X-ray spectra fitted best by LP models, we attempted to obtain synchrotron peak energies as described in \autoref{sec3.1}. However, we found that the peak energies for every one of those X-ray spectra fell below 0.2 keV, the \textit{EPIC PN} lower energy limit.

\subsection{Joint OM+PN fitting results}
Since the peak energy, $E_p$, falls below 0.2 keV, we fit the LP$_R$ model to the combined \textit{PN + OM} data after applying a 3$\%$ systematic error to the \textit{OM} data. Two observations (Obs IDs: 0094380801 and 0902112101) that had only one \textit{OM} filter data point were excluded from this study. Almost all observations, except that of Obs ID: 0790381001, provided a moderate to good spectral fit ($\chi_r^2\simeq 1.04-2.25$) to the combined \textit{PN + OM} data. A sample optical/UV and X-ray spectrum fitted with LP$_R$, along with its ratio, is shown in \autoref{fig2}. Since the LP$_R$ model seems to fit the combined data, we attempted to estimate the synchrotron peak and the isotropic peak luminosity by first fitting with the ELP$_R$ (\textit{tbabs*redden*eplogpar}) model and then using \autoref{eq4} to \autoref{eq10}. Spectral results on fitting both these models, as well as peak energy in the rest frame ($E_{P,R}$ and isotropic peak luminosity ($L_{P,R}$), are reported in \autoref{tab3}. For almost 20 observations, we were able to estimate the synchrotron peak energy, which was found to lie mainly in the UV regime, with $ E_{p,R}\simeq 4.59-48.41$ eV. Previously \cite{Rai15} fitted SEDs of this source using data from \textit{XMM-Newton},  \textit{Whole Earth Blazar Telescope(WEBT)}, \textit{Swift}, and \textit{MAGIC} with a log-cubic fit and suggested a synchrotron peak around $8-33$ eV. This aligns with the values we obtained. For the remaining observations, it was challenging to constrain the peak energies as they appeared to be near the lower energy threshold of the OM instrument. This suggests that, in these epochs, the synchrotron peak is likely located near the low-energy edge of the OM coverage. i.e., within the optical V band or slightly redder, making it difficult to constrain $E_{p,R}$ with OM data alone.  This aligns with the fact that this blazar was indeed known to be an optically selected, and optically bright, BL Lac object since the beginning of the 1990s \citep{Fal90}. Notably, the V band, with an effective wavelength and bandwidth of 543 nm and 70 nm, corresponds to an energy range of roughly $2.02-2.28$ eV,  which limits sensitivity to the synchrotron peak at such low energies. 

In the previous studies \citep[e.g.,][]{Tra07,Mas08,Dev25} on X-ray spectra alone, a positive correlation between synchrotron peak energy $E_p$ and peak height $S_p$ signifying synchrotron emission, and an anti-correlation between curvature parameter, $\beta_{ELP}$, and $E_p$, signifying a statistical/stochastic acceleration mechanism, were reported. However, upon joint fitting of OM and PN data, we did not observe any significant correlations between the parameters. The absence of substantial correlations may be due to the finite number of OM data points and the large spectral gap between OM and X-ray bands, which prevents continuous sampling of the synchrotron curvature profile. Further, though OM extends the spectral coverage and locates the synchrotron peak, this broader coverage reduces the fit's sensitivity to small curvature changes, especially when the peak falls within or near the \textit{OM} band, as in our case. We also tested two additional models on the joint \textit{PN+OM} data. (1) A power law + log parabolic \citep[PLLP;][]{Bha14}, for which a good fit would suggest that the underlying electron energy distribution might consist of a low-energy power law branch that evolves to a curved log parabolic component at higher energies. (2) A broken log parabolic \citep[BLP,][]{Jag18} model, for which a fit would suggest that the escape of electrons from the acceleration region is energy-dependent. However, neither of these models could produce a stable fit with well-constrained parameters for most of the observations.

\begin{longrotatetable}
\begin{deluxetable*}{cccccccccccc}
\tabletypesize{\scriptsize}
\tablewidth{400pt}
\tablecaption{Spectral Properties of PG 1553+113 obtained by joint fitting PN+OM spectra with LP$_R$ and ELP$_R$ models\label{tab3}}
\tablehead{
\colhead{Obs ID} & \colhead{OM filters}&\colhead{{$\alpha_{LP,R}$}} &
\colhead{${\beta_{LP,R}}$} &
\colhead{${\chi_r^2 (dof)}$} &
\colhead{${E_{P,LP,R}}$} &
\colhead{${E_{P,ELP,R}}$} &
\colhead{${\beta_{ELP,R}}$} &
\colhead{${K_{ELP,R}}$} &
\colhead{${S_{P,ELP,R}}$} &
\colhead{${E_{P,R}}$} & \colhead{${L_{P,R}}$}\\
\colhead{} & \colhead{} & \colhead{} & \colhead{} & \colhead{} &\colhead{(eV)} &\colhead{(eV)} &\colhead{} &\colhead{$\times 10^{-3}$}&
\colhead{($\times10^{-12}$ erg cm$^{-2}$ s$^{-1}$)} &\colhead{eV}&\colhead{$(\times 10^{45} erg~s^{-1})$}
}
\colnumbers
\startdata
$0656990101$&1,2,3,4,5,6& $2.77_{-0.01}^{+0.01}$ &$0.21_{-0.01}^{+0.01}$&$1.09(118)$&$14.89$&$14.89_{-0.36}^{+0.36}$&$0.210_{-0.002}^{+0.002}$&$39.692_{-0.379}^{+0.382}$&$63.507_{-0.606}^{+0.611}$&$21.34_{-0.52}^{+0.52}$&$45.951_{-0.439}^{+0.442}$\\
$0727780101$&1,2,3,4,5,6&$2.33_{-0.01}^{+0.01}$&$0.11_{-0.01}^{+0.01}$&$1.72(130)$&$30.57$&$0.93_{-0.91}^{+30.57}$&$0.108_{-0.001}^{+0.001}$&$30.030_{-0.194}^{+0.196}$&$48.048_{-0.310}^{+0.314}$&$43.81_{-1.30}^{+1.33}$&$34.766_{-0.225}^{+0.227}$\\
$0727780201$&1,2,3,4,5,6&$2.36_{-0.01}^{+0.01}$&$0.07_{-0.01}^{+0.01}$&$1.37(128)$&$3.90$&$3.90_{-0.30}^{+0.30}$&$0.075_{-0.002}^{+0.002}$&$26.958_{-0.327}^{+0.334}$&$43.133_{-0.523}^{+0.534}$&$5.60_{-0.44}^{+0.47}$&$31.209_{-0.379}^{+0.387}$\\
$0761100101$&1,2,3,4,5,6&$2.51_{-0.01}^{+0.01}$&$0.09_{-0.01}^{+0.01}$&$1.30(125)$&$<2.00$&$\cdots$&$\cdots$&$\cdots$&$\cdots$&$\cdots$&$\cdots$\\
$0761100201$&1,2,3,4,5,6&$2.50_{-0.01}^{+0.01}$&$0.08_{-0.01}^{+0.01}$&$1.17(125)$&$<2.00$&$\cdots$&$\cdots$&$\cdots$&$\cdots$&$\cdots$&$\cdots$\\
$0761100301$&1,2,3,4,5,6&$2.49_{-0.01}^{+0.01}$&$0.09_{-0.00}^{+0.00}$&$1.25(126)$&$<2.00$&$\cdots$&$\cdots$&$\cdots$&$\cdots$&$\cdots$&$\cdots$\\
$0761100401$&1,2,3,4,5,6&$2.49_{-0.01}^{+0.01}$&$0.09_{-0.01}^{+0.01}$&$1.57(125)$&$<2.00$&$\cdots$&$\cdots$&$\cdots$&$\cdots$&$\cdots$&$\cdots$\\
$0761100701$&1,2,3,4,5,6&$2.54_{-0.01}^{+0.01}$&$0.12_{-0.01}^{+0.01}$&$1.30(129)$&$6.37$&$6.37_{-0.27}^{+0.27}$&$0.123_{-0.001}^{+0.001}$&$21.739_{-0.222}^{+0.225}$&$34.782_{-0.355}^{+0.360}$&$9.13_{-0.39}^{+0.39}$&$25.167_{-0.257}^{+0.260}$\\
$0761101001$&1,2,3,4,5,6&$2.51_{-0.01}^{+0.01}$&$0.12_{-0.01}^{+0.01}$&$1.23(126)$&$6.80$&$0.34_{-0.33}^{+6.80}$&$0.117_{-0.002}^{+0.002}$&$20.410_{-0.207}^{+0.209}$&$32.656_{-0.331}^{+0.334}$&$9.74_{-0.47}^{+0.49}$&$23.629_{-0.240}^{+0.242}$\\
$0727780301$&1,2,3,4,5,6&$2.52_{-0.01}^{+0.01}$&$0.12_{-0.01}^{+0.01}$&$1.40(123)$&$5.94$&$0.35_{-0.34}^{+5.94}$&$0.117_{-0.002}^{+0.002}$&$21.652_{-0.236}^{+0.238}$&$34.643_{-0.378}^{+0.381}$&$8.51_{-0.49}^{+0.50}$&$25.066_{-0.273}^{+0.276}$\\
$0727780401$&1,2,3,4,5,6&$2.72_{-0.01}^{+0.01}$&$0.20_{-0.01}^{+0.01}$&$1.40(120)$&$15.86$&$15.86_{-0.38}^{+0.38}$&$0.199_{-0.002}^{+0.002}$&$30.029_{-0.277}^{+0.279}$&$48.046_{-0.443}^{+0.446}$&$22.73_{-0.54}^{+0.54}$&$34.764_{-0.321}^{+0.323}$\\
$0790380501S$&1,2,3,4,5,6&$2.64_{-0.01}^{+0.01}$&$0.13_{-0.01}^{+0.01}$&$1.04(111)$&$3.73$&$0.34_{-0.33}^{+3.73}$&$0.132_{-0.004}^{+0.004}$&$34.877_{-0.424}^{+0.428}$&$55.803_{-0.678}^{+0.685}$&$5.35_{-0.47}^{+0.49}$&$40.377_{-0.491}^{+0.495}$\\
$0790380501U$&1,2,3,4,5,6&$2.64_{-0.01}^{+0.01}$&$0.13_{-0.01}^{+0.01}$&$1.11(115)$&$3.20$&$0.26_{-0.25}^{+3.20}$&$0.129_{-0.003}^{+0.003}$&$34.882_{-0.445}^{+0.452}$&$55.811_{-0.712}^{+0.723}$&$4.59_{-0.36}^{+0.37}$&$40.383_{-0.515}^{+0.523}$\\
$0790380601$&1,2,3,4,5,6&$2.64_{-0.01}^{+0.01}$&$0.14_{-0.01}^{+0.01}$&$1.25(121)$&$5.27$&$0.27_{-0.26}^{+5.27}$&$0.141_{-0.002}^{+0.002}$&$32.352_{-0.356}^{+0.359}$&$51.763_{-0.570}^{+0.574}$&$7.55_{-0.37}^{+0.39}$&$40.383_{-0.515}^{+0.523}$\\
$0790380801$&1,2,3,4,5,6&$2.59_{-0.01}^{+0.01}$&$0.13_{-0.01}^{+0.01}$&$1.27(124)$&$5.36$&$0.27_{-0.26}^{+5.36}$&$0.131_{-0.002}^{+0.002}$&$31.454_{-0.342}^{+0.346}$&$50.326_{-0.547}^{+0.554}$&$7.68_{-0.37}^{+0.39}$&$36.414_{-0.396}^{+0.401}$\\
$0790380901$&1,2,3,4,5,6&$2.61_{-0.01}^{+0.01}$&$0.11_{-0.01}^{+0.01}$&$1.33(123)$&$<2.00$&$\cdots$&$\cdots$&$\cdots$&$\cdots$&$\cdots$&$\cdots$\\
$0790381401$&1,2,3,4,5,6&$2.65_{-0.01}^{+0.01}$&$0.12_{-0.00}^{+0.00}$&$2.00(122)$&$<2.00$&$\cdots$&$\cdots$&$\cdots$&$\cdots$&$\cdots$&$\cdots$\\
$0790381501$&1,2,3,4,5,6&$2.52_{-0.01}^{+0.01}$&$0.09_{-0.01}^{+0.01}$&$1.76(126)$&$<2.00$&$\cdots$&$\cdots$&$\cdots$&$\cdots$&$\cdots$&$\cdots$\\
$0727780501$&1,2,3,4,5,6&$2.38_{-0.01}^{+0.01}$&$0.13_{-0.01}^{+0.01}$&$1.29(129)$&$33.78$&$0.83_{-0.82}^{+33.78}$&$0.129_{-0.001}^{+0.001}$&$51.090_{-0.326}^{+0.328}$&$81.744_{-0.522}^{+0.525}$&$48.41_{-1.18}^{+1.19}$&$51.090_{-0.326}^{+0.328}$\\
$0810830101$&1,2,3,4,5,6&$2.58_{-0.01}^{+0.01}$&$0.16_{-0.01}^{+0.01}$&$1.56(127)$&$14.29$&$14.29_{-0.40}^{+0.40}$&$0.157_{-0.002}^{+0.002}$&$35.095_{-0.296}^{+0.298}$&$56.152_{-0.474}^{+0.477}$&$20.48_{-0.57}^{+0.57}$&$40.629_{-0.343}^{+0.345}$\\
$0810830201$&1,2,3,4,5,6&$2.46_{-0.01}^{+0.01}$&$0.11_{-0.01}^{+0.01}$&$1.49(128)$&$7.96$&$0.37_{-0.36}^{+7.96}$&$0.111_{-0.002}^{+0.002}$&$37.120_{-0.359}^{+0.363}$&$59.392_{-0.574}^{+0.581}$&$11.41_{-0.52}^{+0.53}$&$42.974_{-0.416}^{+0.420}$\\
$0810830801$&1,2,3,4,5,6&$2.54_{-0.01}^{+0.01}$&$0.13_{-0.01}^{+0.01}$&$1.46(128)$&$8.13$&$0.33_{-0.32}^{+8.13}$&$0.130_{-0.002}^{+0.002}$&$35.865_{-0.346}^{+0.349}$&$57.384_{-0.554}^{+0.558}$&$11.65_{-0.46}^{+0.47}$&$41.521_{-0.401}^{+0.404}$\\
$0810831201$&1,2,3,4,5,6&$2.13_{-0.01}^{+0.01}$&$0.04_{-0.01}^{+0.01}$&$1.27(131)$&$20.26$&$1.93_{-1.83}^{+20.26}$&$0.039_{-0.001}^{+0.001}$&$29.575_{-0.207}^{+0.211}$&$47.320_{-0.331}^{+0.338}$&$29.03_{-2.62}^{+2.77}$&$34.239_{-0.240}^{+0.244}$\\
$0923770201$&1,2,3,4,5,6&2.39&0.12&2.25(131)&22.97&22.97&0.118&72.01&115.22&32.92&83.365\\
$0810831401$&1,2,3,4,5,6&$2.45_{-0.01}^{+0.01}$&$0.13_{-0.01}^{+0.01}$&$1.53(129)$&$17.37$&$0.54_{-0.53}^{+17.37}$&$0.127_{-0.001}^{+0.001}$&$40.883_{-0.313}^{+0.315}$&$65.413_{-0.501}^{+0.504}$&$24.89_{-0.76}^{+0.77}$&$47.330_{-0.362}^{+0.365}$\\
$0810831601$&1,2,3,4,5,6&$2.64_{-0.01}^{+0.01}$&$0.14_{-0.01}^{+0.01}$&$1.18(124)$&$4.71$&$4.71_{-0.24}^{+0.24}$&$0.137_{-0.002}^{+0.002}$&$34.760_{-0.396}^{+0.401}$&$55.616_{-0.634}^{+0.642}$&$6.75_{-0.34}^{+0.34}$&$40.241_{-0.458}^{+0.464}$\\
$0810831801$&1,2,3,4,5,6&$2.60_{-0.01}^{+0.01}$&$0.14_{-0.01}^{+0.01}$&$1.26(125)$&$7.28$&$7.28_{-0.30}^{+0.30}$&$0.140_{-0.002}^{+0.002}$&$44.139_{-0.446}^{+0.449}$&$70.622_{-0.714}^{+0.718}$&$10.43_{-0.43}^{+0.43}$&$51.099_{-0.516}^{+0.520}$\\
\enddata
\tablecomments{ (1) : Observation IDs. Note that we neglected  Obs ID: 0790381001 which has ($\chi_r^2>30$). (2) : OM filters used for fitting. 1-UVW2, 2-UVM2, 3-UVW1 ,4-U, 5-V ,6-B. 
(3) : Local photon index at 1 keV obtained on fitting LP$_{R}$ model on combined \textit{EPIC-PN} + \textit{OM} data. (4) : Curvature parameter obtained by the LP$_{R}$ fit. (5) : Reduced chi-square and degrees of freedom obtained on fitting the $ELP_R$ model on the combined \textit{EPIC-PN} + \textit{OM} data. (6) : synchrotron peak energy in eV calculated with \autoref{eq4} using  $\alpha_{LP,R}$, and $\beta_{LP,R}$ values. (7) : Synchrotron peak energy in eV obtained by fitting the $ELP_R$ model on the combined \textit{EPIC-PN} + \textit{OM} data. (8) : Curvature parameter obtained by the $ELP_R$ fit. (9) :  Flux in $\nu F_{\nu}$ units at peak energy $E_{P,ELP,R}$ obtained from the $ELP_R$ fit. (10): SED peak value obtained by modifying \autoref{eq7} in terms of the ELP$_R$ model. (11) : Peak energy in rest frame obtained by modifying \autoref{eq8} in terms of  ELP$_R$ model. (12) : Isotropic peak luminosity calculated in the rest frame by modifying \autoref{eq9} in terms of the ELP$_R$ model.}
\end{deluxetable*}
\end{longrotatetable}

\begin{deluxetable*}{cccc|c}
\tabletypesize{\scriptsize}
\tablewidth{400pt}
\tablecaption{Spectral results of joint PN+OM data using LP  model with free Galactic absorption and reddening\label{tab4}}
\tablehead{
\colhead{Obs ID$^*$} & \colhead{${n_{H}}$} &
\colhead{{E(B-V)}} &
\colhead{{$\chi^2_r$ (dof)}} &
\colhead{{$r/r_{MW}$}} \\
\nocolhead{}&\colhead{$10^{20} {\rm cm}^{-2}$} & \colhead{10$^{-2}$ mag} & \colhead{} & \colhead{} 
}
\colnumbers
\startdata
$0656990101$ & $4.16 \pm 0.92$ & $2.31 \pm 0.43$ & $0.89(116)$ & $1.13$ \\
$0727780101$ & $3.33 \pm 0.90$ & $4.26 \pm 0.22$ & $1.48(128)$ & $2.60$ \\
$0727780201$ & $2.35 \pm 0.89$ & $3.76 \pm 0.30$ & $1.25(126)$ & $3.24$ \\
$0761100101$ & $2.71 \pm 0.87$ & $3.15 \pm 0.36$ & $1.22(123)$ & $2.36$ \\
$0761100201$ & $2.55 \pm 0.88$ & $2.83 \pm 0.38$ & $1.04(123)$ & $2.25$ \\
$0761100301$ & $2.54 \pm 0.88$ & $2.62 \pm 0.36$ & $1.06(124)$ & $2.09$ \\
$0761100401$ & $1.71 \pm 0.88$ & $2.11 \pm 0.35$ & $1.11(123)$ & $2.50$ \\
$0761100701$ & $4.20 \pm 0.88$ & $2.72 \pm 0.24$ & $1.03(127)$ & $1.32$ \\
$0761101001$ & $3.29 \pm 0.88$ & $2.82 \pm 0.35$ & $1.13(124)$ & $1.74$ \\
$0727780301$ & $2.22 \pm 0.92$ & $2.42 \pm 0.41$ & $1.18(121)$ & $2.21$ \\
$0727780401$ & $4.25 \pm 0.92$ & $2.73 \pm 0.40$ & $1.30(118)$ & $1.30$ \\
$0790380501S$ & $2.69 \pm 0.94$ & $2.13 \pm 0.77$ & $0.93(109)$ & $1.60$ \\
$0790380501U$ & $2.77 \pm 0.91$ & $2.99 \pm 0.61$ & $0.98(112)$ & $2.19$ \\
$0790380601$ & $3.55 \pm 0.89$ & $3.35 \pm 0.43$ & $1.24(119)$ & $1.92$ \\
$0790380801$ & $3.71 \pm 0.88$ & $3.49 \pm 0.38$ & $1.27(122)$ & $1.91$ \\
$0790380901$ & $1.82 \pm 0.87$ & $2.00 \pm 0.44$ & $0.97(121)$ & $2.23$ \\
$0790381401$ & $1.59 \pm 0.88$ & $1.55 \pm 0.45$ & $1.49(120)$ & $1.97$ \\
$0790381501$ & $1.26 \pm 0.86$ & $2.27 \pm 0.33$ & $1.28(124)$ & $3.66$ \\
$0727780501$ & $5.06 \pm 0.88$ & $3.43 \pm 0.24$ & $1.28(127)$ & $1.38$ \\
$0810830101$ & $2.79 \pm 0.87$ & $2.34 \pm 0.29$ & $1.16(125)$ & $1.70$ \\
$0810830201$ & $2.65 \pm 0.88$ & $2.14 \pm 0.29$ & $0.95(126)$ & $1.63$ \\
$0810830801$ & $3.30 \pm 0.88$ & $2.09 \pm 0.30$ & $0.96(126)$ & $1.28$ \\
$0810831201$ & $4.04 \pm 0.89$ & $3.52 \pm 0.19$ & $1.29(129)$ & $1.77$ \\
$0923770201$ & $3.48 \pm 0.83$ & $3.06 \pm 0.11$ & $1.72(129)$ & $1.78$ \\
$0810831401$ & $3.24 \pm 0.88$ & $2.30 \pm 0.25$ & $0.97(127)$ & $1.44$ \\
$0810831601$ & $3.23 \pm 0.89$ & $2.38 \pm 0.39$ & $0.98(122)$ & $1.49$ \\
$0810831801$ & $4.19 \pm 0.89$ & $2.34 \pm 0.34$ & $0.96(123)$ & $1.13$
\enddata
\tablecomments{
(1):Observation ID. (2): Galactic absorption value obtained in units of 10$^{20}$\,cm$^{-2}$.  (3): Galactic reddening value obtained in units of 10$^{-2}$ mag. (4): Reduced chi-square and degrees of freedom of respective fits. (5): Gas to dust ratio compared to that of the Milky Way galaxy; $r_{MW}=4.93\times10^{21}\,{\rm cm}^{-2}\,{\rm mag}^{-1}$}.\\
$^*$Obs ID 0790381001 ($\chi_r^2>30$ for the LP model) is neglected for this study.
\end{deluxetable*}

Previously \cite{Rai17} performed joint \textit{EPIC+OM} spectral fitting of six of these observations (Obs IDs: 0761100101,0761100201, 0761100301, 0761100401, 0761100701, 0761101001) in a fashion similar to ours. However, unlike our approach, they did not give any systematic errors to the \textit{OM} data. On analyzing the six joint spectra, they suggested a gas/dust ratio $r=n_{H}/E(B-V)$ larger than that of the Milky Way \citep[r$_{MW}=4.93 \times 10^{21}\hspace{0.1cm}{\rm cm}^{-2}\hspace{0.1cm} {\rm mag}^{-1}$;][]{Dip94}. We have extended this study to 27 of our observations. Like \cite{Rai17}, we found that the reduced chi-square improves greatly when both the reddening parameter E(B-V) and galactic absorption $n_H$ are set free. These results are reported in \autoref{tab4}. These analyses yield gas-to-dust ratios ranging from 1.13 to 3.66 times that of our Galaxy. 

\subsection{Results from the Search for Signatures of the IC Component}
On fitting X-ray spectra of 30 EPIC \textit{PN} spectra with LP models, significant negative curvatures ($|\beta|>\beta_{err}$) were exhibited by a few of them, as can be seen from \autoref{tab2}. Additionally, we investigated any potential breaks in the entire energy range of 0.6--10 keV, for all 30 observations, by fitting all of their four {\it EPIC PN}, {\it EPIC MOS1}, {\it EPIC MOS2}, and joint {PN+MOS} X-ray spectra with LP and BPL models. We identified seven candidates with breaks in their X-ray spectra, suggesting possible spectral hardening. These are Obs IDs 0761100401, 0790380501S, 0790380901, 0810831601, 0790381401, 0790381501, and 0790381001. Among these, the first four listed showed spectral hardening only in one data set, either \textit{PN}, \textit{MOS1}, or \textit{MOS2}, and failed to show it in the other three data sets. Thus, these four observations with inconclusive evidence are not considered further. The remaining three candidate observations (IDs 0790381401, 0790381501, and 0790381001) were observed in February 2017 and showed possible spectral hardening in \textit{PN, MOS1, MOS2,} and joint \textit{PN-MOS} data, and will therefore be investigated further. The higher energy limit of each of these spectra is decided by how much background contaminates the source count at these energies, and their spectral results together with the studied energy range are reported in \autoref{tab5}. Below, we shall discuss each of these three observations in detail.

\begin{figure*}
{\vspace{-0.5cm} \includegraphics[width=8.5cm, height=7.5cm]{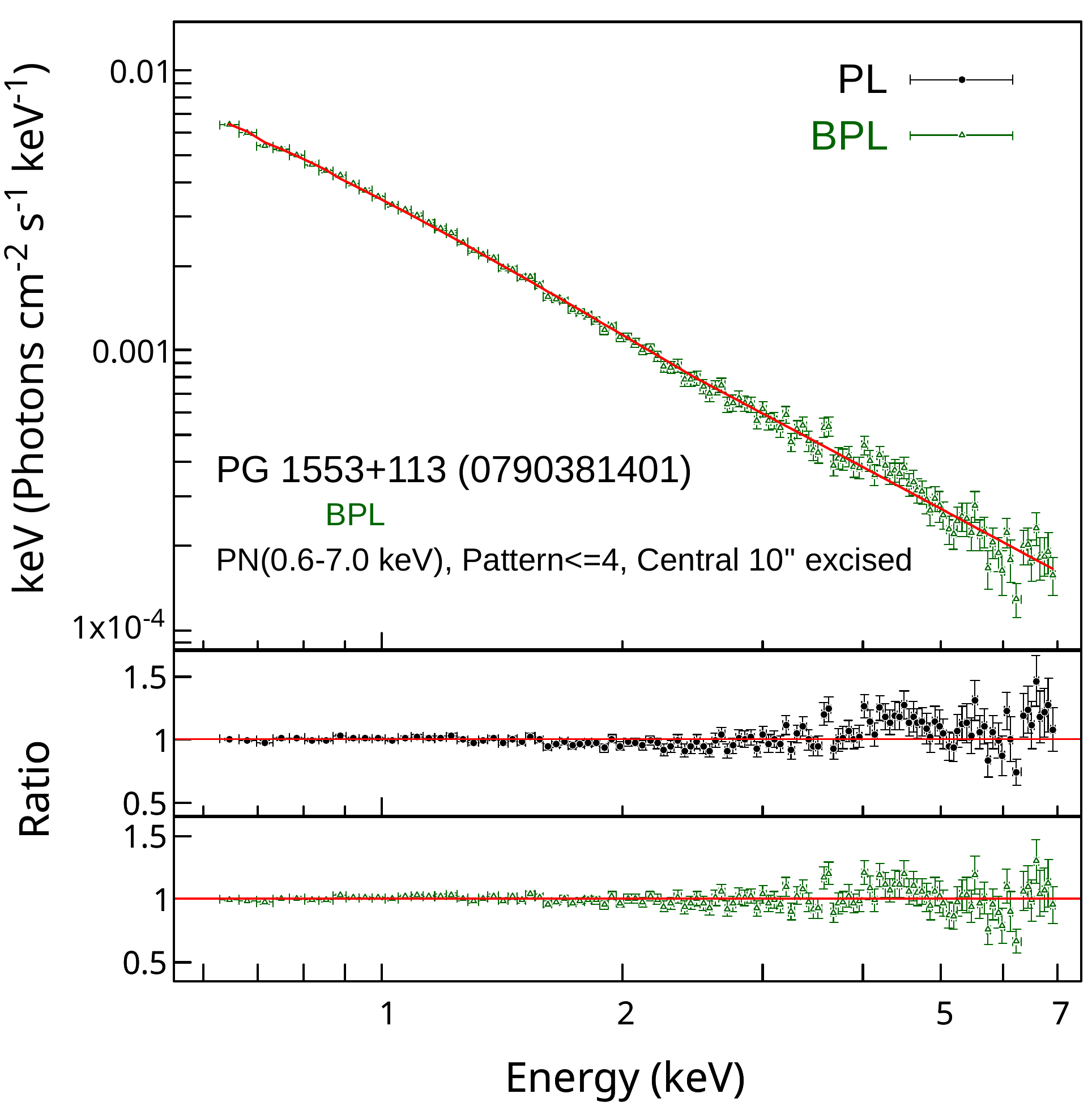}}
\includegraphics[width=8.5cm, height=7.5cm]{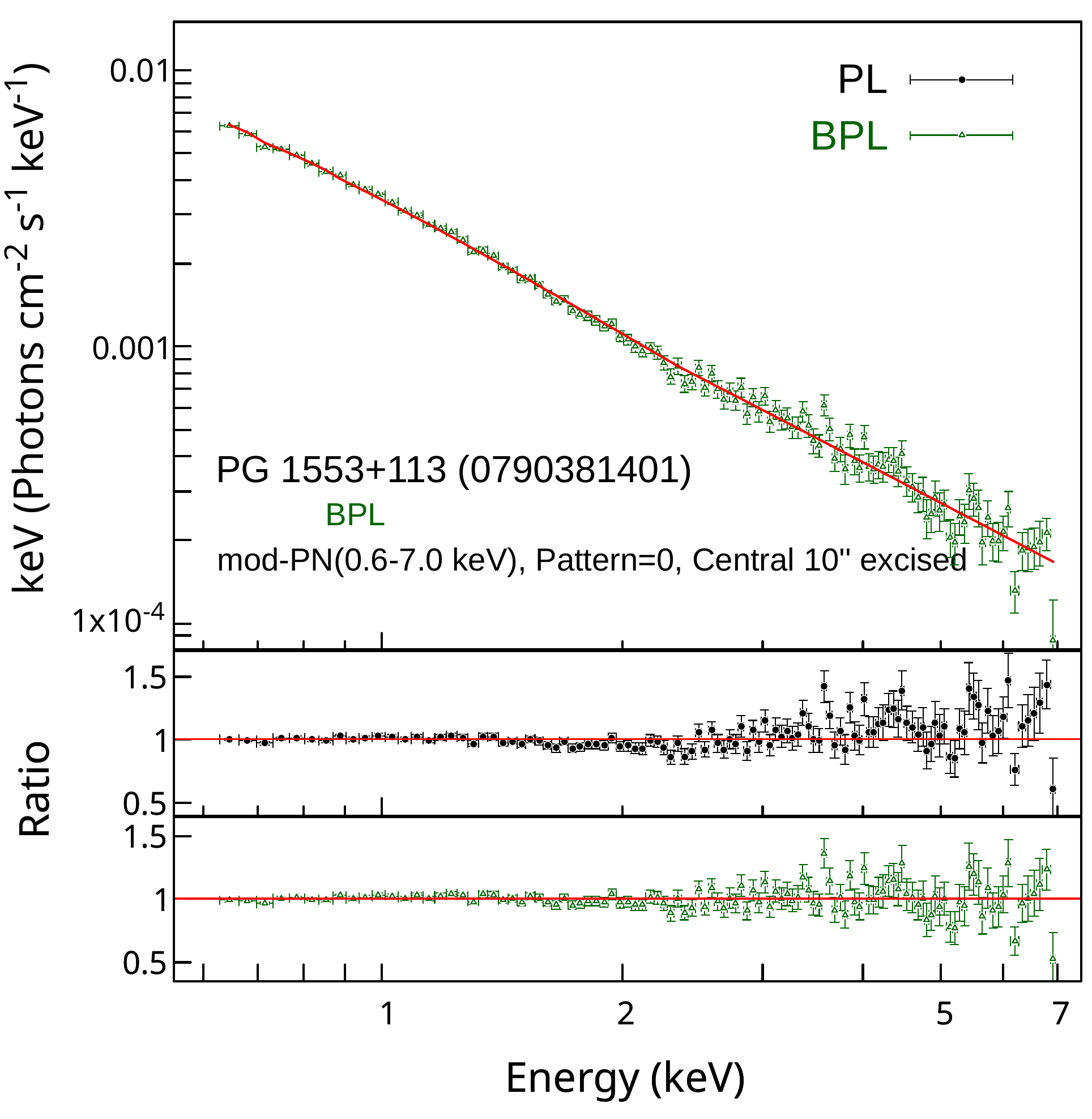}
{\vspace{-0.17cm} \includegraphics[width=8.5cm, height=7.5cm]{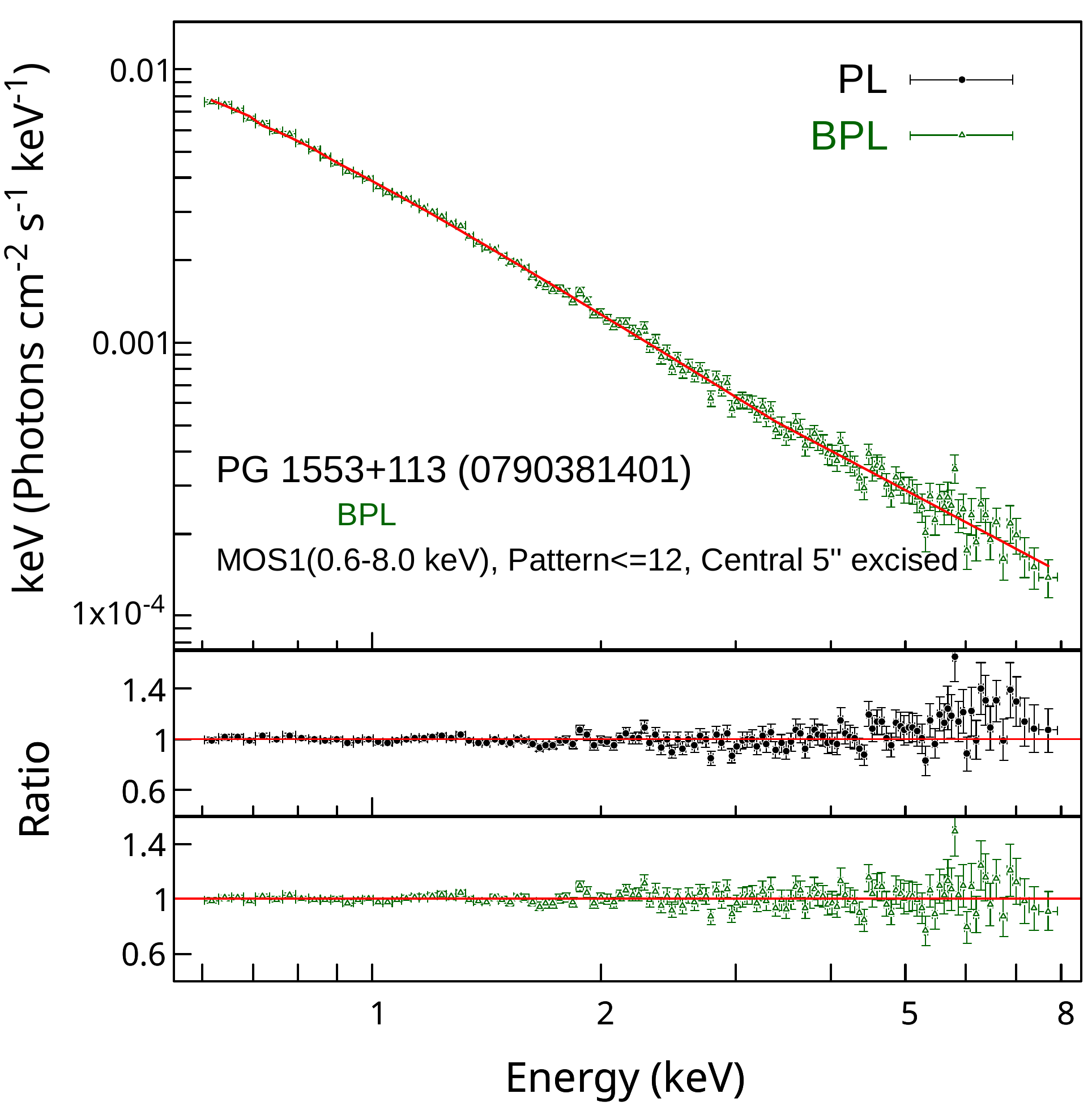}}
\includegraphics[width=8.5cm, height=7.5cm]{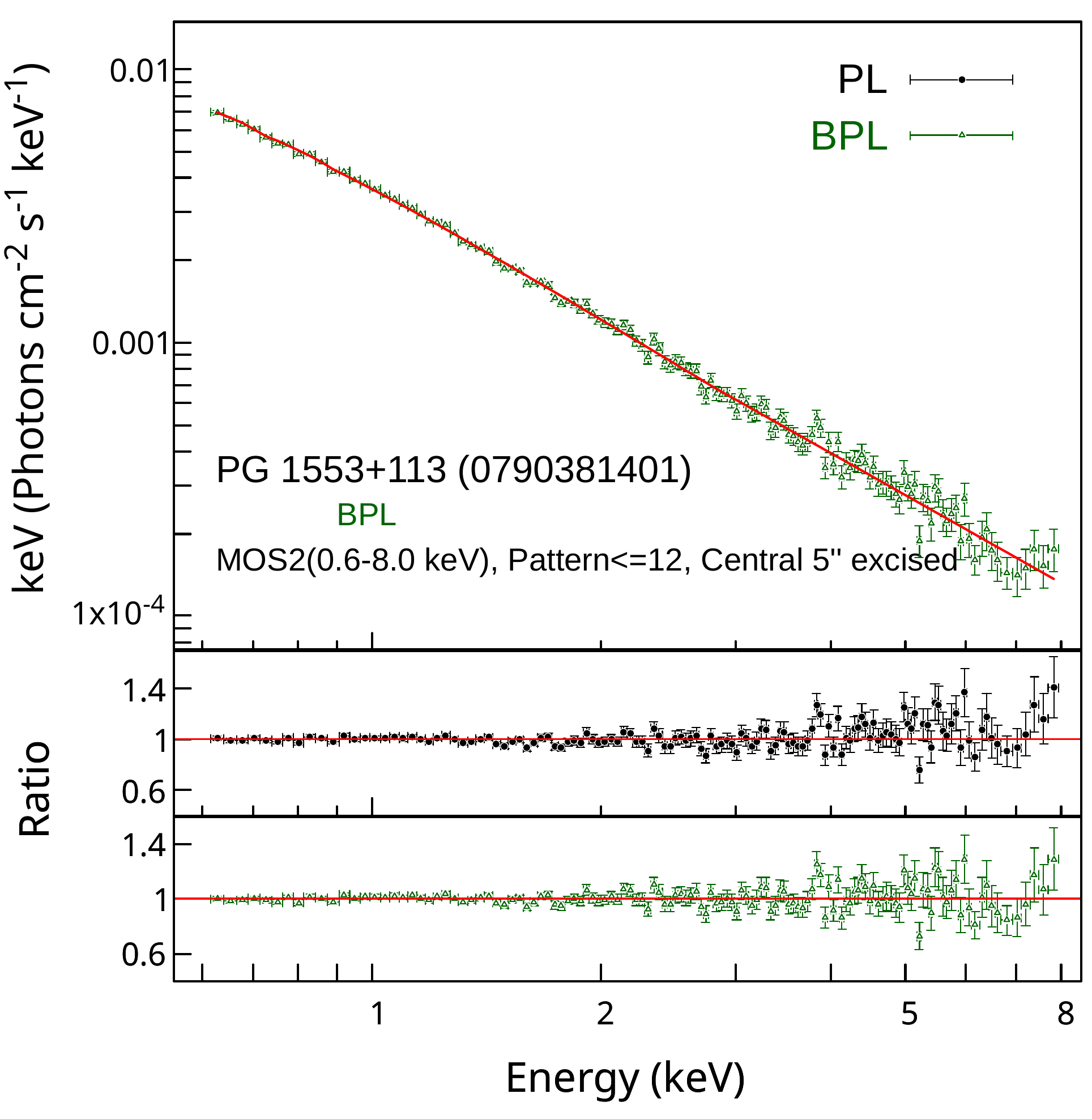}
{\vspace{-0.17cm} 
{\hspace{0.2cm}\includegraphics[width=8.5cm, height=7.5cm]{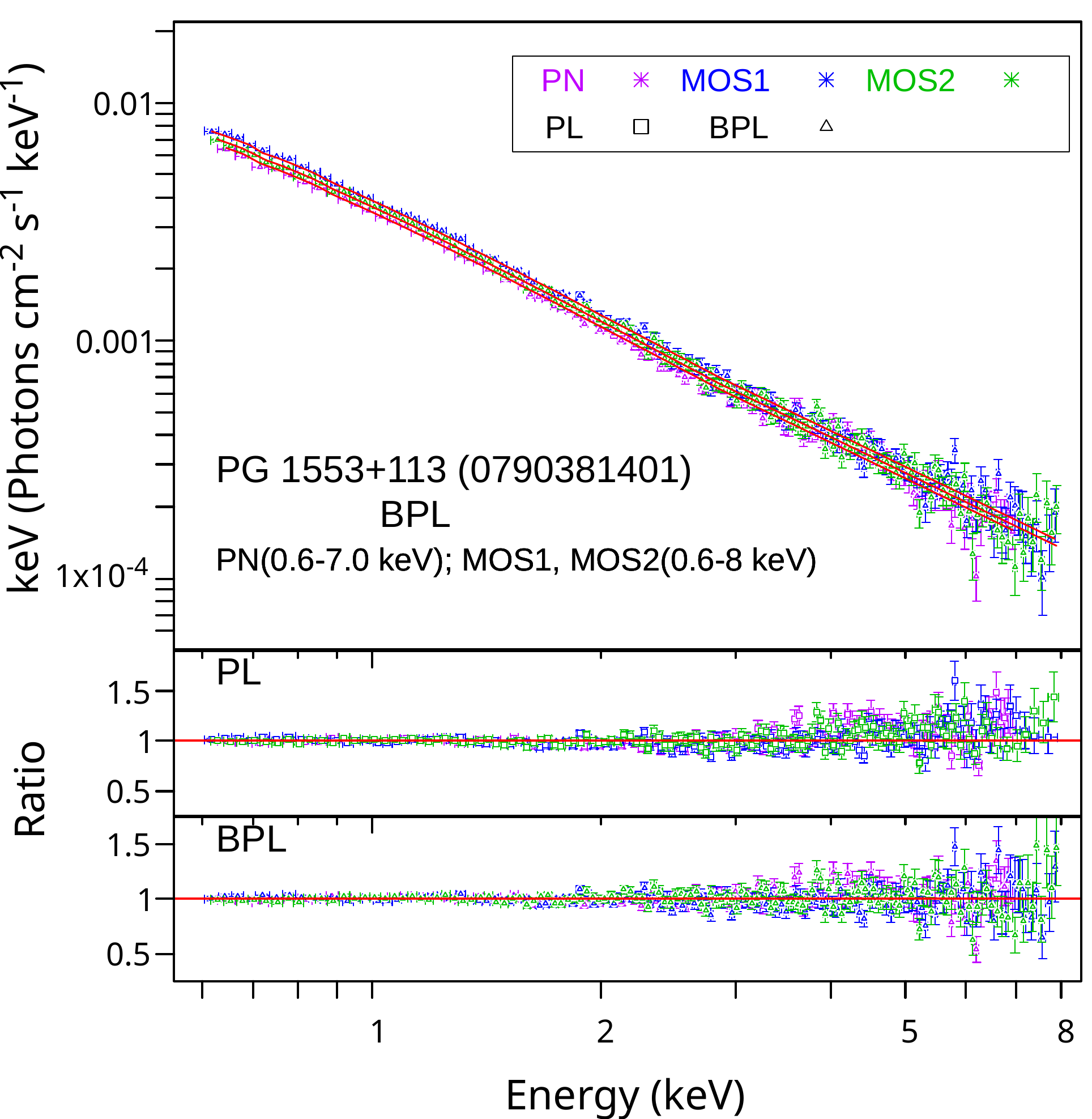}}
{\hspace{1cm}\includegraphics[width=8.5cm, height=7.7cm]{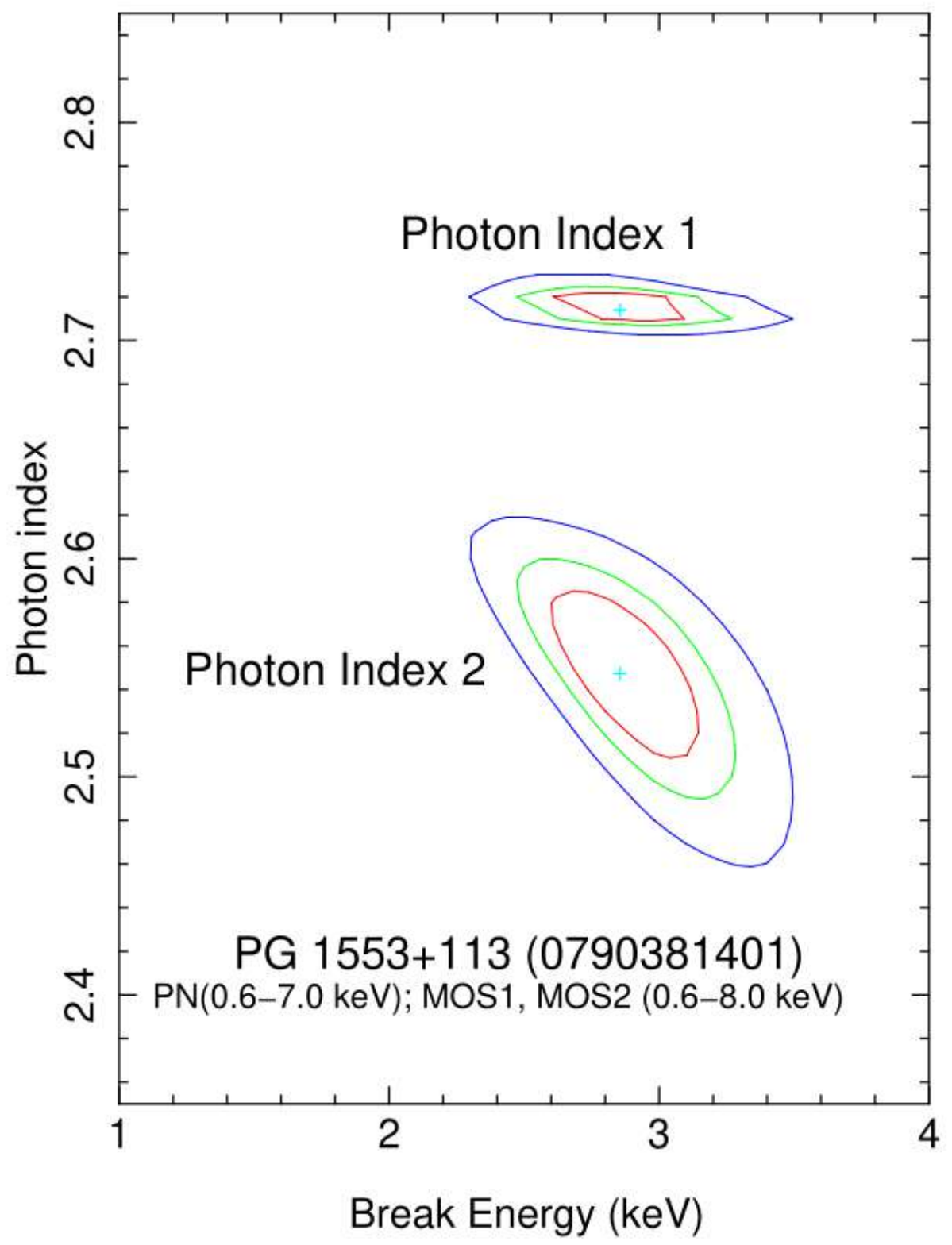}}}
\caption{ X-ray spectral fit plots and contour plots for Obs ID: 0790381401. In the top four plots, the PL and BPL models are represented by black-filled circles and green-filled triangles, respectively. Observation ID, source name, EPIC camera, energy range, patterns used, and the central region, if excised, are displayed on each plot. The bottom left plot
is a joint spectral fit using data from all  3 EPIC cameras. In this plot, PL and BPL model fits are shown by squares and triangles, respectively, with \textit{PN}, \textit{MOS1}, \textit{MOS2} data points, respectively, shown in dark-magenta, blue, and green colors. The last plot is broken power law model confidence contour (Chi-squared) plots at 68$\%$ (red),  90$\%$ (green), and 99$\%$ (blue) confidence for photon index 1, photon index 2, and break energy. A similar figure for the other observations, which we tested for the signature of the IC component, is in the appendix \autoref{A3}.\label{fig3}} 
\end{figure*}

The observation with ID 0790381401 (hereafter, Obs A) was conducted on 13 Feb 2017 and showed a significant negative curvature $|\beta/\beta_{err}|\simeq2$ on fitting {\it PN} spectra with an LP model. A PL fit gives a high energy tail as can be seen in the top left plot of \autoref{fig3}, which gets rectified by a BPL fit. The \textit{mod-PN} X-ray spectra generated by taking single events alone also showed significant negative curvature, and a BPL model provided a stable spectral fit. The observation yielded similar results for {\it MOS1}, {\it MOS2}, and joint{\it PN+MOS} fits and their spectral results are reported in \autoref{tab5}.  It is to be noted that {\it PN} and {\it MOS1, MOS2} spectra were studied in energy ranges of 0.6--7.0 and 0.6--8.0 keV, respectively. A BPL model fit to joint {\it PN+MOS} spectra resulted in a stable fit with $\Gamma_1\simeq2.72$, $\Gamma_2\simeq2.55$, and $E_{break}\simeq2.46$ keV, indicating a spectral hardening of $\Delta \Gamma \simeq 0.17$.
The confidence contour plots generated from the BPL fit to the same indicate that they are independent at the 99$\%$ confidence level. These spectral fits, as well as contour plots, are displayed in \autoref{fig3}.  From \autoref{tab5}, we see that the reduced chi-square significantly improves when we go from PL to BPL for all four datasets. The BPL parameters of \textit{mod-PN, MOS1, MOS2,} and \textit{PN} also agree with those of the joint \textit{PN+MOS} fit within the error limits.

\begin{deluxetable*}{ccccccc}
\tabletypesize{\scriptsize}
\tablewidth{400pt}
\tablecaption{Spectral parameters for the  observations that we search for an IC component.\label{tab5}}
\tablehead{
\colhead{Obs ID} & \colhead{Camera} &
\colhead{$\Gamma/\alpha/\Gamma_{1}$}&
\colhead{$E_{break}(keV)$}&\colhead{$\beta/\Gamma_2$}&\colhead{Energy(keV)}&\colhead{${\chi^2_r(dof)}$}
}
\colnumbers
\startdata
% &&Power Law model (\textit{tbabs*pow})&&\\
\multicolumn{7}{c}{Power$\,$Law$\,$model$\,$(\textit{tbabs*pow})}\\
\hline
$0790381401$&PN&$2.69\pm0.01$&$\cdots$&$\cdots$&0.6-7.0&1.38(117)\\
&mod-PN&$2.69\pm0.01$&$\cdots$&$\cdots$&0.6-7.0&1.48(114)\\
&MOS1&$2.72\pm0.01$&$\cdots$&$\cdots$&0.6-8.0&1.09(135)\\
&MOS2&$2.69\pm0.01$&$\cdots$&$\cdots$&0.6-8.0&1.16(139)\\
&PN+MOS1+MOS2$^*$&$2.70\pm0.01$&$\cdots$&$\cdots$&0.6-7, 0.6-8, 0.6-8&1.23(393)\\
$0790381501$&PN&$2.55\pm0.01$&$\cdots$&$\cdots$&0.6-8.0&1.10(134)\\
&mod-PN&$2.55\pm0.01$&$\cdots$&$\cdots$&0.6-8.0&1.16(129)\\
&MOS1&$2.58\pm0.01$&$\cdots$&$\cdots$&0.6-10.0&1.53(148)\\
&MOS2&$2.55\pm0.01$&$\cdots$&$\cdots$&0.6-10.0&0.96(145)\\
&PN+MOS1+MOS2&$2.56\pm0.01$&$\cdots$&$\cdots$&0.6-8, 0.6-10, 0.6-10&1.25(429)\\
$0790381001$ &PN&$2.68\pm0.01$&$\cdots$&$\cdots$&0.6-8.0&1.36(137)\\
&mod-PN&$2.69\pm0.01$&$\cdots$&$\cdots$&0.6-8.0&1.38(131)\\
&MOS1&$2.70\pm0.01$&$\cdots$&$\cdots$&0.6-8.0&1.14(127)\\
&MOS2&$2.68\pm0.01$&$\cdots$&$\cdots$&0.6-10.0&1.40(153)\\
&PN+MOS1+MOS2&$2.69\pm0.01$&$\cdots$&$\cdots$&0.6-8, 0.6-8, 0.6-10&1.31(419)\\
\hline
\multicolumn{7}{c}{Log$\,$Parabolic$\,$model$\,$ (\textit{tbabs*logpar})}\\
\hline
$0790381401$ & PN&$2.71\pm0.02$&$\cdots$&$-0.08\pm0.04$&0.6-7.0&1.27(116) \\
&mod-PN&$2.72\pm0.02$&$\cdots$&$-0.09\pm0.04$&0.6-7.0&1.38(113) \\
&MOS1&$2.75\pm0.02$&$\cdots$&$-0.09\pm0.04$&0.6-8.0&0.99(134)\\
&MOS2&$2.71\pm0.02$&$\cdots$&$-0.06\pm0.04$&0.6-8.0&1.13(138)\\
&PN+MOS1+MOS2$^*$&$2.72 \pm 0.01$&$\cdots$&$-0.07 \pm 0.02$&0.6-7, 0.6-8, 0.6-8&1.16(392)\\
$0790381501$ & PN&$2.56\pm0.01$&$\cdots$&$-0.04\pm0.02$&0.6-8.0&1.07(133) \\
&mod-PN&$2.56\pm0.01$&$\cdots$&$-0.03\pm0.03$&0.6-8.0&1.14(128) \\
&MOS1&$2.62\pm0.02$&$\cdots$&$-0.09\pm0.03$&0.6-10.0&1.39(147)\\
&MOS2&$2.57\pm0.02$&$\cdots$&$-0.06_{-0.03}^{+0.04}$&0.6-10.0&0.92(144)\\
&PN+MOS1+MOS2&$2.58 \pm 0.01$&$\cdots$&$-0.05 \pm 0.02$&0.6-8, 0.6-10, 0.6-10&1.20(428)\\
$0790381001$  & PN&$2.70\pm0.01$&$\cdots$&$-0.03\pm0.02$&0.6-8.0&1.30(136) \\
&mod-PN&$2.71\pm0.01$&$\cdots$&$-0.07\pm0.03$&0.6-8.0&1.26(130) \\
&MOS1&$2.72\pm0.02$&$\cdots$&$-0.05\pm0.05$&0.6-8.0&1.12(126)\\
&MOS2&$2.69\pm0.01$&$\cdots$&$-0.02_{-0.03}^{+0.03}$&0.6-10.0&1.40(152)\\
&PN+MOS1+MOS2&$2.70 \pm 0.01$&$\cdots$&$-0.03 \pm 0.01$&0.6-8, 0.6-8, 0.6-10&1.29(418)\\
\hline
\multicolumn{7}{c}{Broken$\,$Power$\,$Law$\,$model$\,$(\textit{tbabs*bknpow})}\\
\hline
$0790381401$ & PN&$2.71\pm0.01$&$2.46_{-0.26}^{+0.45}$&$2.54_{-0.07}^{+0.05}$&0.6-7.0&1.09(115)\\
&mod-PN&$2.72\pm0.02$&$2.35_{-0.27}^{+0.30}$&$2.52_{-0.06}^{+0.06}$&0.6-7.0&1.20(112)\\
&MOS1&$2.74\pm0.01$&$3.39_{-0.59}^{+0.64}$&$2.48_{-0.13}^{+0.10}$&0.6-8.0&0.93(133)\\
&MOS2&$2.70\pm0.01$&$2.84_{-1.11}^{+0.68}$&$2.58\pm0.07$&0.6-8.0&1.10(137)\\
&PN+MOS1+MOS2$^*$&$2.72\pm0.01$&$2.87_{-0.30}^{+0.31}$&$2.55\pm0.04$&0.6-7, 0.6-8, 0.6-8&1.08(391)\\
$0790381501$ & PN&$2.56\pm0.01$&$3.92_{-1.01}^{+0.93}$&$2.38_{-0.16}^{+0.09}$&0.6-8.0&0.97(132)\\
&mod-PN&$2.57\pm0.01$&$3.03_{-0.81}^{+1.27}$&$2.45_{-0.14}^{+0.06}$&0.6-8.0&1.08(127)\\
&MOS1&$2.60_{-0.02}^{+0.01}$&$3.40_{-0.63}^{+1.29}$&$2.39_{-0.25}^{+0.07}$&0.6-10.0&1.33(146)\\
&MOS2&$2.57\pm0.02$&$1.95_{-0.48}^{+1.44}$&$2.51_{-0.07}^{+0.03}$&0.6-10.0&0.91(143)\\
&PN+MOS1+MOS2&$2.57\pm0.01$&$4.26_{-1.25}^{+0.33}$&$2.33_{-0.07}^{+0.13}$&0.6-8, 0.6-10, 0.6-10&1.14(427)\\
$0790381001$ & PN&$2.70\pm0.01$&$1.42_{-0.3}^{+0.70}$&$2.66_{-0.02}^{+0.1}$&0.6-8.0&1.29(135)\\
&mod-PN&$2.71\pm0.01$&$1.70_{-0.36}^{+0.34}$&$2.64_{-0.02}^{+0.02}$&0.6-8.0&1.25(129)\\
&MOS1&$2.70_{-0.02}^{+0.01}$&$4.83_{-2.0}^{+1.54}$&$2.25_{-1.04}^{+0.37}$&0.6-8.0&1.09(125)\\
&MOS2&$2.70\pm0.01$&$1.70_{-0.26}^{+0.37}$&$2.66_{-0.02}^{+0.02}$&0.6-10.0&1.38(151)\\
&PN+MOS1+MOS2&$2.70\pm0.01$&$1.76_{-0.24}^{+0.24}$&$2.66_{-0.01}^{+0.01}$&0.6-8, 0.6-8, 0.6-10&1.27(417)
\enddata
\tablecomments{(1) : Observation IDs that show spectral hardening and are investigated for signatures of an additional IC component. (2) : PN-{\it EPIC PN}, MOS1-{\it EPIC MOS1}, MOS2-{\it EPIC MOS2}, mod PN-modified {EPIC PN}. (3):$\Gamma-$PL photon index, $\alpha$-LP local photon index(at 1 keV), $\Gamma_1$-BPL soft photon index. (4)$E_{break}$ : BPL break energy in keV. (5) :$ \beta$ - LP curvature parameter, $\Gamma_2$-BPL hard photon index.(6) Energy range chosen for respective spectral fit based on background contamination.  (7) ${\chi^2_r(dof)}$: Reduced chi-square and degrees of freedom for the respective fits.\\
$^*$: During joint spectral fitting {\it(PN+MOS1+MOS2)}, the cross calibration constant for PN was fixed at 1, and that of  MOS1 and MOS2 took values in the range 0.99-1.12. }
\end{deluxetable*}

The observation with ID 0790381501 (hereafter Obs B) happened eight hours after Obs A and consisted of a lengthy observation of 34 hours, similar to Obs A. The {\it PN} spectrum and both {\it MOS} spectra were studied in an energy range of 0.6--8 keV and 0.6--10 keV, respectively. A LP fit to the X-ray spectra of Obs B resulted in significant negative curvatures, and a PL fit to them resulted in a high-energy tail. A BPL fit of all the data sets yielded a stable fit, with their photon indices and break energies in agreement with the BPL parameters of the joint {\it PN+MOS} fit. The joint fit  yielded $\gamma_1\simeq2.57$, $\Gamma_2\simeq2.33$ and a break energy, $E_{Break}\simeq4.26$ keV, suggesting a spectral hardening of $\Delta \Gamma\simeq0.24$. It is to be noted that slightly higher high-energy photon index ($\Gamma_2 \simeq 2.51$) observed in the MOS2 data when fitting a BPL model, compared to the values ($\Gamma_2 \simeq 2.29 – 2.45$) obtained using other data, may be due to the smaller source extraction region (an annulus with an outer radius of \SI{20}{\arcsecond}), which was necessary since the the source was located closer to the edge of the CCD during this observation, and which can lead to a lower S/N ratio.

The observation with ID 0790381001 (hereafter Obs C) was taken on 21 Feb 2017, a few days after Obs B. {\it PN} and  \textit{MOS1} spectra were studied in an energy range of 0.6--8 keV, while \textit{MOS2}was studied over 0.6--10 keV. Obs C yielded significant negative curvatures on fitting a LP model to different spectra and also resulted in stable BPL fits. However, a high-energy tail is not evident from a PL fit in their spectral plots . A BPL fit to joint {\it PN+MOS} spectrum yields $\gamma_1\simeq2.70$, $\Gamma_2\simeq2.66$ and a break energy, $E_{Break}\simeq1.76$ keV, suggesting a small spectral hardening of $\Delta \Gamma\simeq0.04$. The BPL confidence contours, though close to each other, appear independent at 99$\%$ confidence.

 Since Obs A (ID:0790381401) satisfies all our criteria as discussed in \autoref{sec3.3}, we infer a strong possibility that there is an intermixing of the synchrotron component with the IC component, though the former remains dominant. Further from \autoref{tab2}, we can see that among our 30 X-ray spectra, the lowest spectral flux (in the energy range of $0.6-7.0$ keV), $F_{0.6-7} \simeq 10.3 \times 10^{-12} {\rm erg~cm}^{-2}~{\rm s}^{-1}$, was found for Obs A. This fact strengthens our claim for the presence of an IC component. Observation B (ID:0790381501), made approximately 8 hours after Observation A, also showed significant negative curvature on fitting LP models and yielded stable BPL fits with spectral hardening. A BPL fit to joint {\it PN+MOS} spectra for Obs A and Obs B resulted in spectral hardening, $\Delta\Gamma$, of  0.17 and  0.24, respectively.  A hard energy tail observed when fitting a PL model, which is rectified when fitting the BPL model, is evident from the spectral plots of both observations. Thus, we present these two observations as strong candidates for having signatures of an additional IC component. On the contrary, observation C (Obs ID 0790381001) made just a few days after Obs B, yields a small value of $\Delta\Gamma\simeq0.04$. This is consistent with the high-energy tail in the PL model fit, which is not very evident in the spectra. Thus, we cannot assert that Obs C is a strong candidate for signatures of an additional IC component, as we can for Obs A and Obs B.  We can, however, assert that there may well be an IC component in Obs C, though it clearly is not as strong as in Obs A or B.

For HBLs, a concave X-ray spectrum on fitting  LP models, and spectral hardening ($\Gamma_2<\Gamma_1$) on fitting  BPL models has been interpreted as a mixture of a dominant synchrotron component and an additional IC component. Such behavior has been occasionally reported in the literature. For instance, \cite{Zha08} reported such behavior in PKS 2155-304 in a study of two {\it XMM-Newton} observations from 2006. In Paper I, we reported such signatures of IC components in one {\it XMM-Newton} observation each of PKS 0548$-$322 and MRK 501. In the case of PG 1553+113, previous X-ray spectral studies done using data from \textit{BeppoSAX}, \textit{RXTE}, \textit{Chandra}, \textit{Swift}, \textit{Suzaku}, and \textit{NuSTAR}, described the X-ray spectra with PL or LP models with convex curvatures. The strong spectral hardening observed in two of our observations may therefore indicate a temporary spectral state where the high-energy tail of the synchrotron component overlaps with the low-energy tail of the IC component. Further, such behavior could suggest that during these epochs, HBL PG 1553+113 temporarily exhibited spectral characteristics similar to those of an IBL.

\section{Conclusions}\label{sec5}
In this work, we have studied X-ray spectra of  30 \textit{EPIC PN} observations made by {\it XMM-Newton} throughout its operational period. The \textit{EPIC PN} X-ray spectra were studied in the energy range of $0.6-7.0$ keV, because we found that at higher energies, background counts were comparable to the source counts. Among the 30 observations, 21 spectra were analyzed for the first time in our study. We found that 14 out of 30 X-ray spectra were best fitted with LP models, with parameters in the range $\alpha\simeq2.13-2.80$ and $\beta\simeq0.04-0.18$. Energy-dependent particle acceleration scenarios combined with radiative cooling can describe these curvatures in spectra \citep{Mas04,Tra07,Mas08}. One other spectrum favored a LP model with negative curvature($\beta<0$).

However, 15 of the X-ray spectra were well described by a simple PL model. This probably means that during these times, the $0.6-7.0$ keV energy band samples the essentially linear part of the synchrotron continuum, where the intrinsic curvature is weak or statistically unconstrained due to a low S/N ratio. In such scenarios, any subtle curvature that might arise from particle acceleration, for example, is not significant enough to justify the additional complexity of an LP model. The PL photon indices had values in the range $\Gamma\simeq2.53-2.69$. We attempted to estimate the synchrotron peak and peak luminosities from X-ray spectra but found the peak energy to be significantly below the instrumental energy range ($<0.2$ keV) for all \textit{PN} observations.

An interesting  X-ray spectrum of the observation taken on 13 Feb 2017 (Obs ID:0790381401 or Obs A), when fitted in the energy range of 0.6--7.0 keV with an LP model, results in significant negative curvature. When we fit its joint \textit{PN+MOS} X-ray spectrum with a BPL model, we obtain these parameters: $\Gamma_1=2.72\pm0.01$; $\Gamma_2=2.55_{-0.04}^{+0.04}$; and $E_{break}=2.87_{-0.30}^{+0.31}$ keV. Another observation (Obs ID:0790381501 or Obs B), taken approximately eight hours after Obs A, also showed significant negative curvature in its X-ray spectra. A BPL fit to combined PN-MOS spectra resulted  in $\Gamma_1\simeq2.57\pm0.01$, $\Gamma_2\simeq2.33_{-0.07}^{+0.13}$, and $E_{break}=4.28_{-1.27}^{+0.32}$ keV. The confidence contours for the BPL fits to both plots showed that the parameters are independent at the 99$\%$ confidence level. A high-energy tail was left over when PL models were fit, but this tail was modeled when a BPL was used. These facts lead us to consider these observations to be strong candidates for showing signatures of an inverse Compton component, in addition to the dominant synchrotron component. Additionally, Obs ID 0790381001 (or Obs C) with BPL parameters $\Gamma_1\simeq2.70\pm0.01$, $\Gamma_2\simeq2.66_{-0.01}^{+0.01}$, and $E_{break}=1.76_{-0.24}^{+0.24}$ keV from joint {\it PN+MOS} spectral fit also suggested a mild spectral hardening of $\Delta\Gamma\simeq0.04$. However, no high-energy tails were evident on fitting PL model, and so we consider this observation to be only a mild candidate for any signature of an additional IC component.

Because most of the synchrotron peaks appeared to be below 0.2 keV, we also attempted to fit the joint spectral data of \textit{PN} and \textit{OM} using a LP model convolved with galactic absorption and reddening corrections. Although the reduced chi-square was initially large, upon applying a $3\%$ systematic error to the \textit{OM} data, most of the observations were well-fitted by this model, and we were able to estimate the rest frame synchrotron peak energy in the range of 4.59 to 48.61 eV. We searched for correlations between spectral parameters for the LP fits to both X-ray alone and X-ray + Optical/UV. The only significant correlation we found was a correlation between spectrum hardening and brightness, a typical behavior of HBLs. 
With five other \textit{XMM-Newton} observations of PG 1553+113 expected to become publicly available in 2026, the temporal coverage and spectral sampling of the source will be significantly improved. This will allow a more detailed investigation into the long-term spectral evolution of this source, potentially revealing new details about the evolution of its synchrotron and inverse Compton components.

\begin{acknowledgments}
We thank the anonymous referee for their constructive comments and suggestions, which improved this manuscript. P.U.D. would like to thank Sitha K.\ Jagan, and Riya Bhowmick for their valuable insights during the discussion of the work.  P.U.D. would also like to acknowledge the XMM-Newton help desk for their prompt response to queries. This research is based on observations obtained with XMM-Newton, an ESA science mission with instruments and contributions directly funded by ESA Member States and NASA. This research has made use of data obtained through the High Energy Astrophysics Science Archive Research Center Online Service, provided by the NASA/Goddard Space Flight Center. The Open Universe initiative for blazars was available thanks to the web portal developed at the Italian Space Agency, ASI. V.J. acknowledges the
support provided by the Department of Science and Technology (DST) under the “Fund for Improvement of S$\&$T Infrastructure (FIST)” program (SR/FST/PS-I/2022/208). V.J. also thanks the Inter-University Centre for Astronomy and Astrophysics
(IUCAA), Pune, India, for a visiting associateship.

\end{acknowledgments}

\software{HEASoft \citep{HEA14}, SAS \citep{Gab04}, Xspec \citep{Arn96}.}

% \begin{contribution}

% All authors contributed equally to the Terra Mater collaboration.

% \end{contribution}

\appendix

\section{Appendix information}

\renewcommand{\thefigure}{A\arabic{figure}}
\setcounter{figure}{0}
\begin{figure*}
\centering
{\vspace{-0.14cm} \includegraphics[width=8.5cm, height=7.5cm]{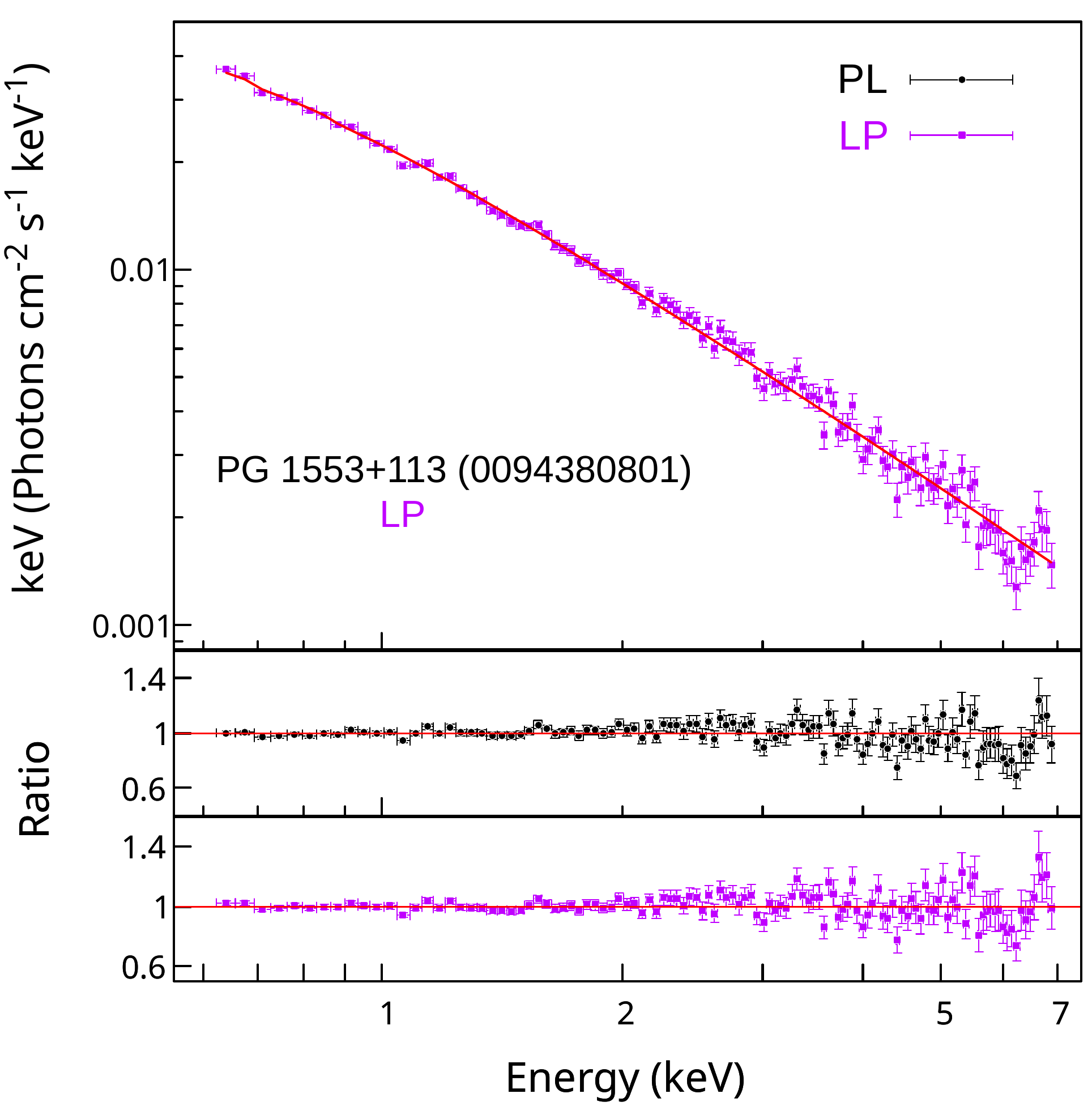}}
\includegraphics[width=8.5cm, height=7.5cm]{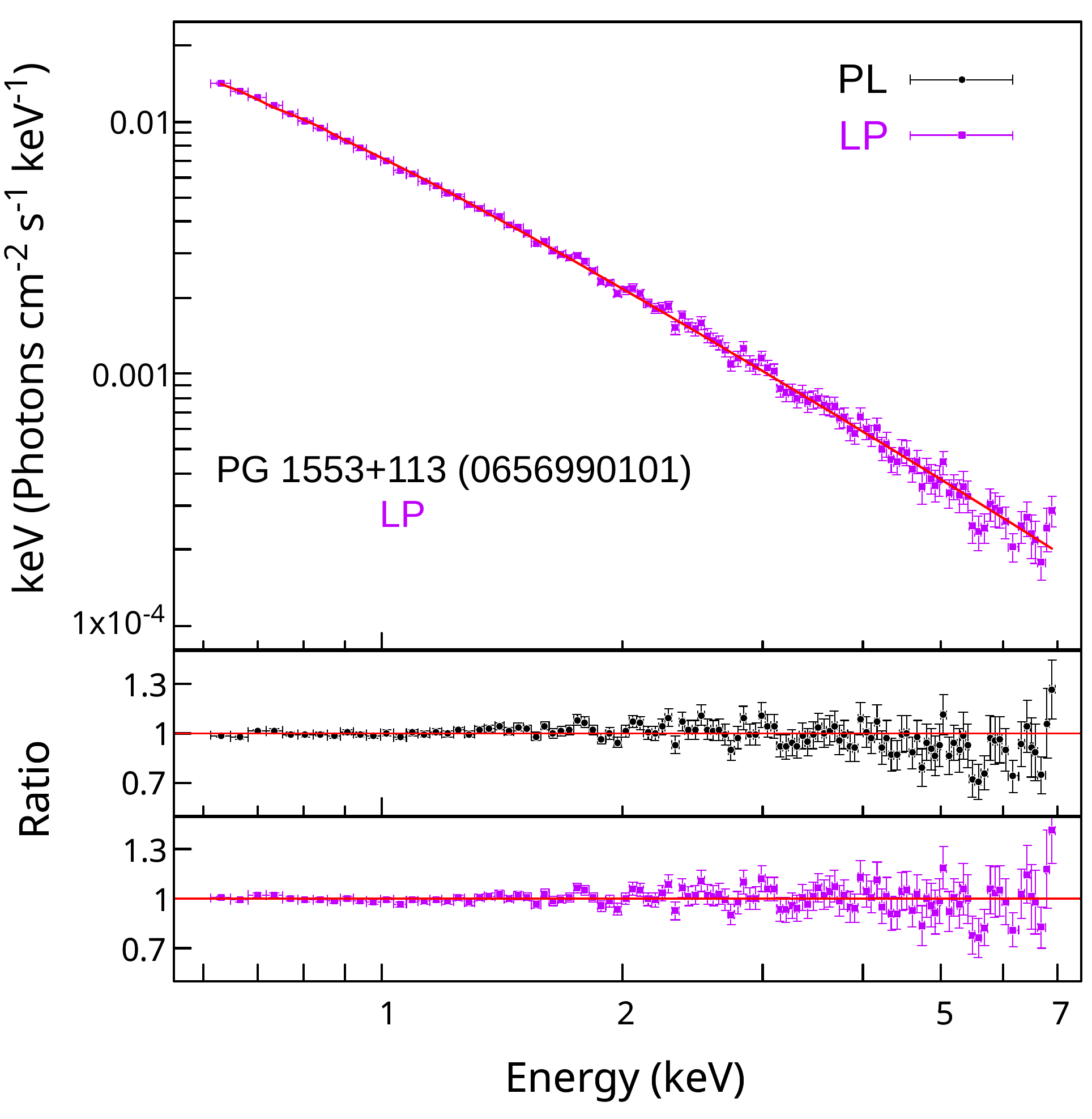}

{\vspace{-0.14cm} \includegraphics[width=8.5cm, height=7.5cm]{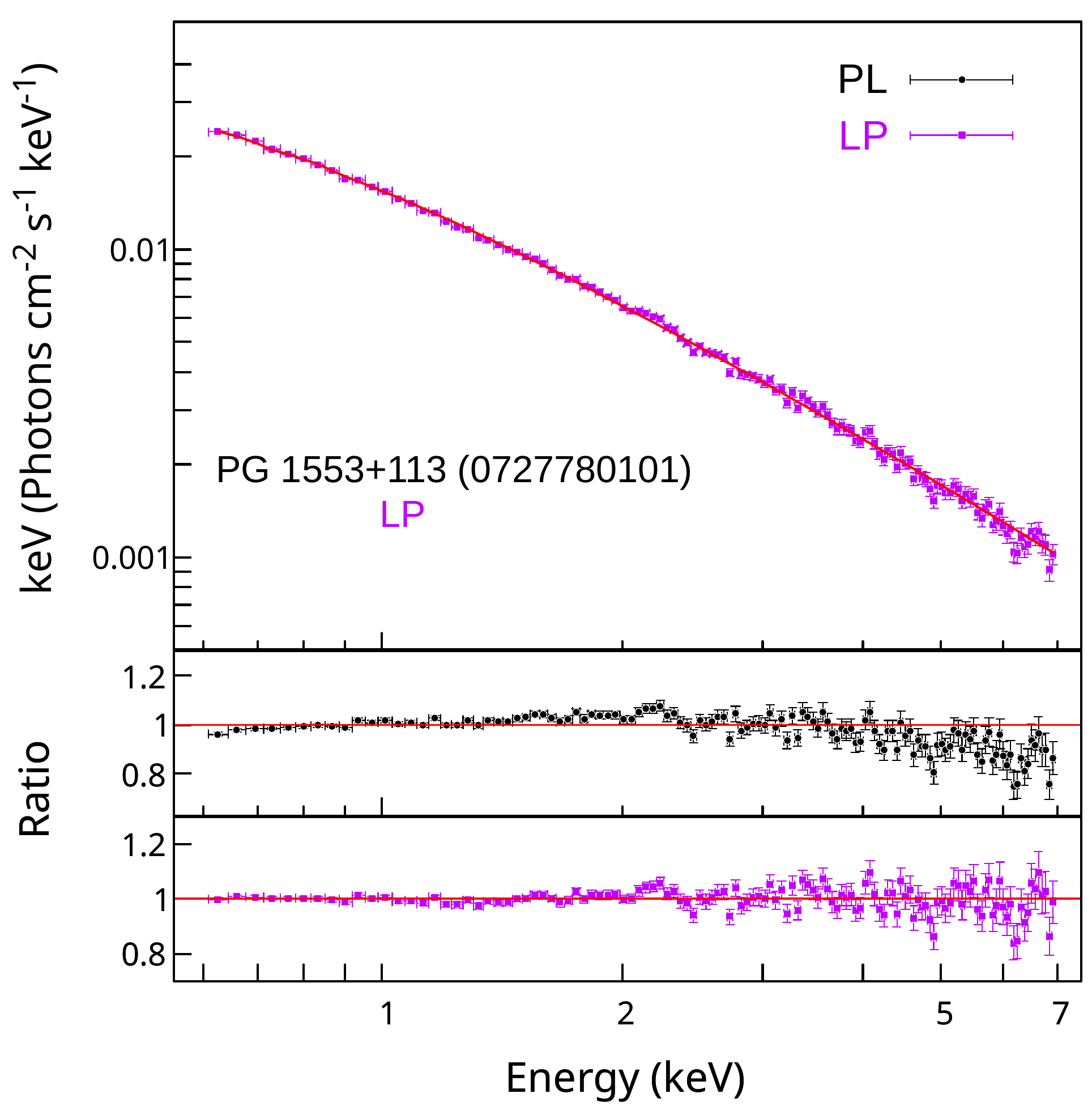}}
\includegraphics[width=8.5cm, height=7.5cm]{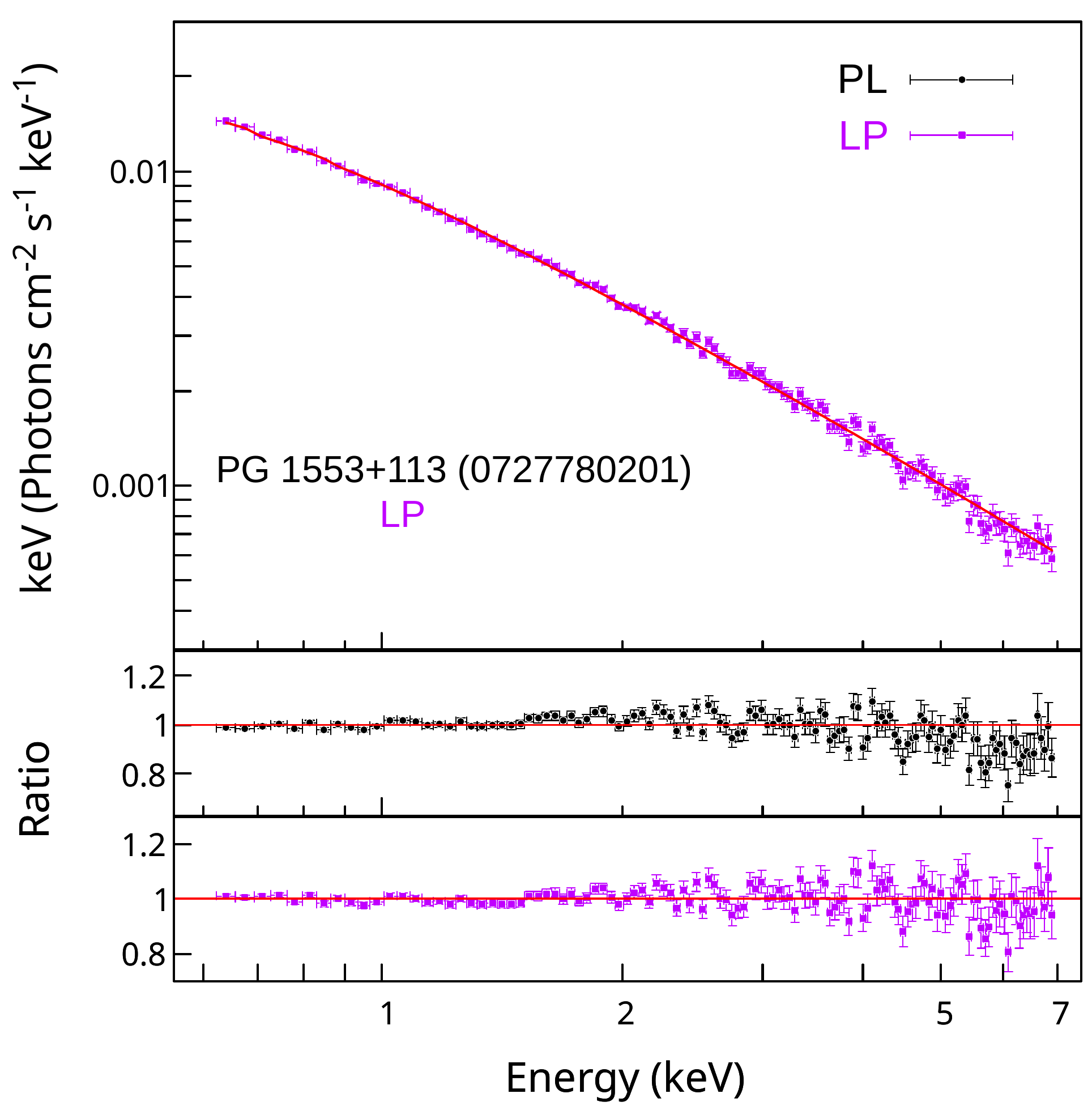}

{\vspace{-0.14cm} \includegraphics[width=8.5cm, height=7.5cm]{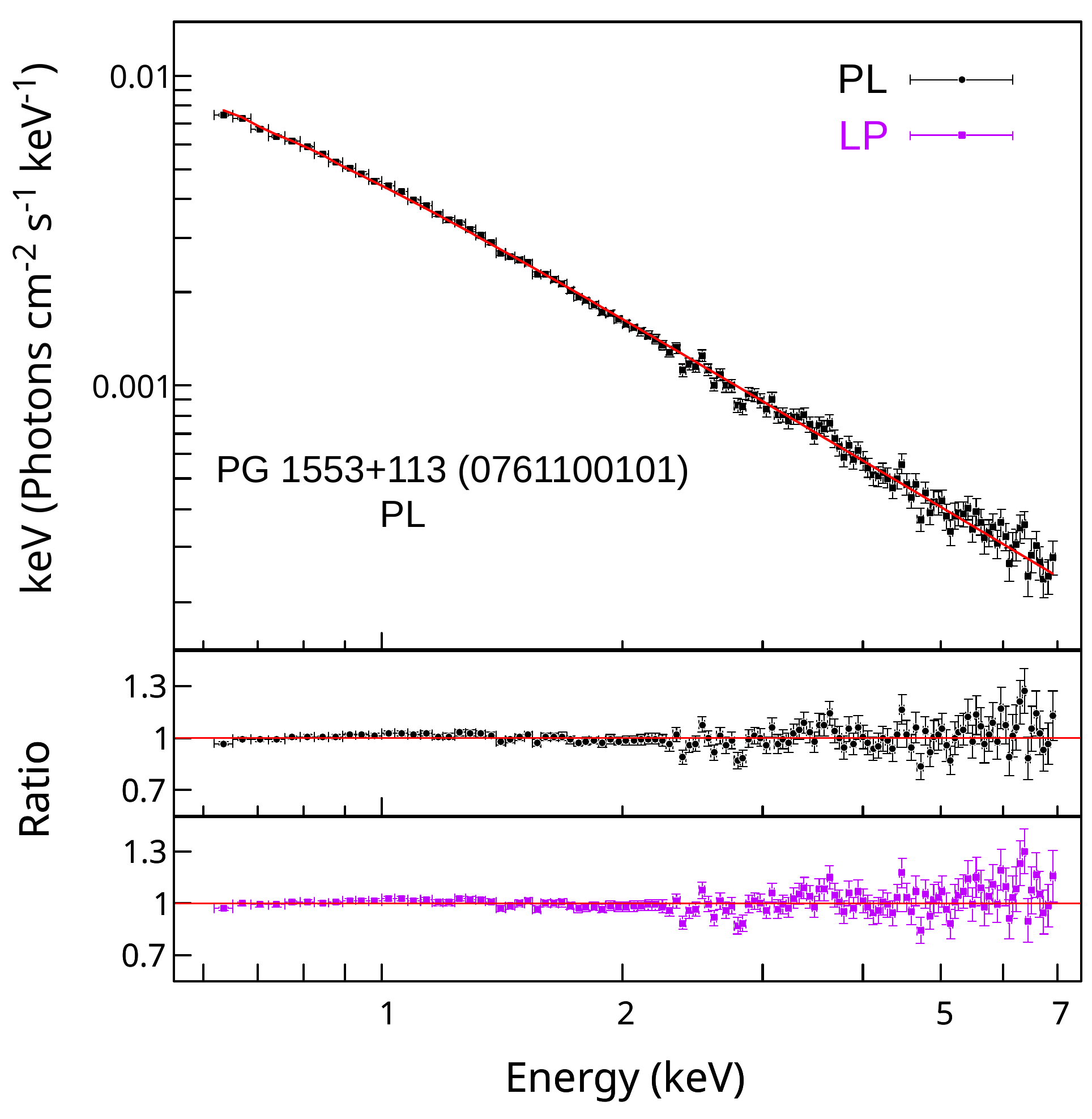}}
\includegraphics[width=8.5cm, height=7.5cm]{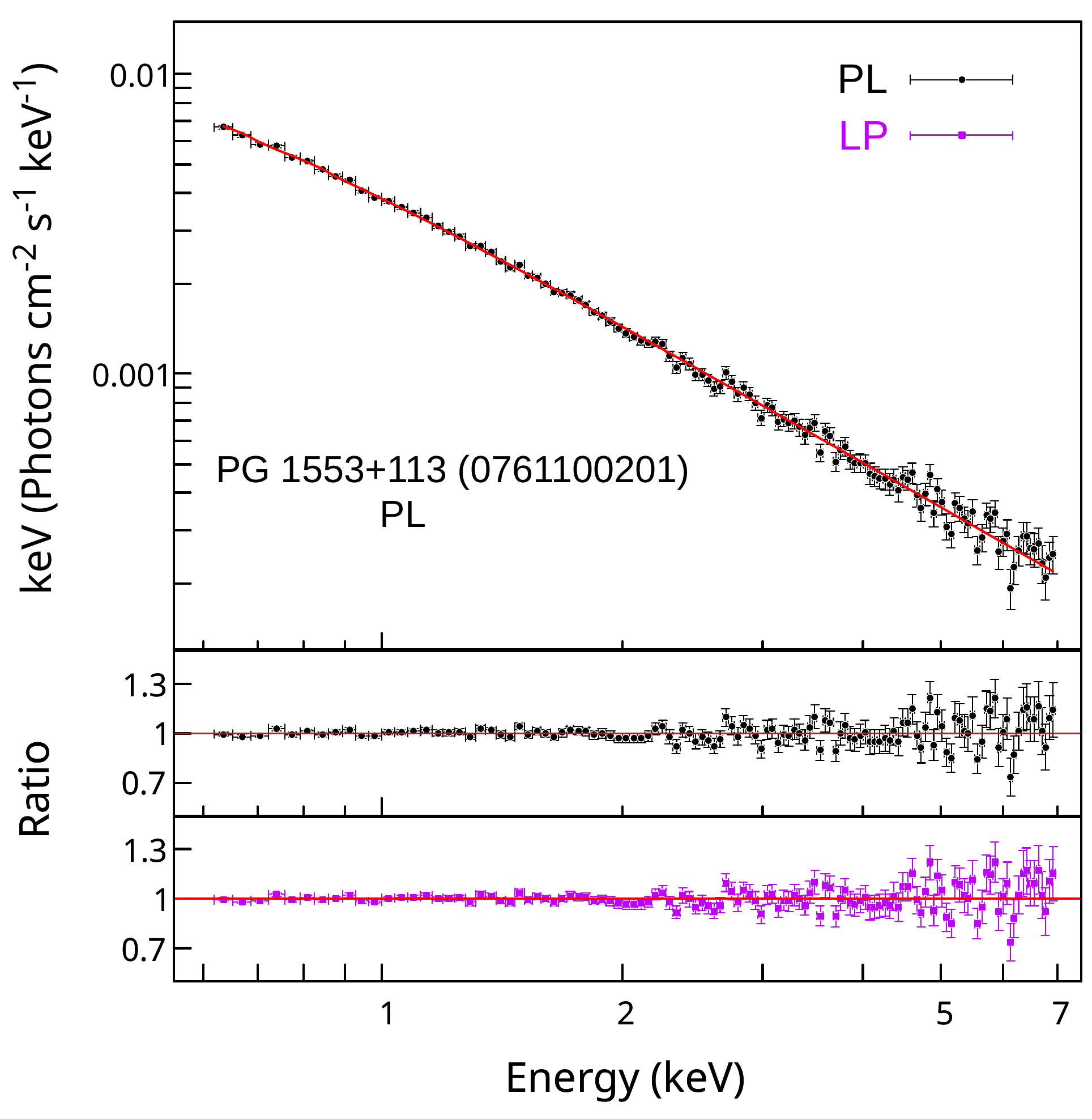}
\vspace{-0.1cm}
\caption{X-ray spectral fit of 30 {\it XMM-Newton EPIC PN} pointed observations of PG 1553+113 in energy range 0.6 - 7.0 keV. The observation ID, source, and best-fit model are shown in the figure. The Power Law (PL) model is represented by black circles, and the Log Parabolic (LP) model by dark-magenta squares, Data-to-model ratio plots for each model are shown in additional panels.\label{A1}}   
\end{figure*}

\clearpage
\setcounter{figure}{0}
\begin{figure*}
\centering
{\vspace{-0.14cm} \includegraphics[width=8.5cm, height=7.5cm]{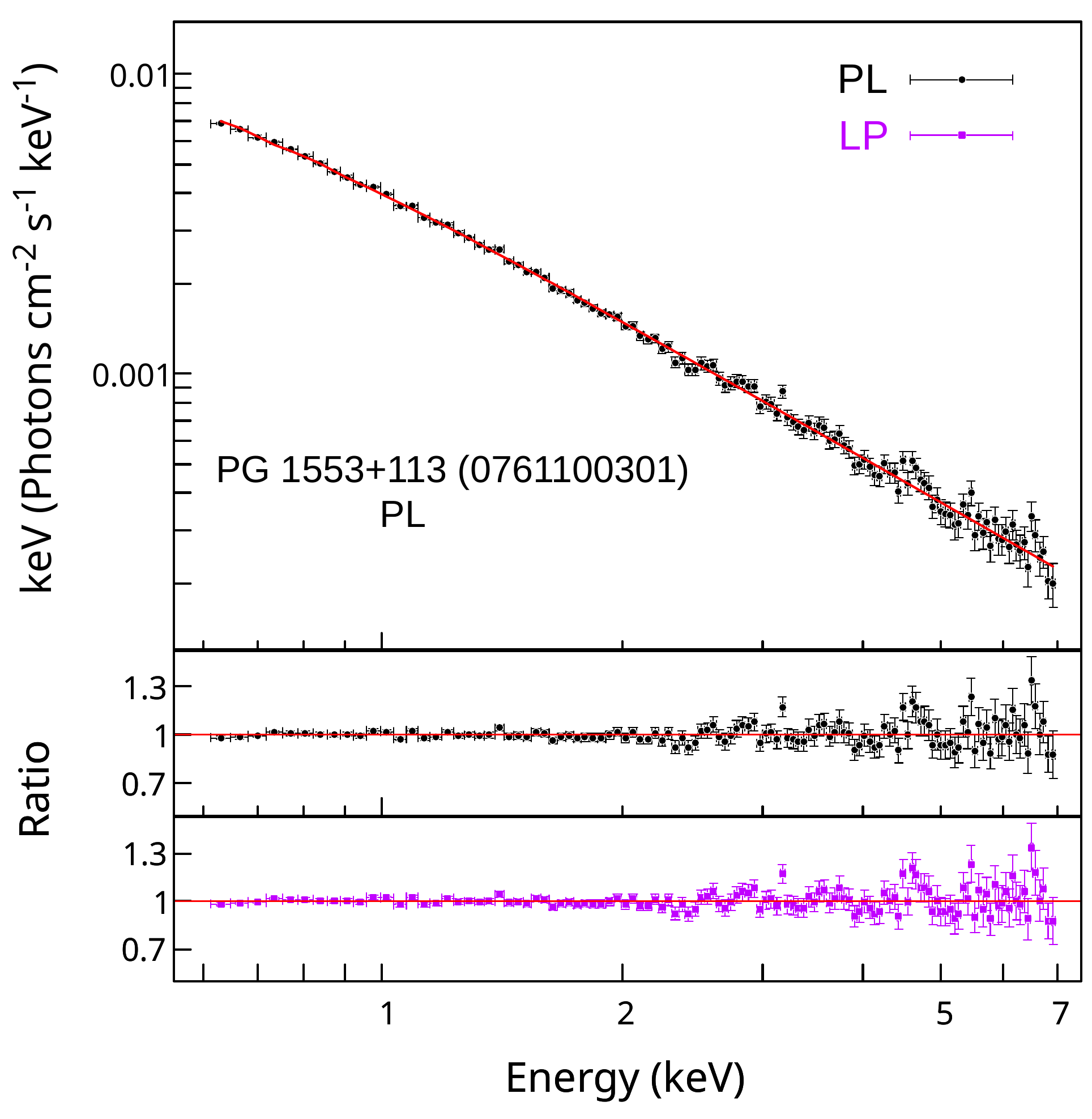}}
\includegraphics[width=8.5cm, height=7.5cm]{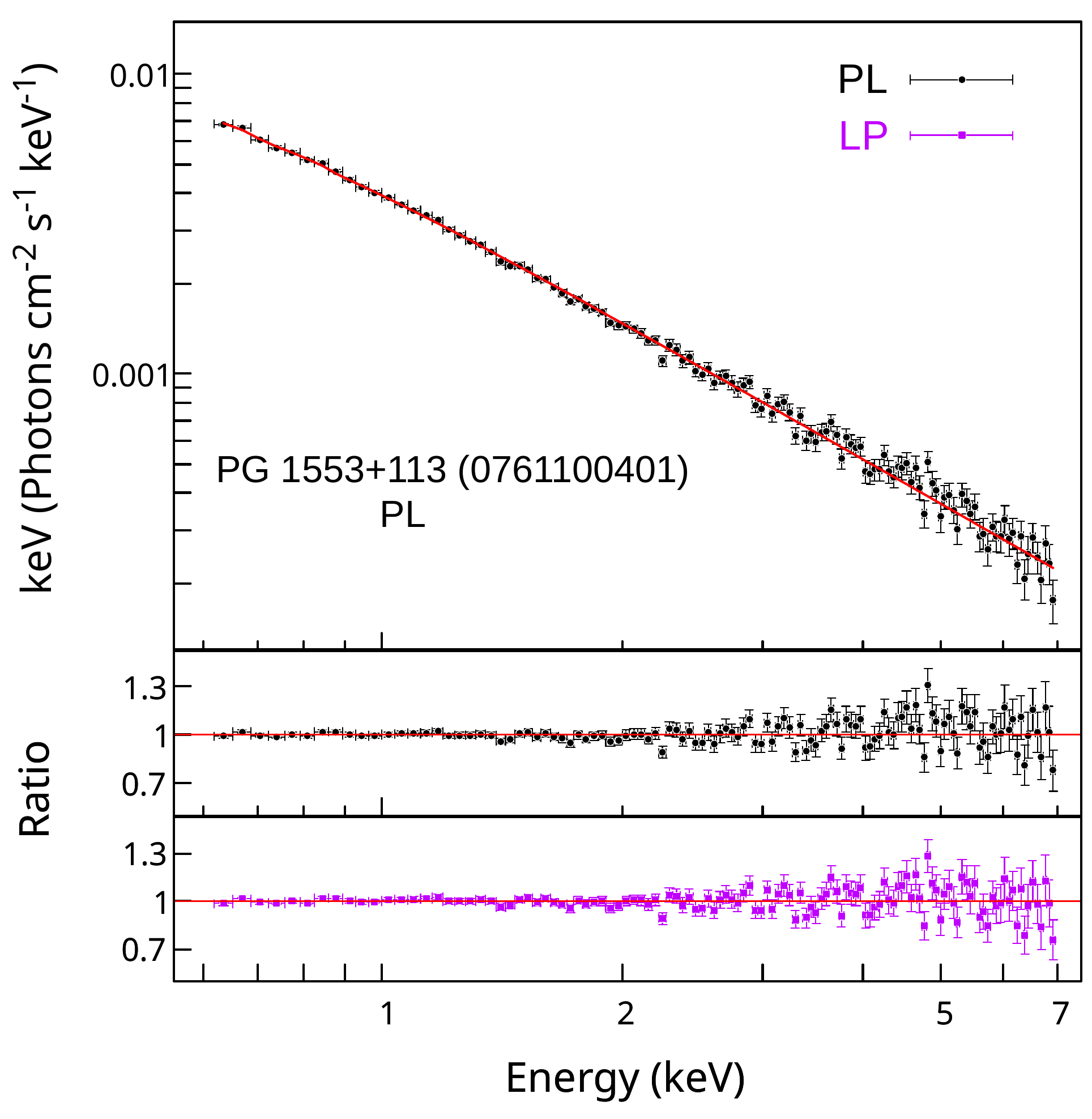}

{\vspace{-0.14cm} \includegraphics[width=8.5cm, height=7.5cm]{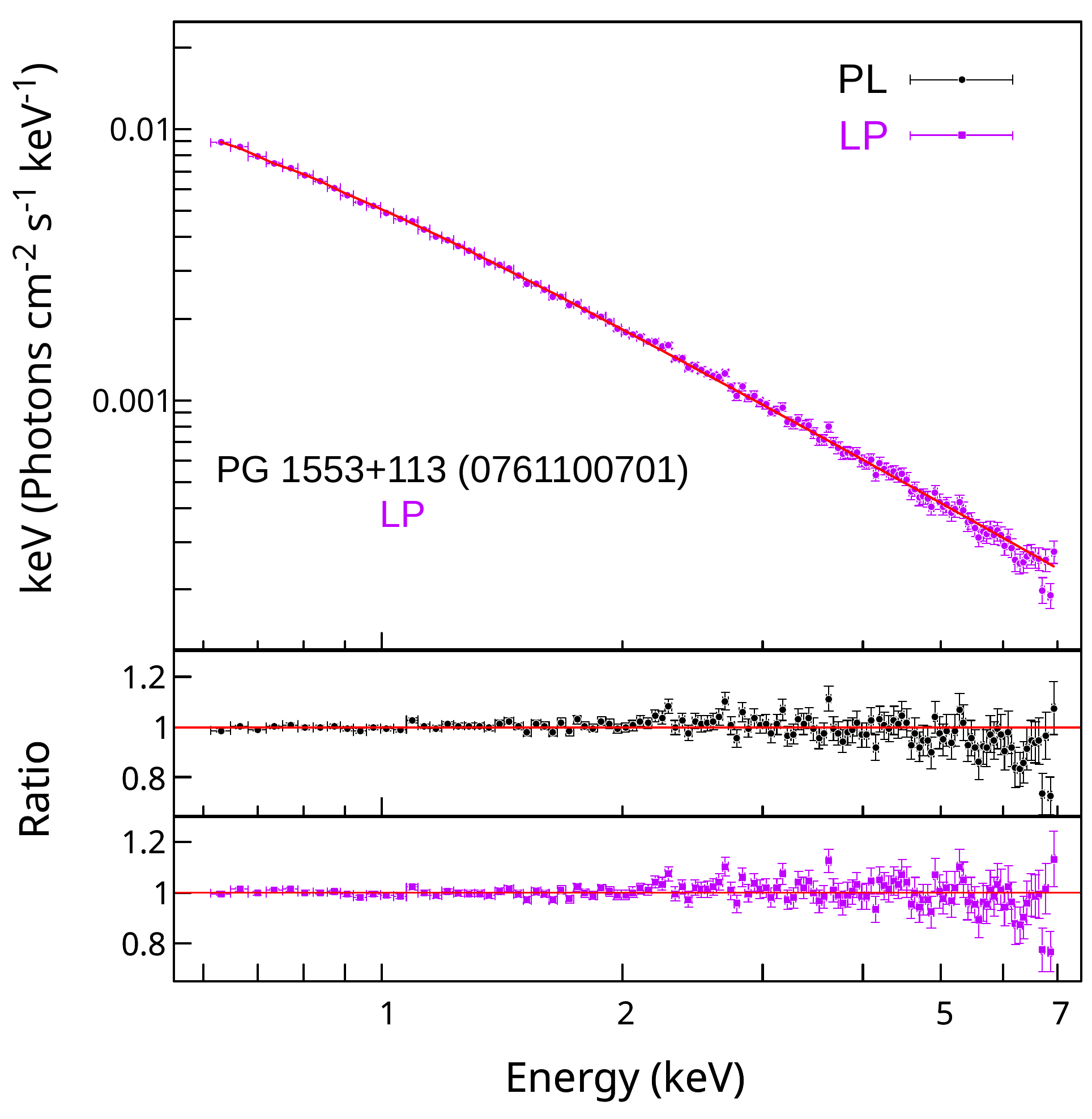}}
\includegraphics[width=8.5cm, height=7.5cm]{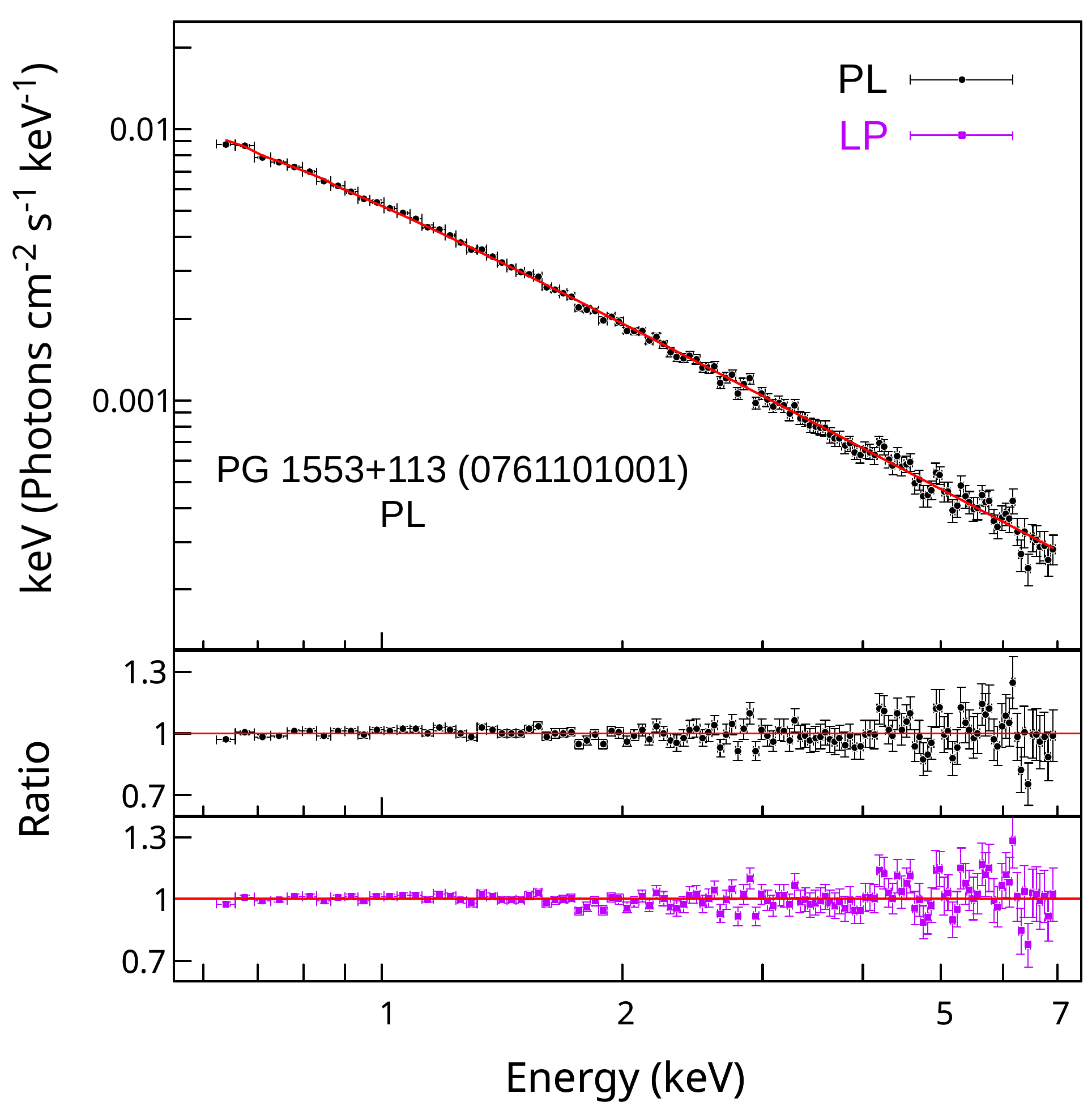}

{\vspace{-0.14cm} \includegraphics[width=8.5cm, height=7.5cm]{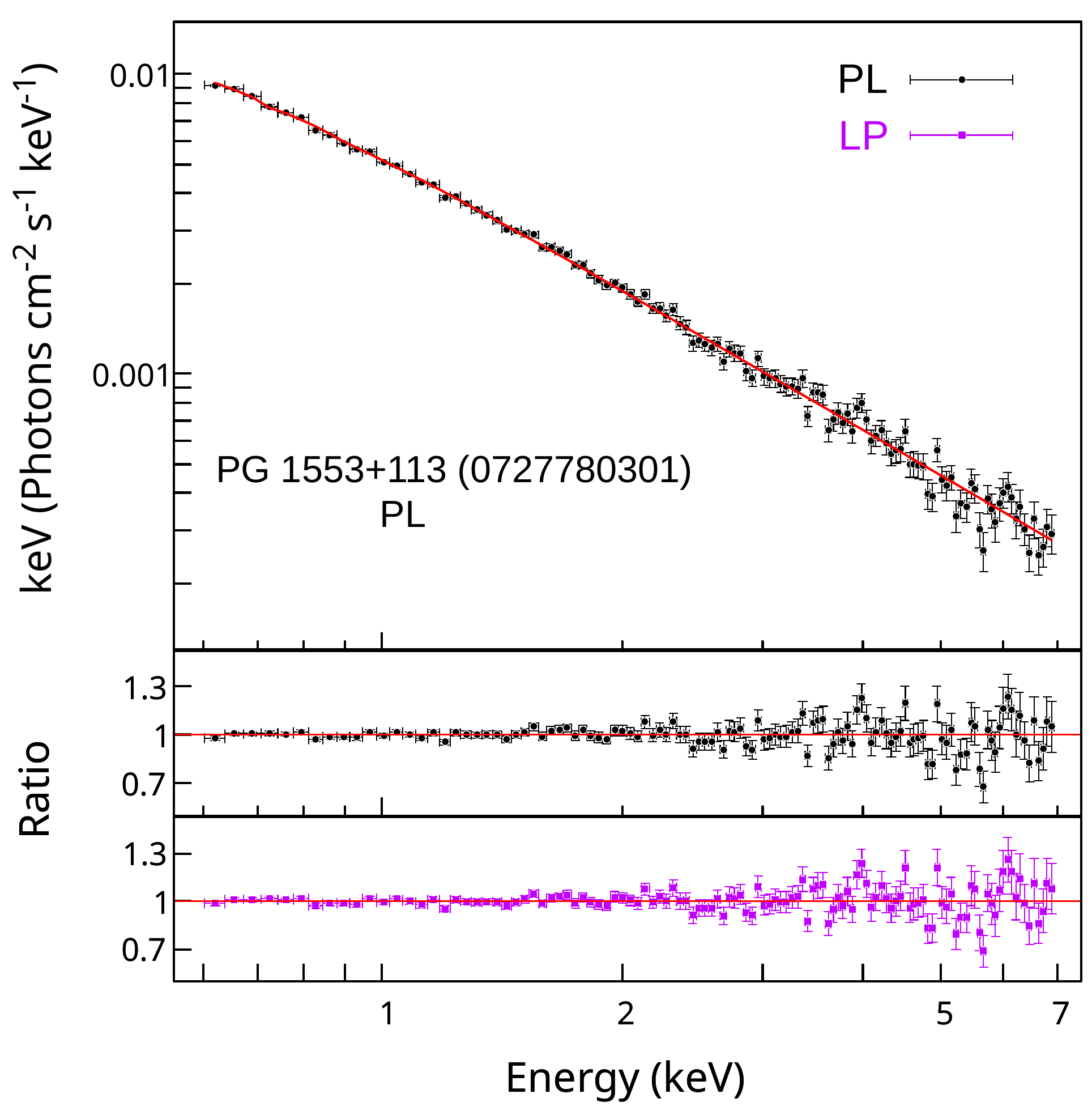}}
\includegraphics[width=8.5cm, height=7.5cm]{fig1_12.pdf}
\caption{Continued} 
\end{figure*}

\clearpage
\setcounter{figure}{0}
\begin{figure*}
\centering
{\vspace{-0.14cm} \includegraphics[width=8.5cm, height=7.5cm]{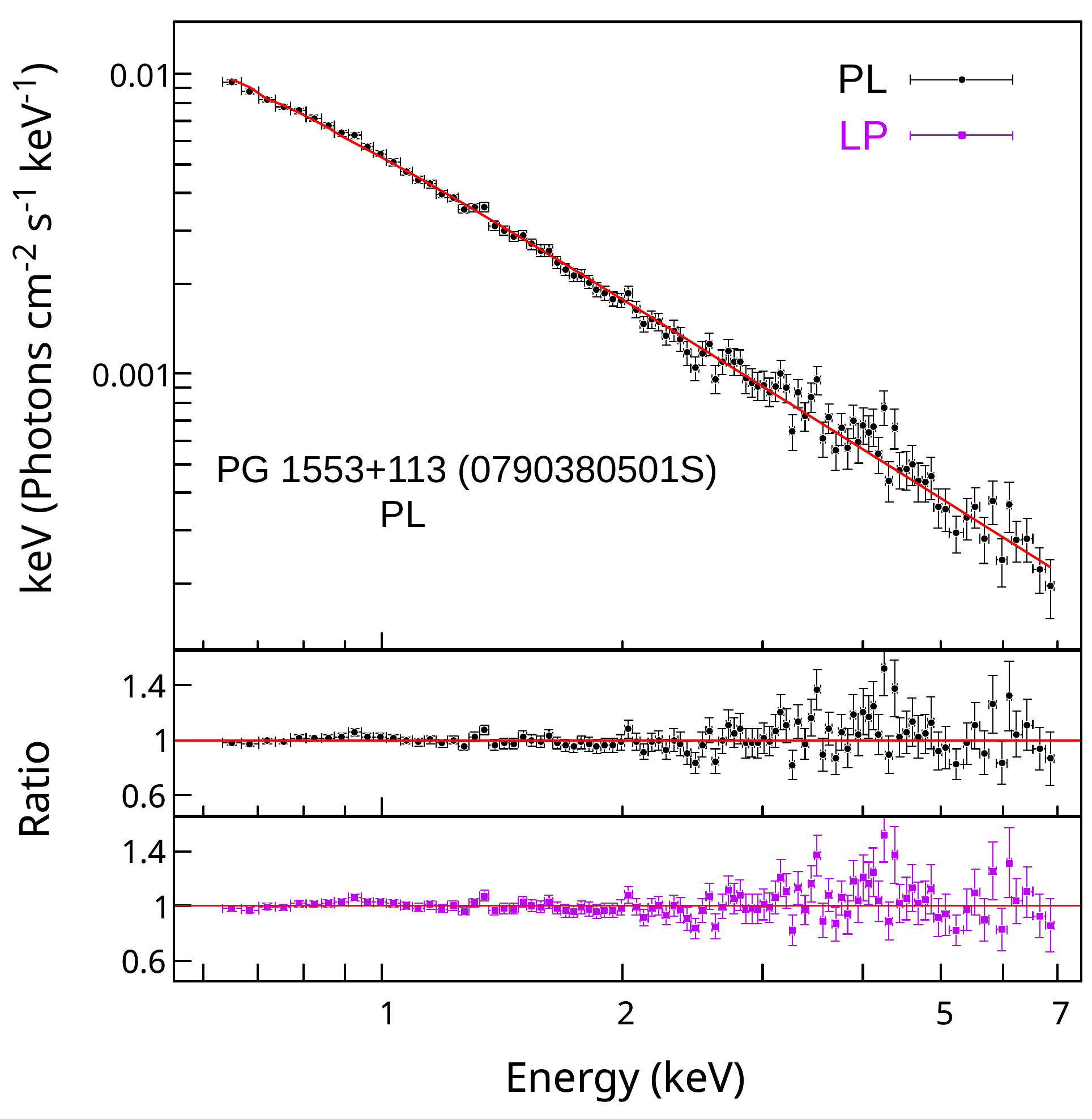}}
\includegraphics[width=8.5cm, height=7.5cm]{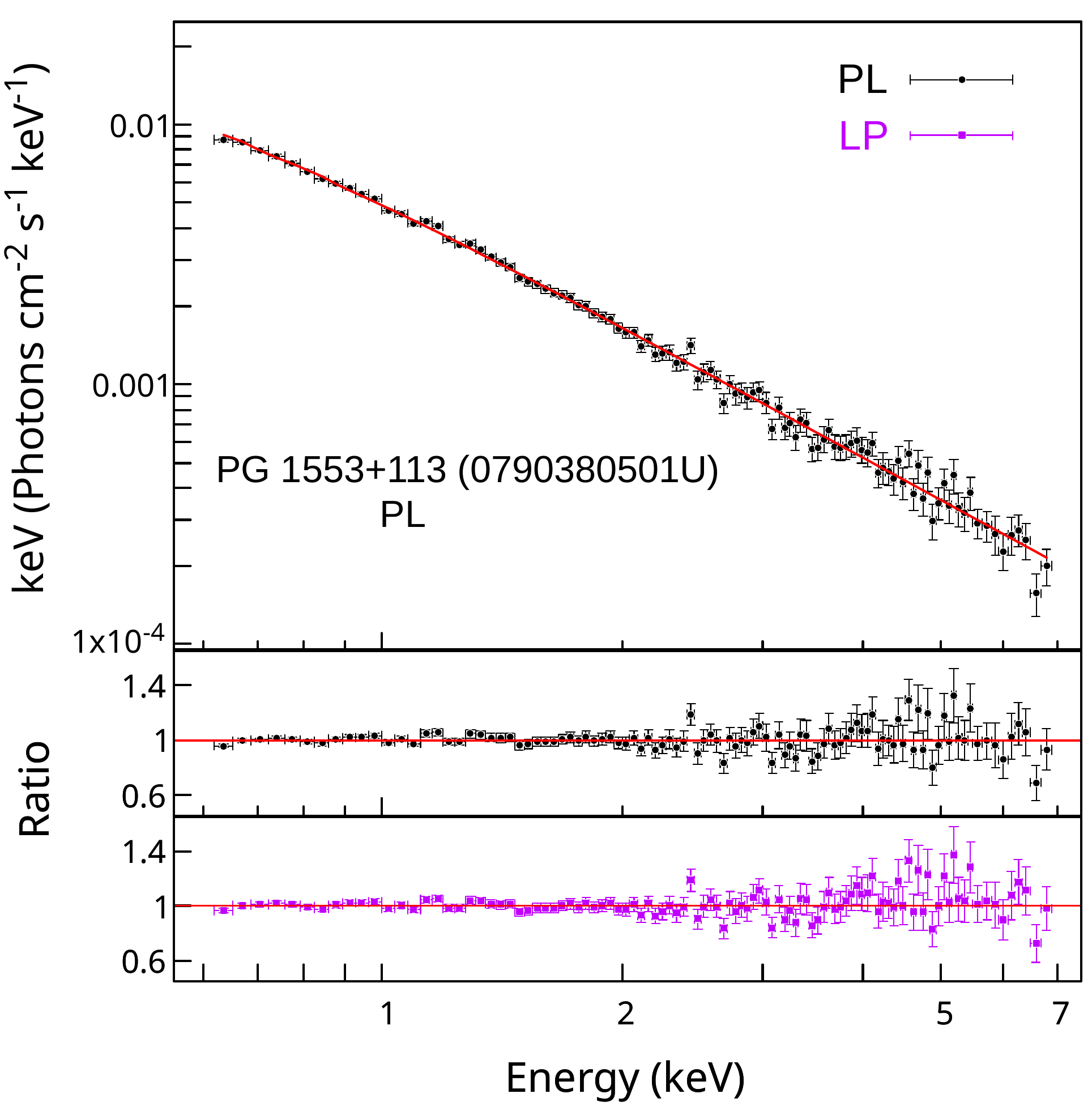}

{\vspace{-0.14cm} \includegraphics[width=8.5cm, height=7.5cm]{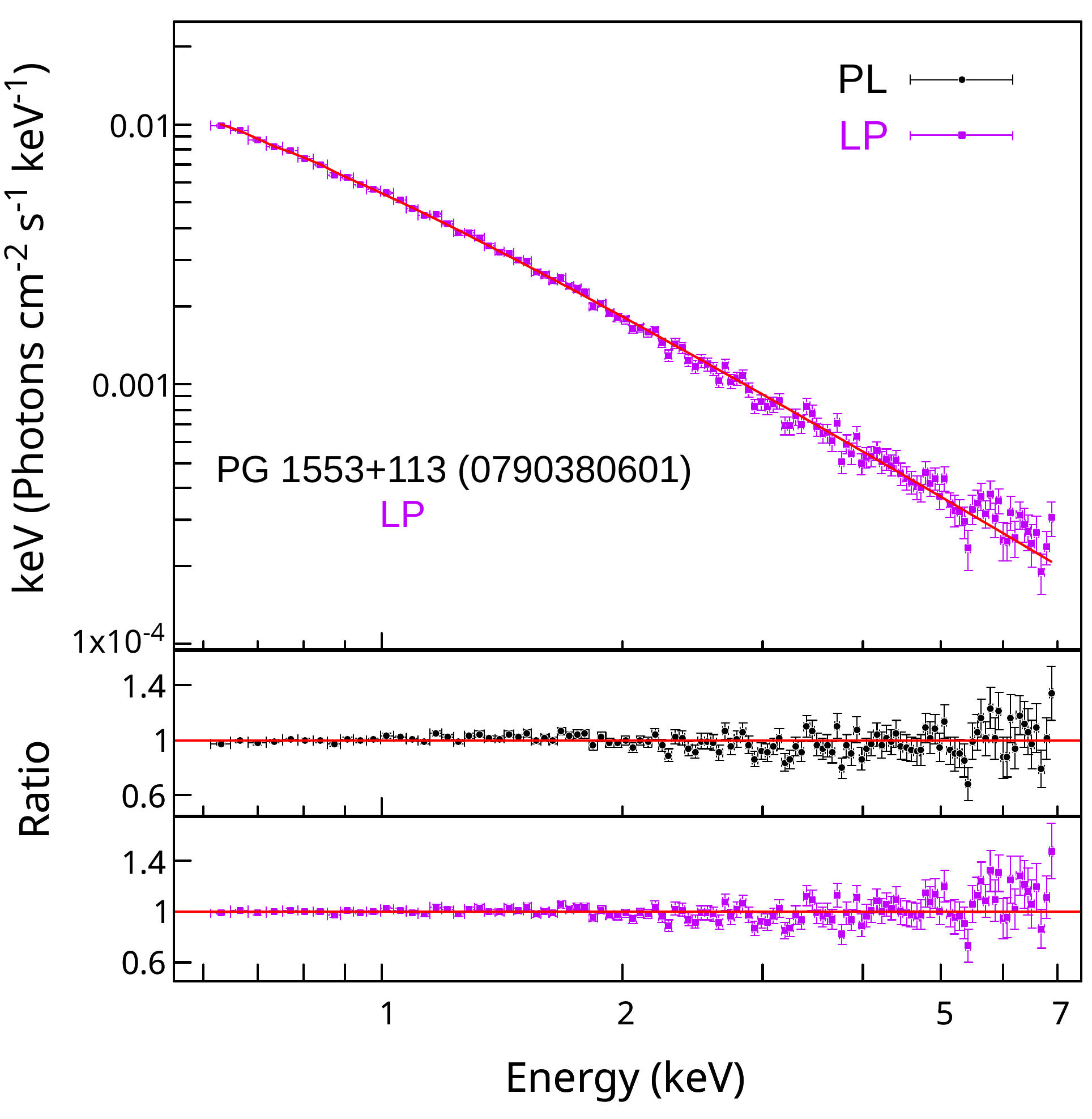}}
\includegraphics[width=8.5cm, height=7.5cm]{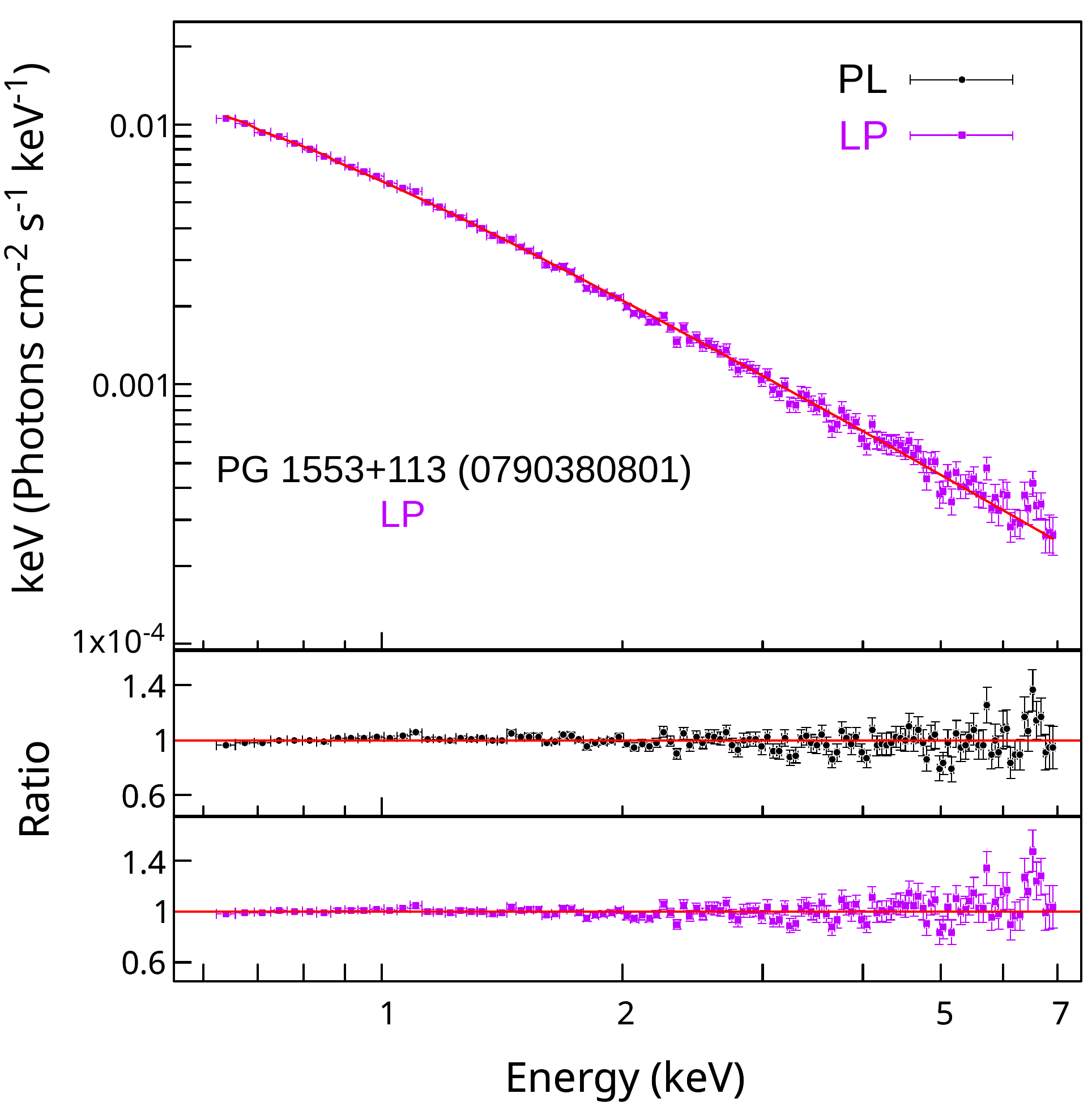}

{\vspace{-0.14cm} \includegraphics[width=8.5cm, height=7.5cm]{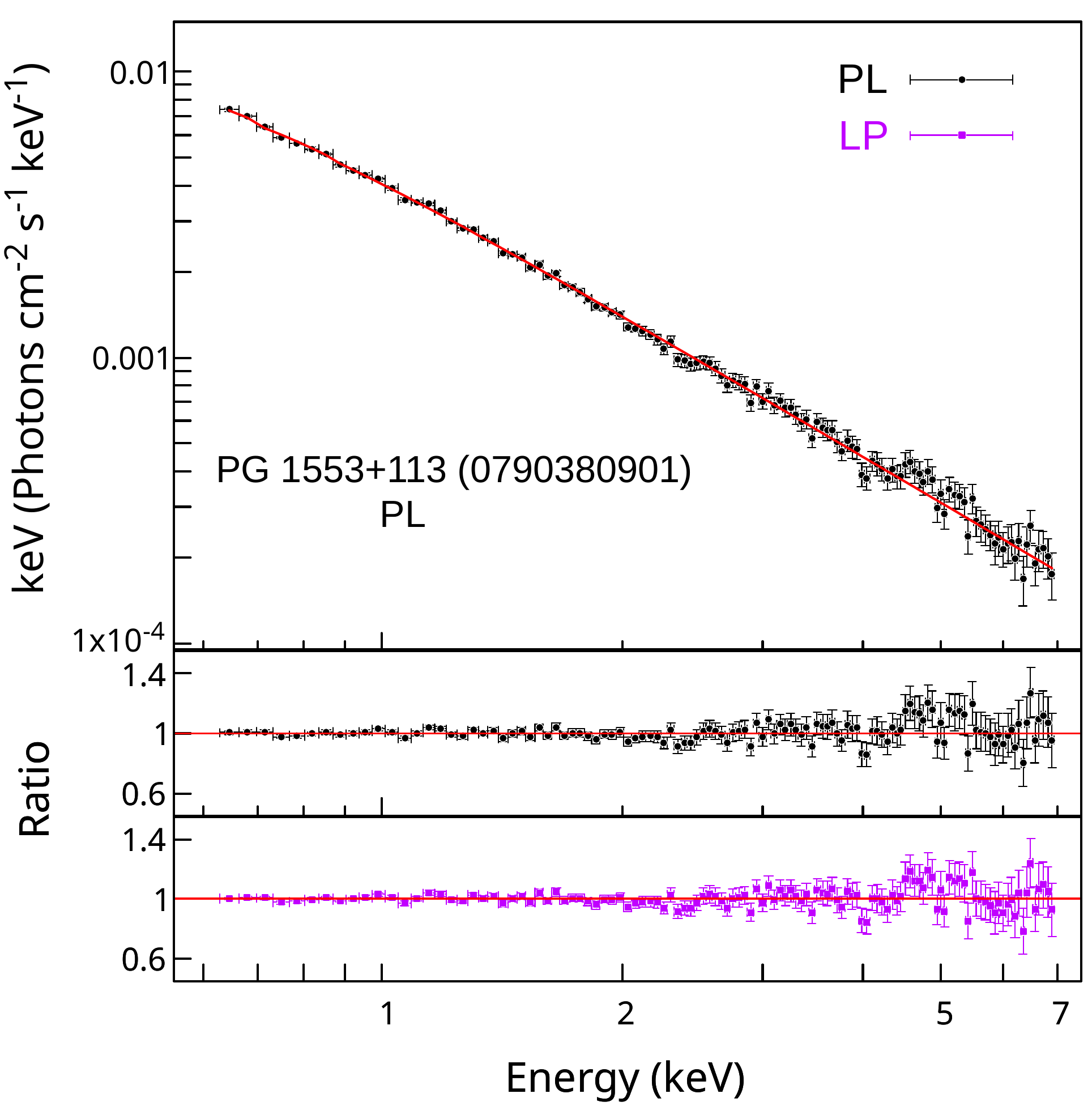}}
\includegraphics[width=8.5cm, height=7.5cm]{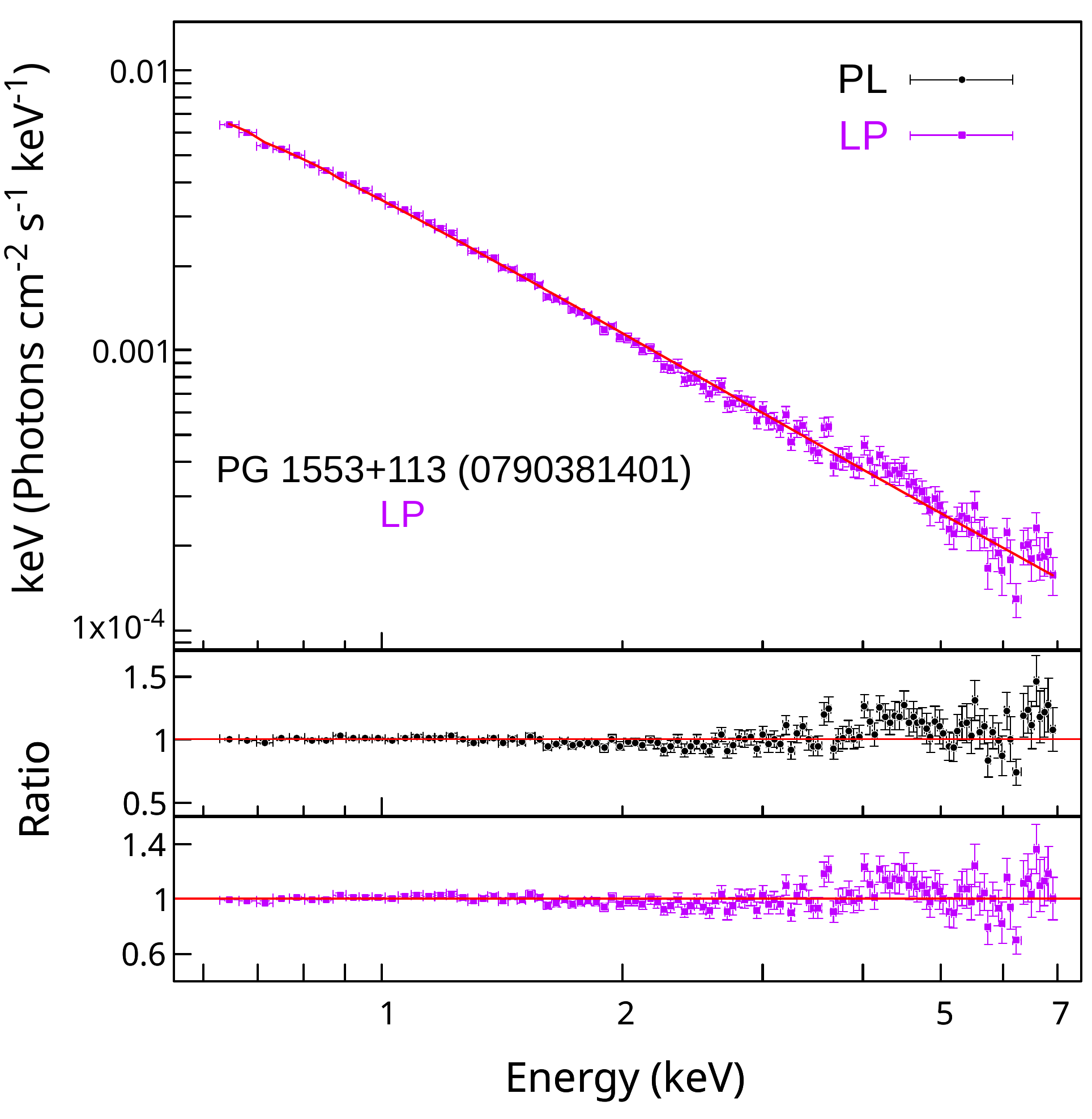}
\caption{Continued} 
\end{figure*}

\clearpage
\setcounter{figure}{0}
\begin{figure*}
\centering
{\vspace{-0.14cm} \includegraphics[width=8.5cm, height=7.5cm]{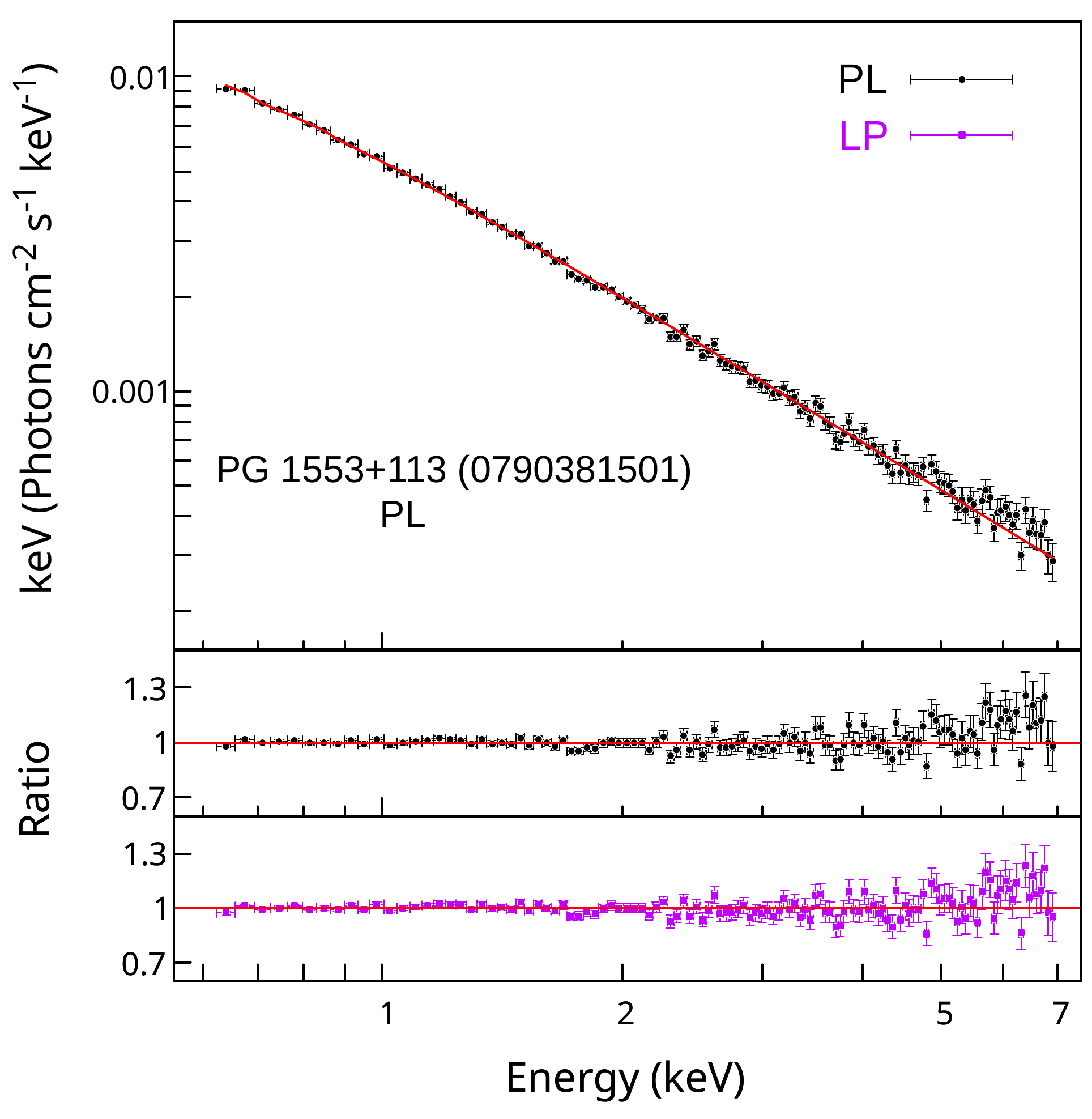}}
\includegraphics[width=8.5cm, height=7.5cm]{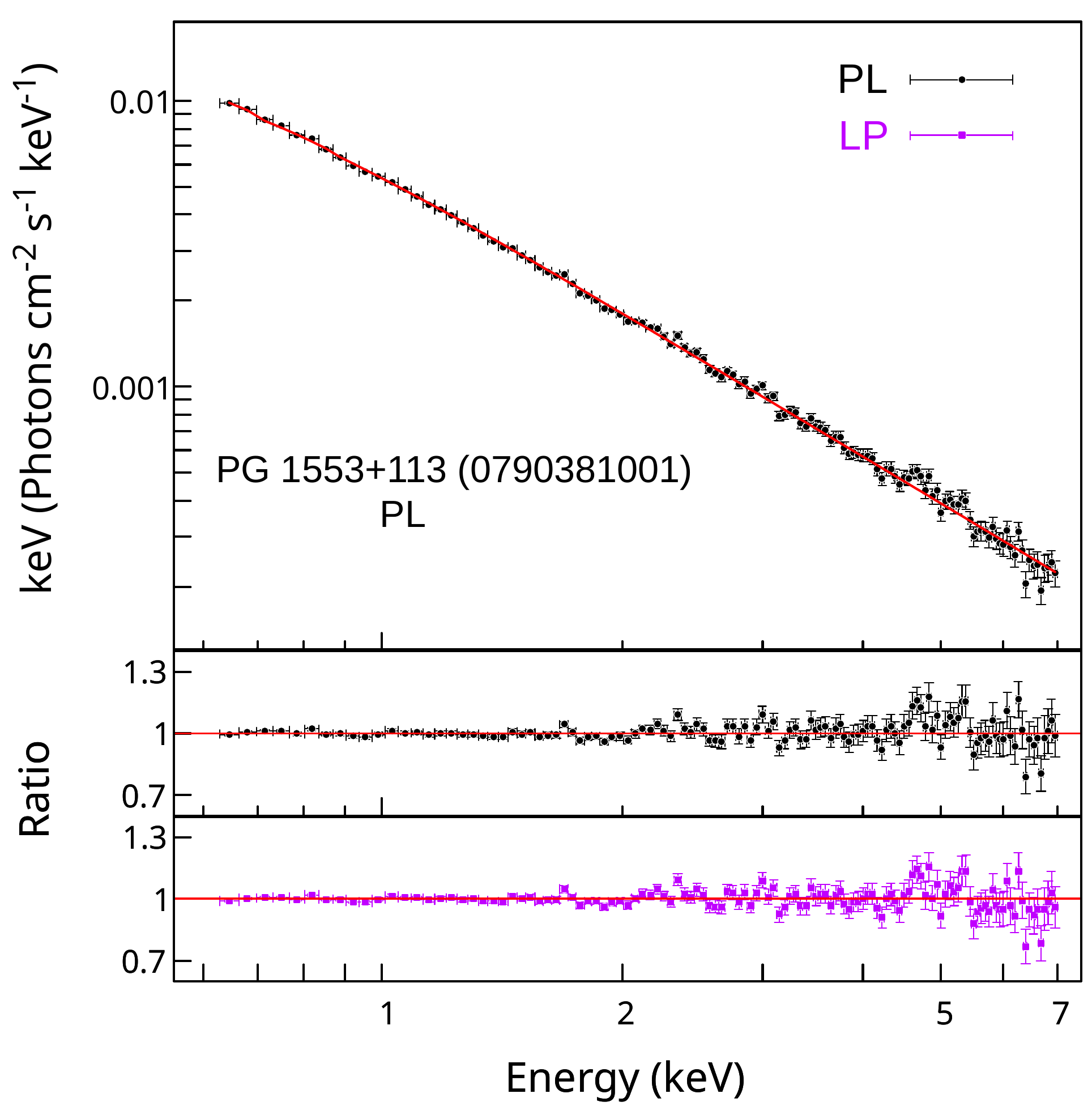}

{\vspace{-0.14cm} \includegraphics[width=8.5cm, height=7.5cm]{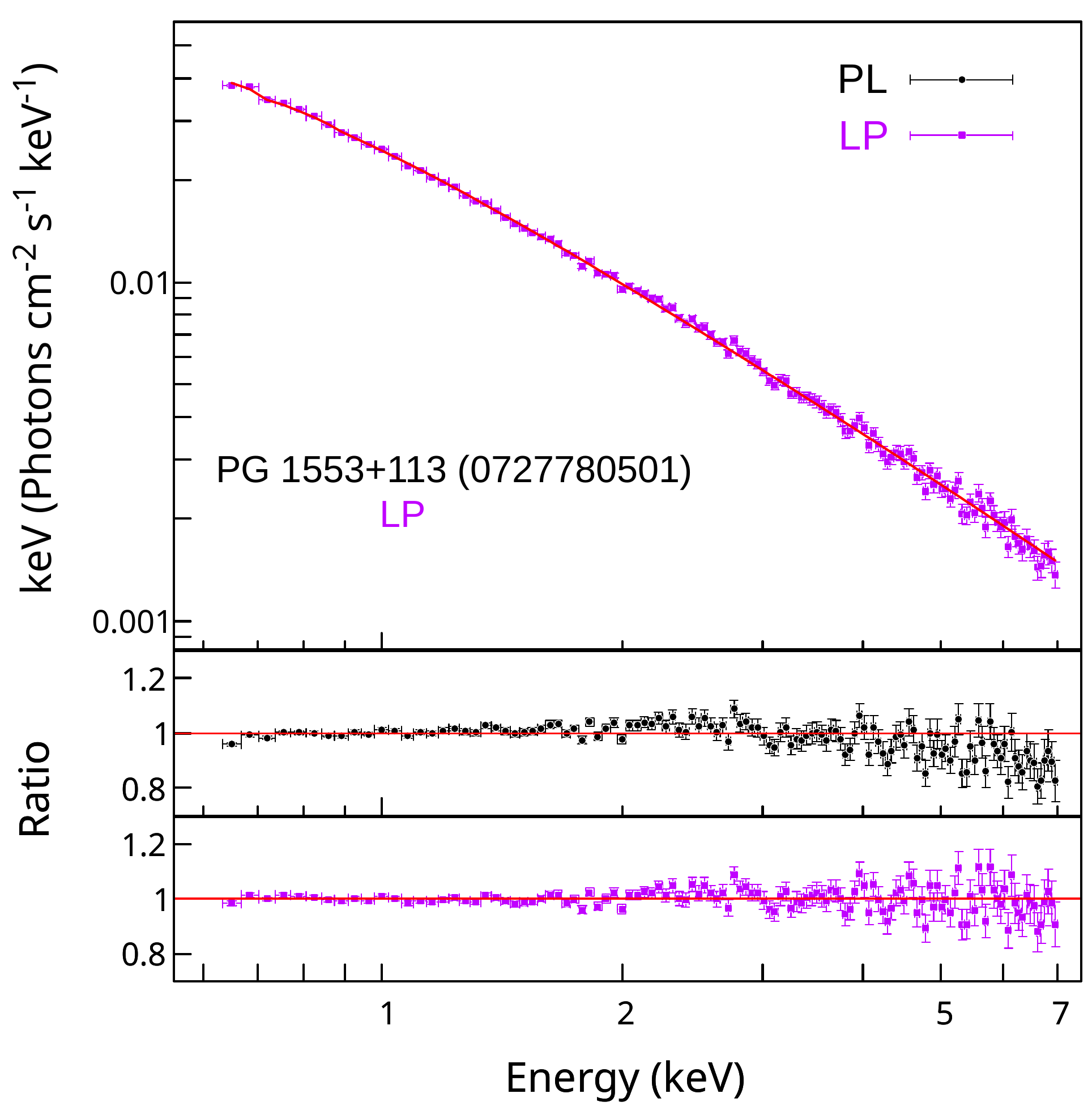}}
\includegraphics[width=8.5cm, height=7.5cm]{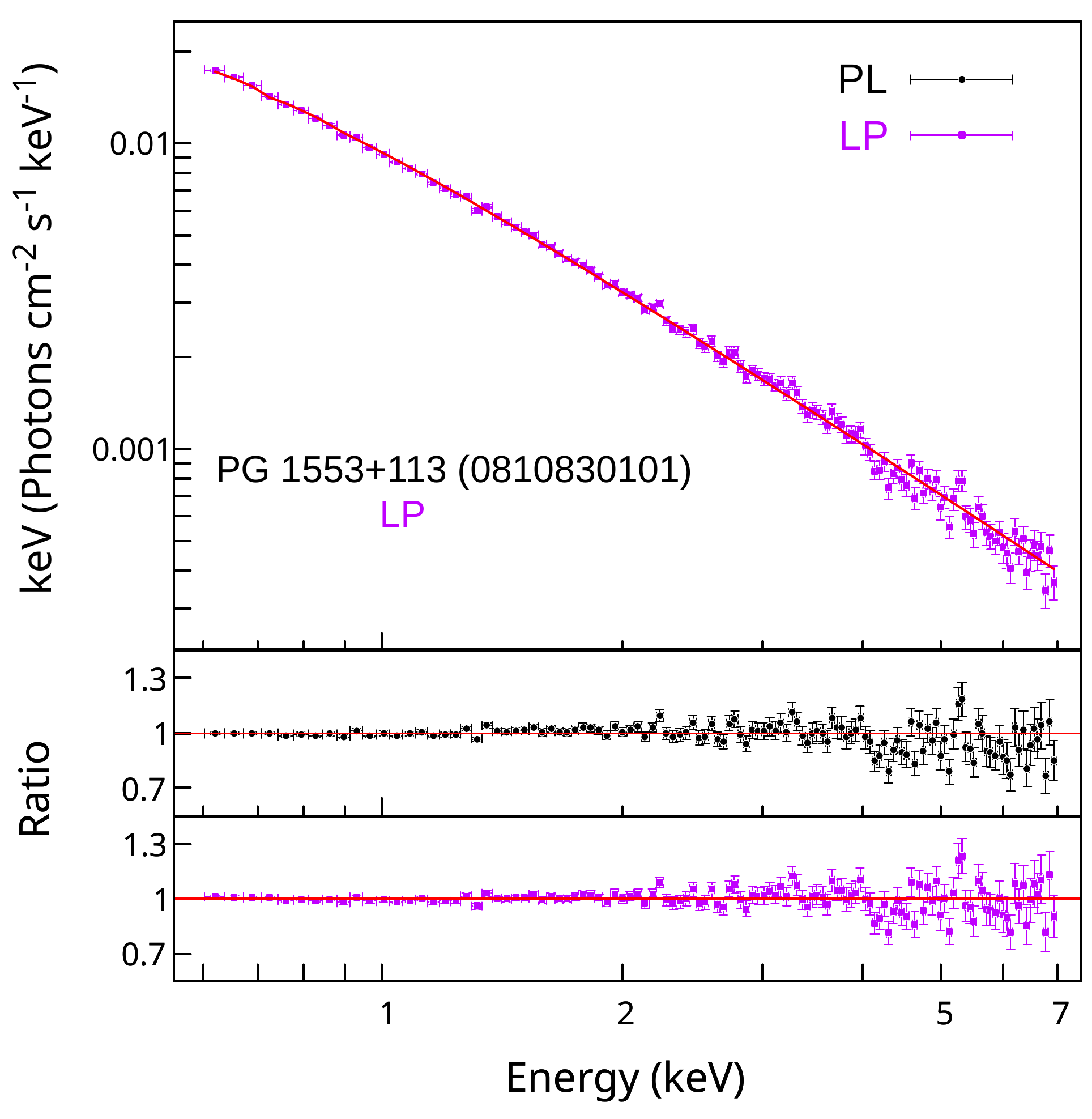}

{\vspace{-0.14cm} \includegraphics[width=8.5cm, height=7.5cm]{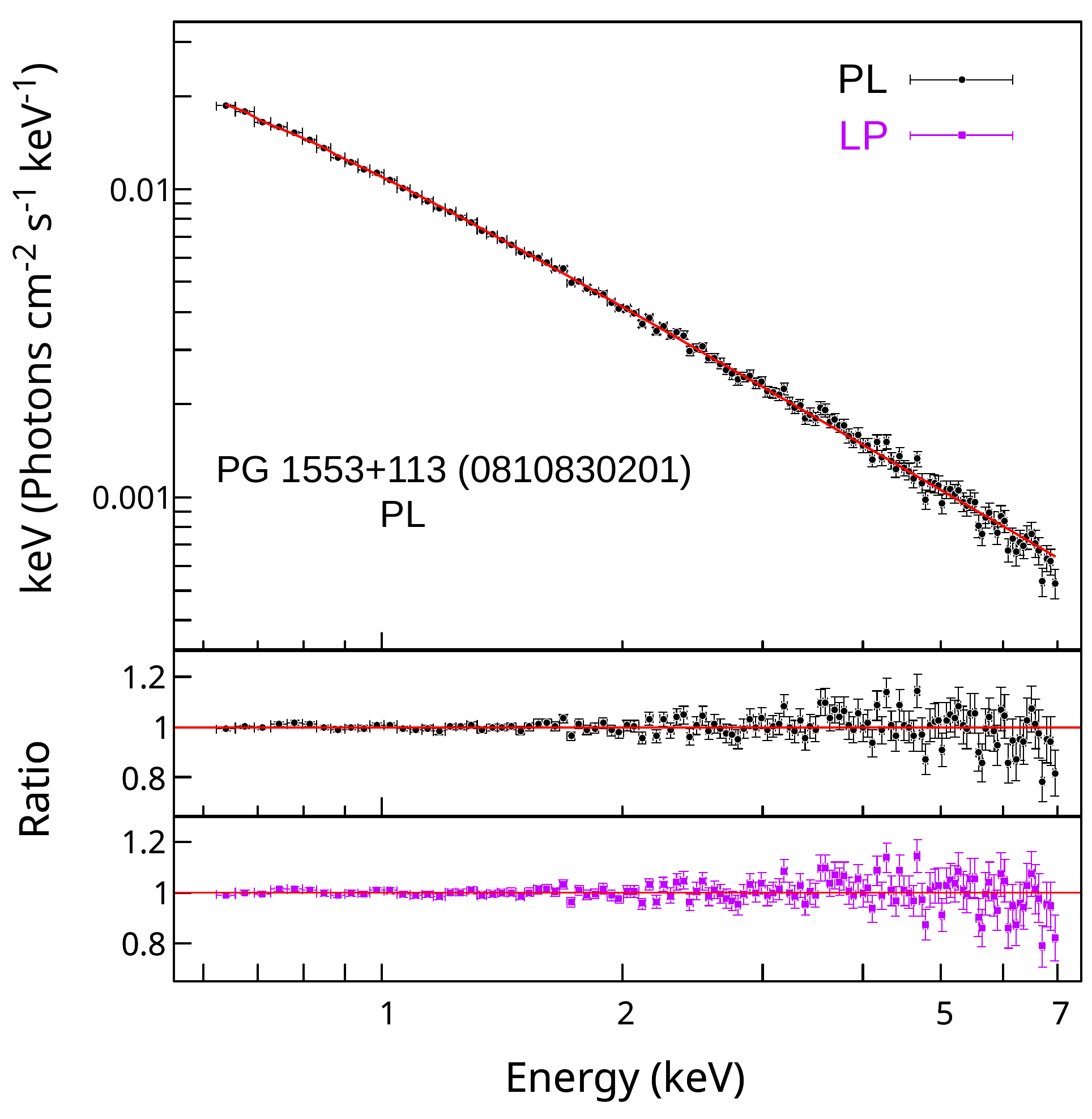}}
\includegraphics[width=8.5cm, height=7.5cm]{fig1_24.pdf}
\caption{Continued} 
\end{figure*}

\clearpage
\setcounter{figure}{0}
\begin{figure*}
\centering
{\vspace{-0.14cm} \includegraphics[width=8.5cm, height=7.5cm]{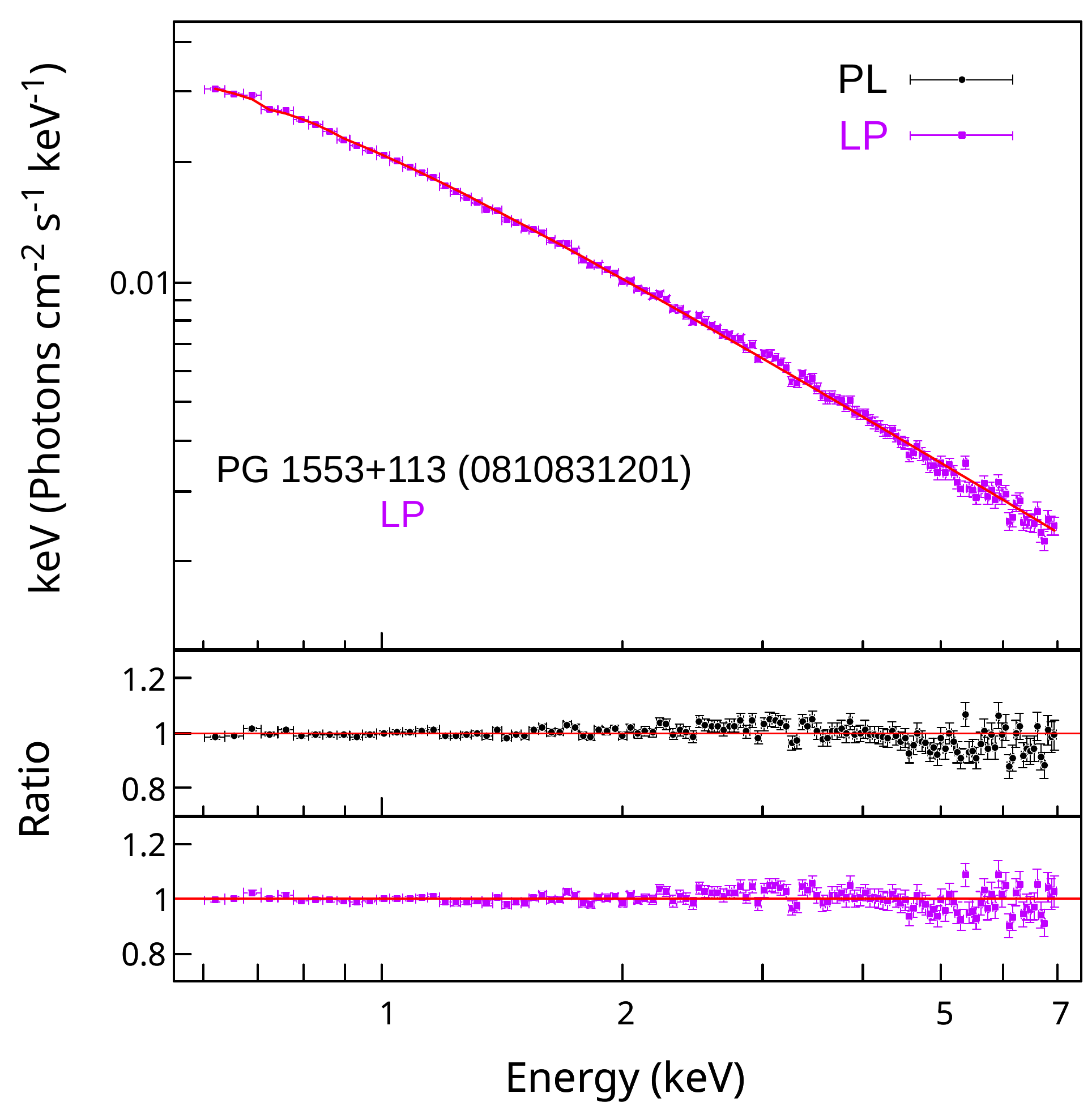}}
\includegraphics[width=8.5cm, height=7.5cm]{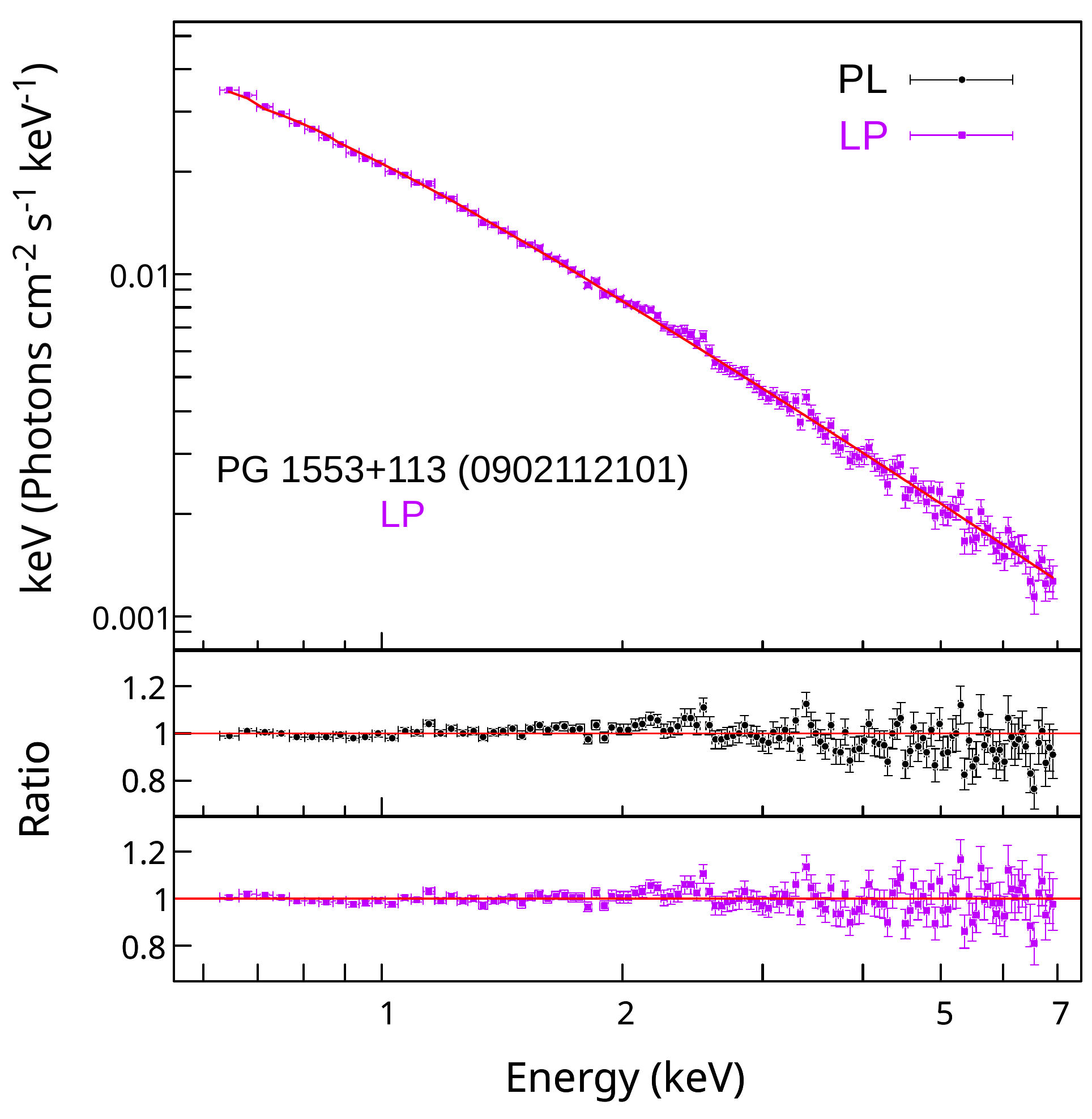}

{\vspace{-0.14cm} \includegraphics[width=8.5cm, height=7.5cm]{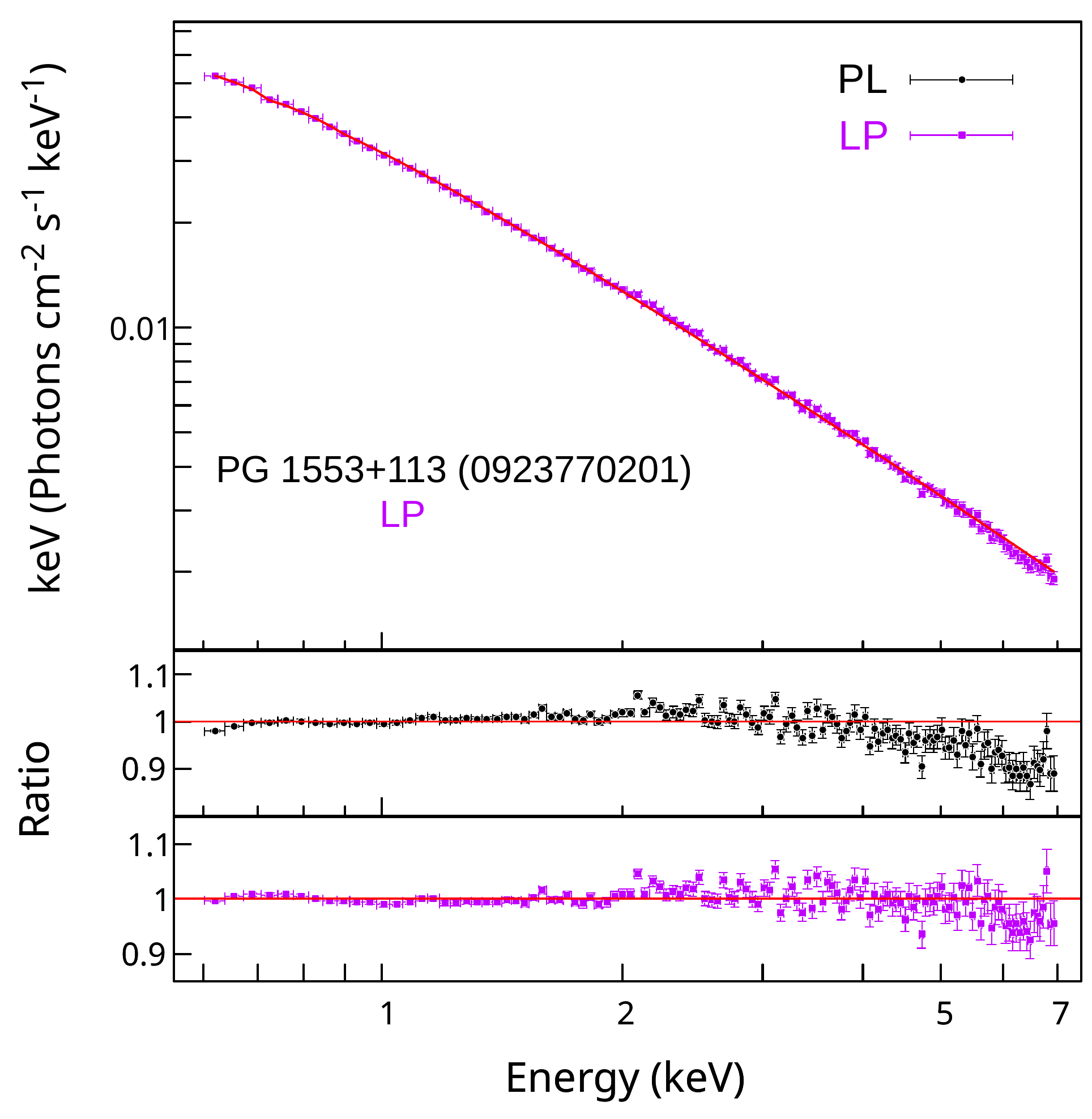}}
\includegraphics[width=8.5cm, height=7.5cm]{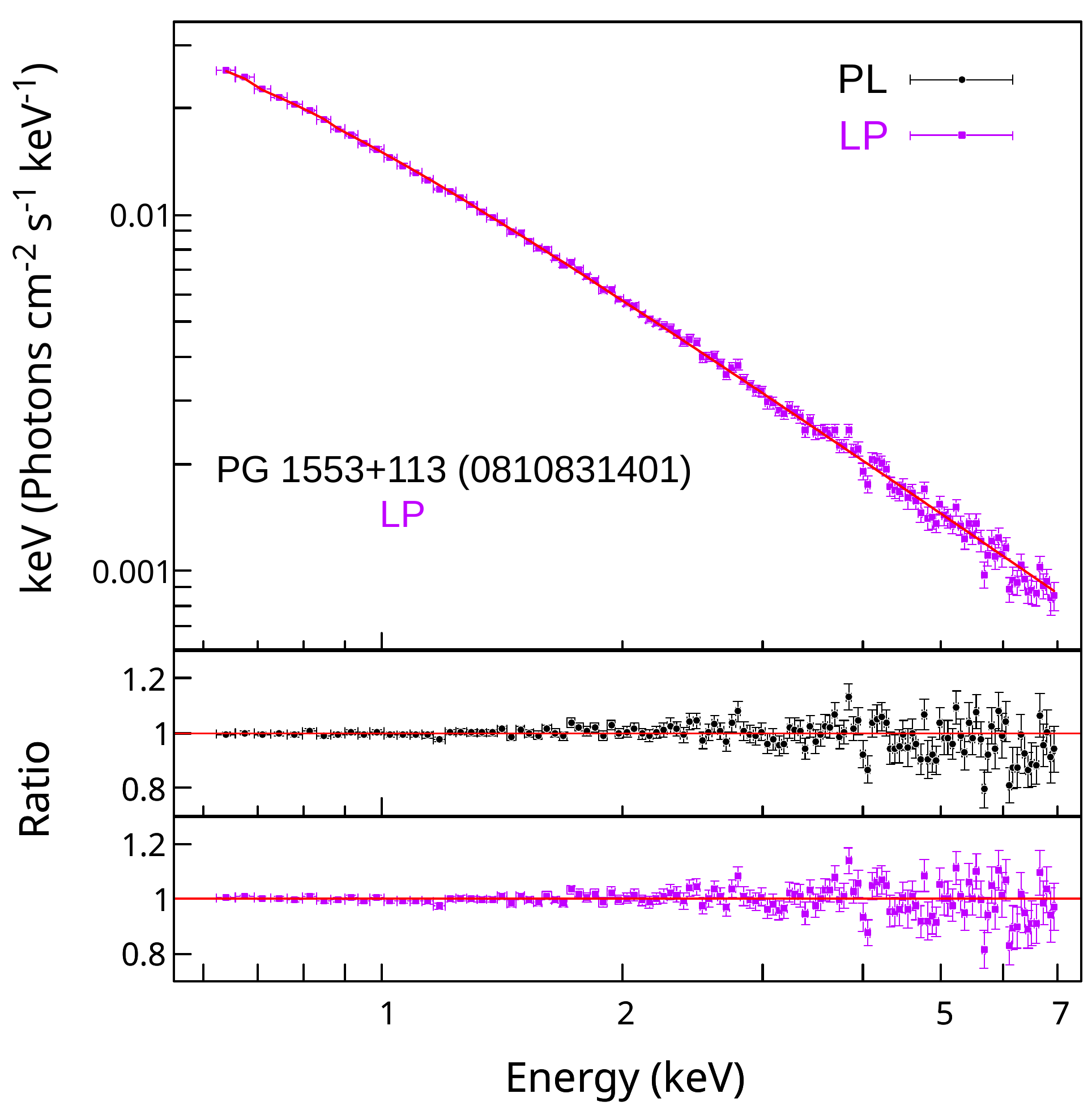}

{\vspace{-0.14cm} \includegraphics[width=8.5cm, height=7.5cm]{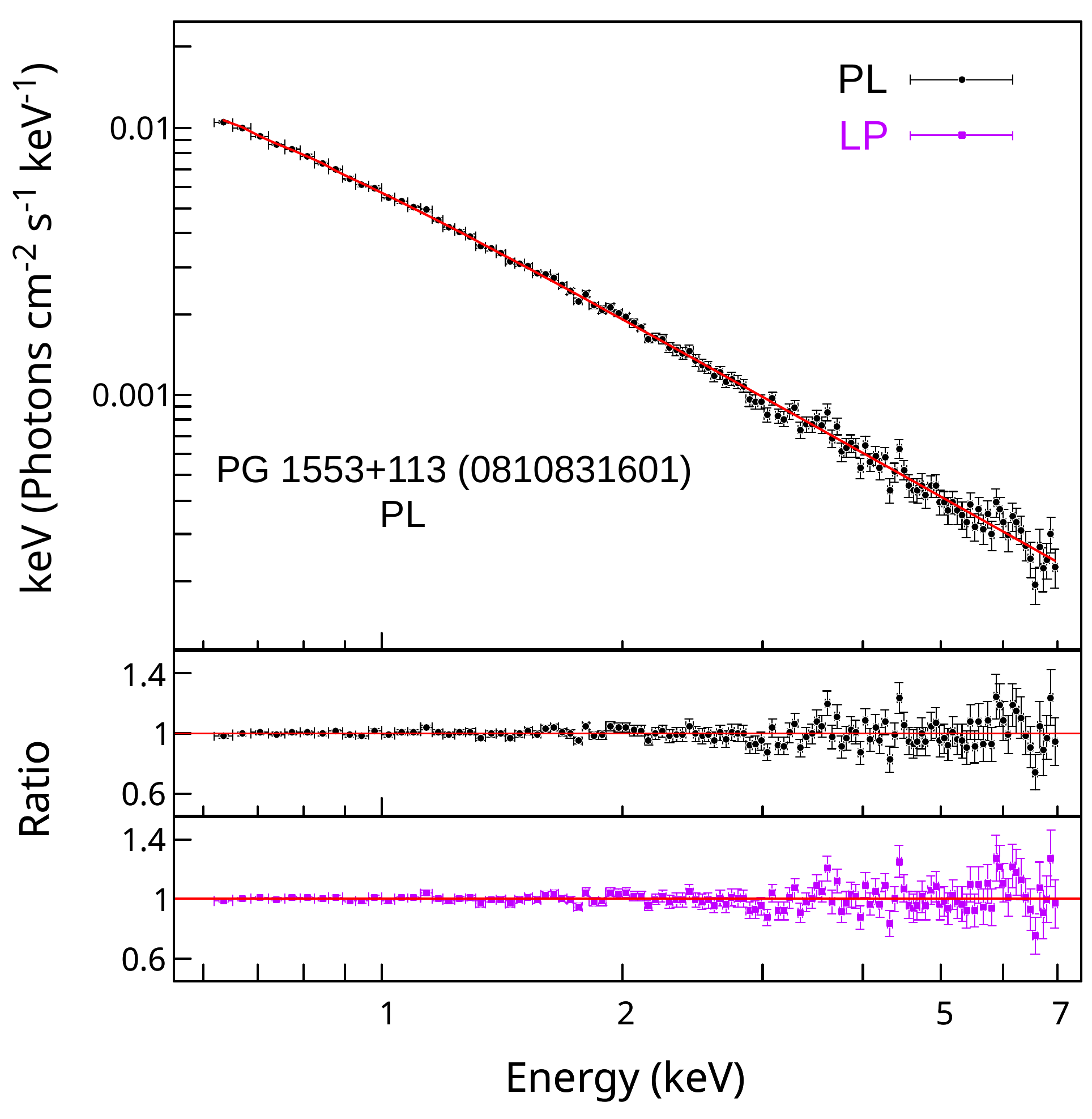}}
\includegraphics[width=8.5cm, height=7.5cm]{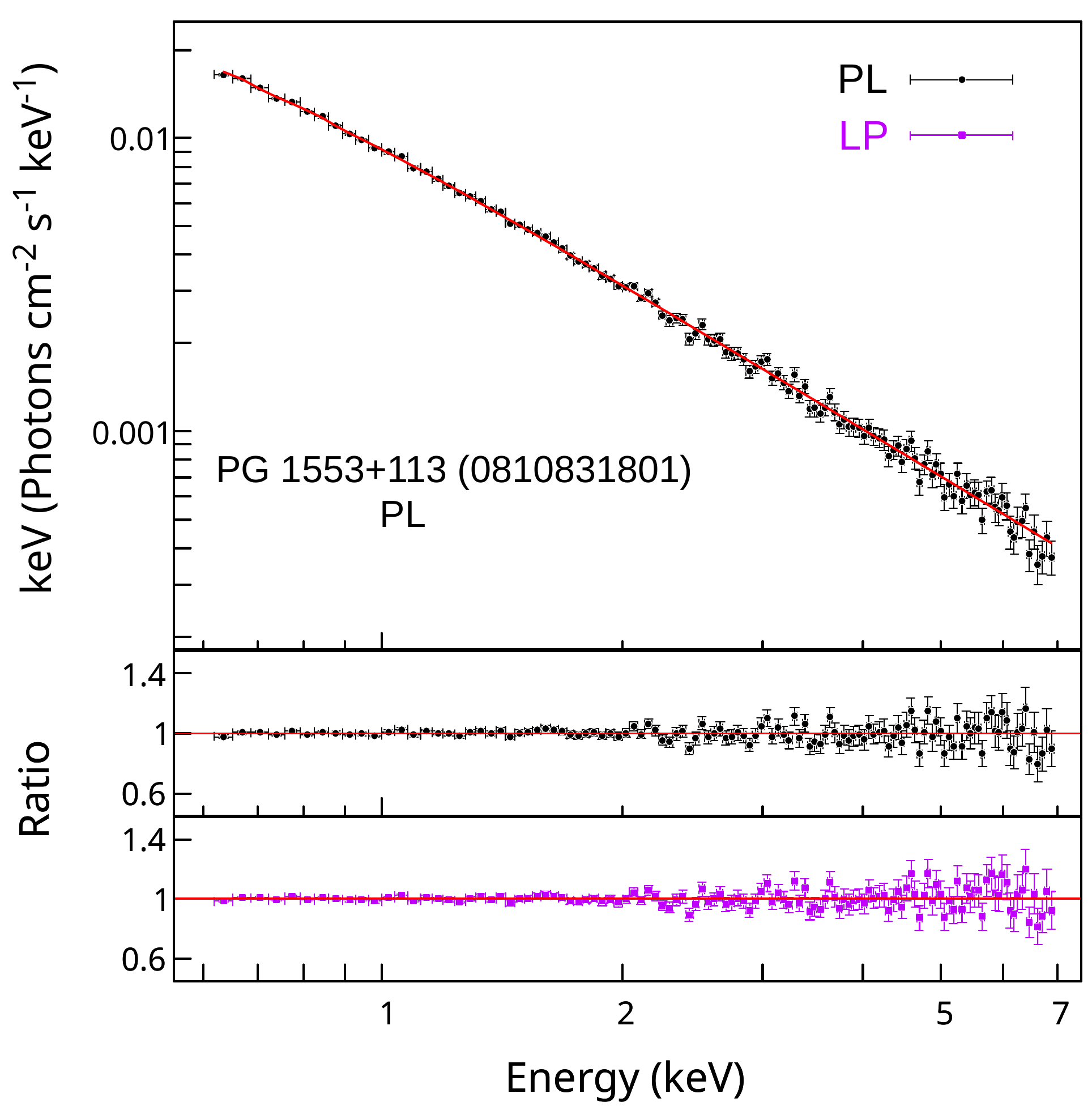}
\caption{Continued} 
\end{figure*}

\newpage

\setcounter{figure}{1}
\begin{figure*}
\centering
{\vspace{-0.14cm} \includegraphics[width=8.5cm, height=7.5cm]{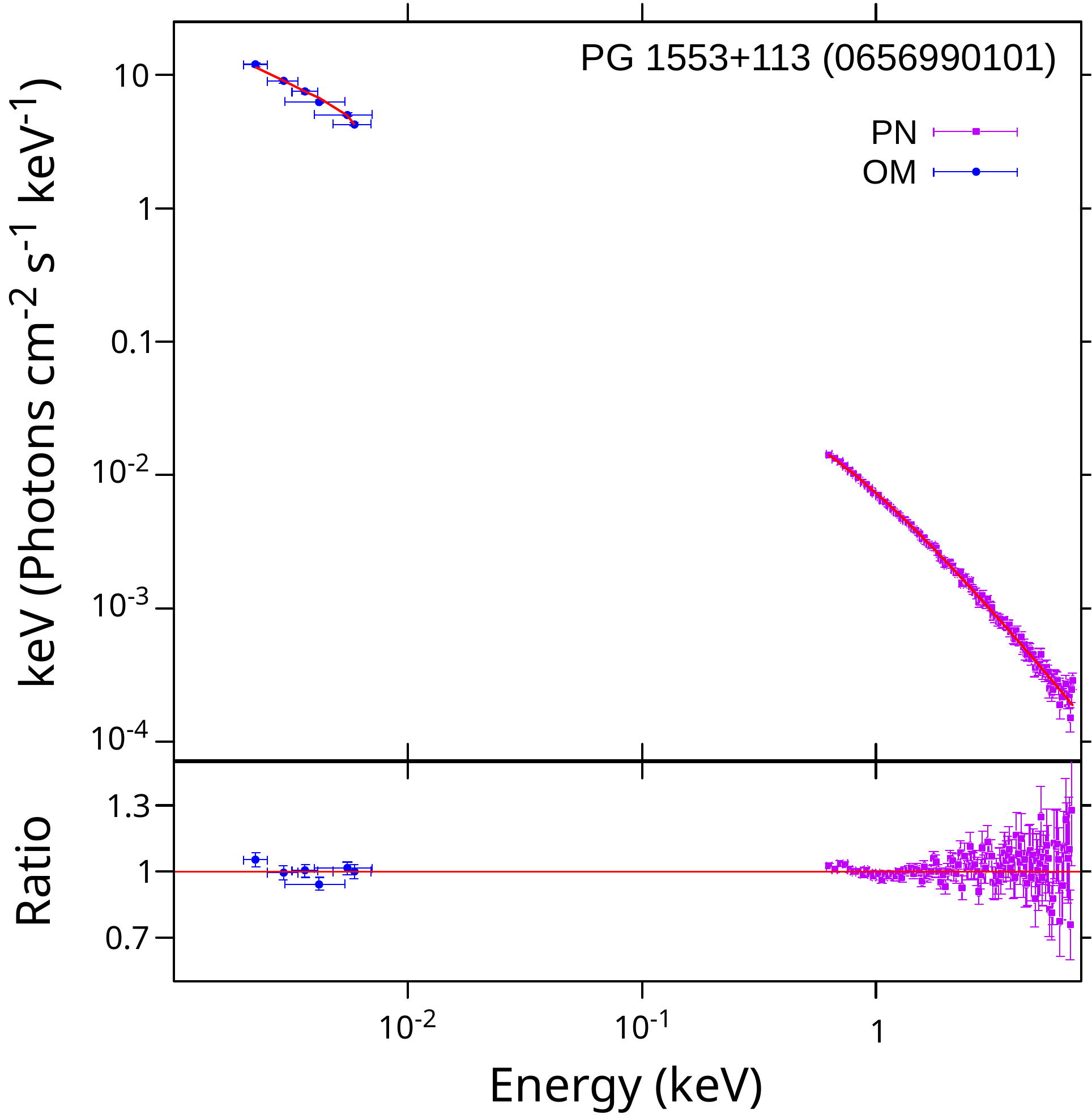}}
\includegraphics[width=8.5cm, height=7.5cm]{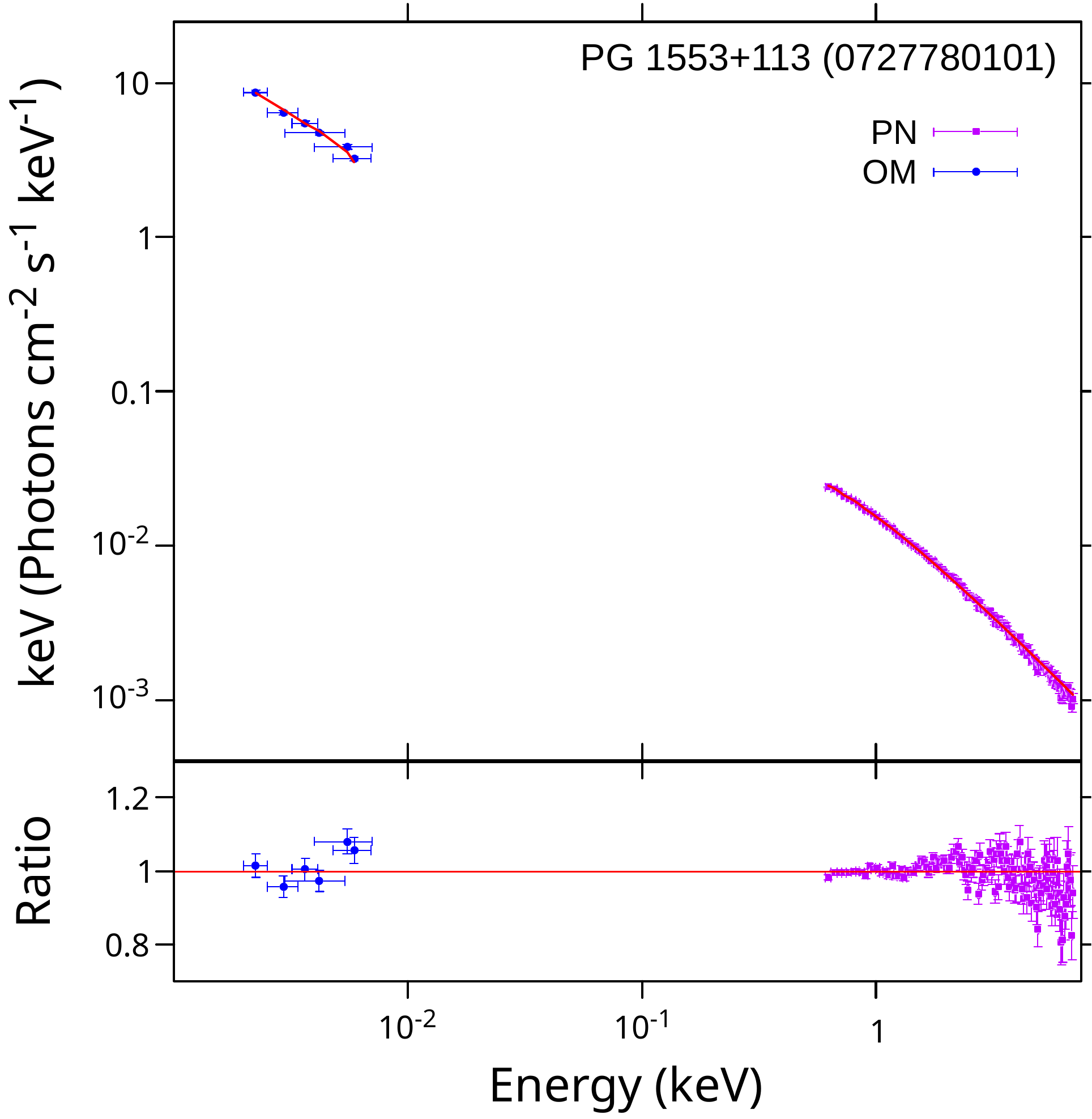}

{\vspace{-0.14cm} \includegraphics[width=8.5cm, height=7.5cm]{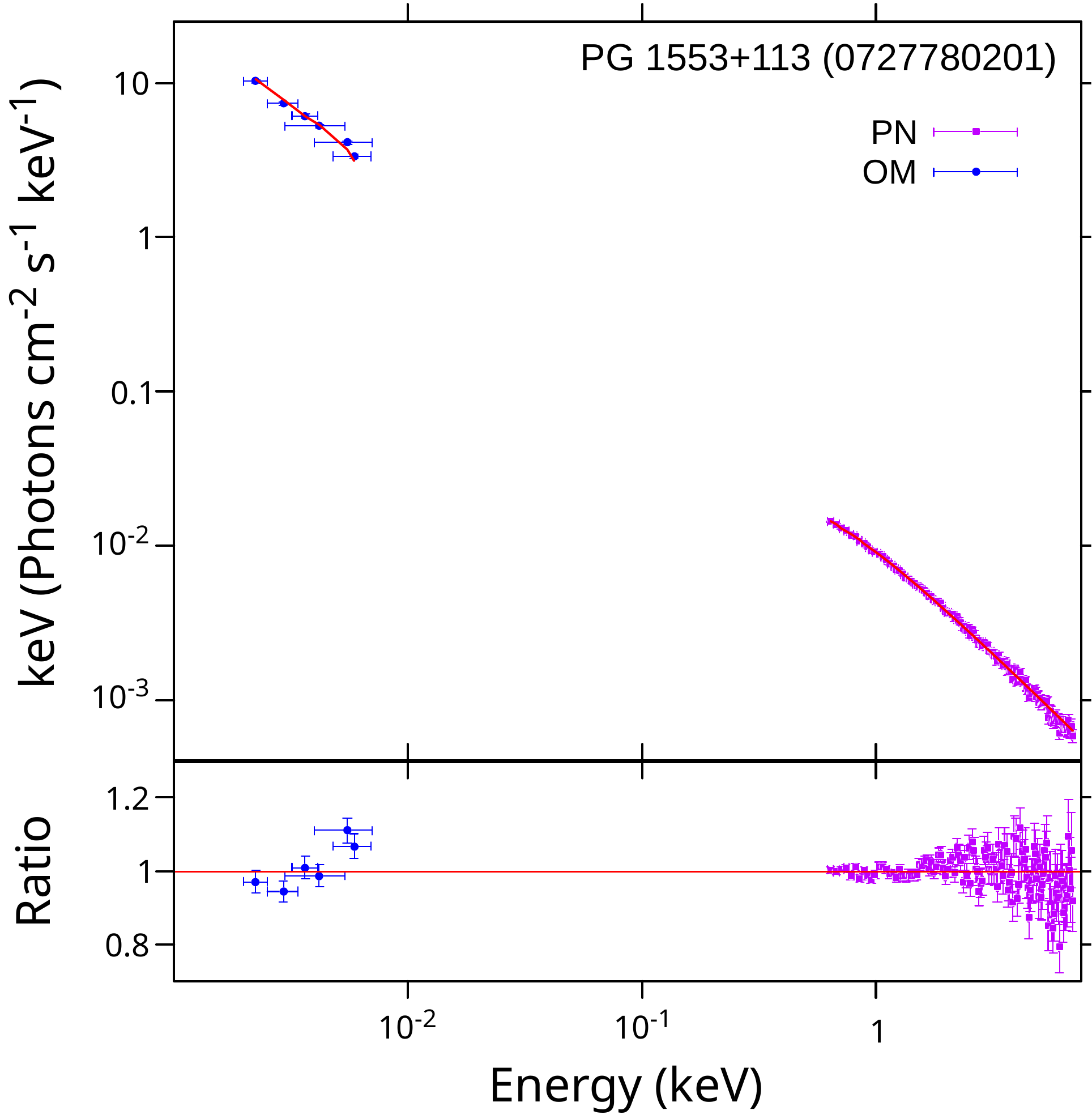}}
\includegraphics[width=8.5cm, height=7.5cm]{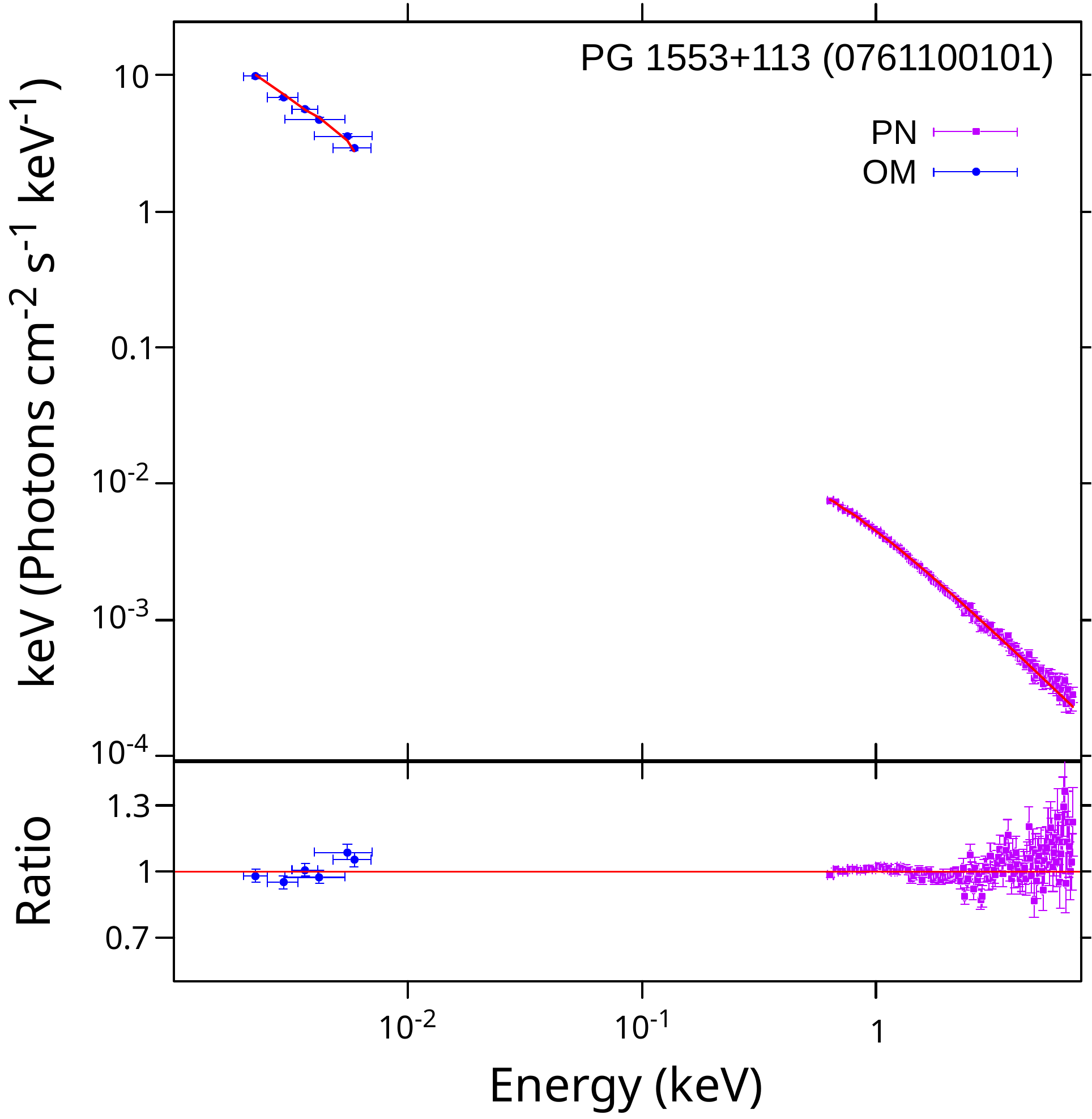}

{\vspace{-0.14cm} \includegraphics[width=8.5cm, height=7.5cm]{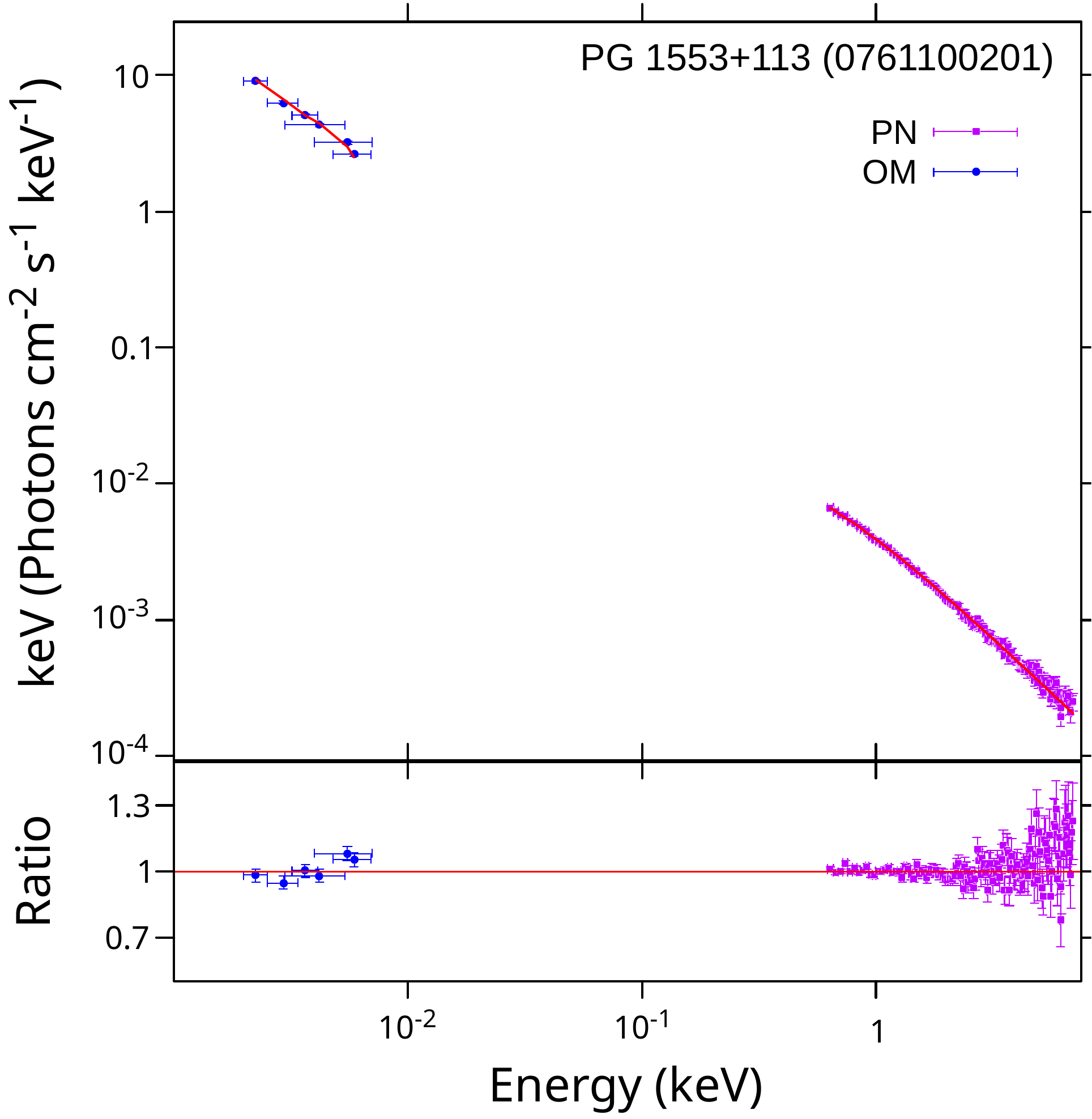}}
\includegraphics[width=8.5cm, height=7.5cm]{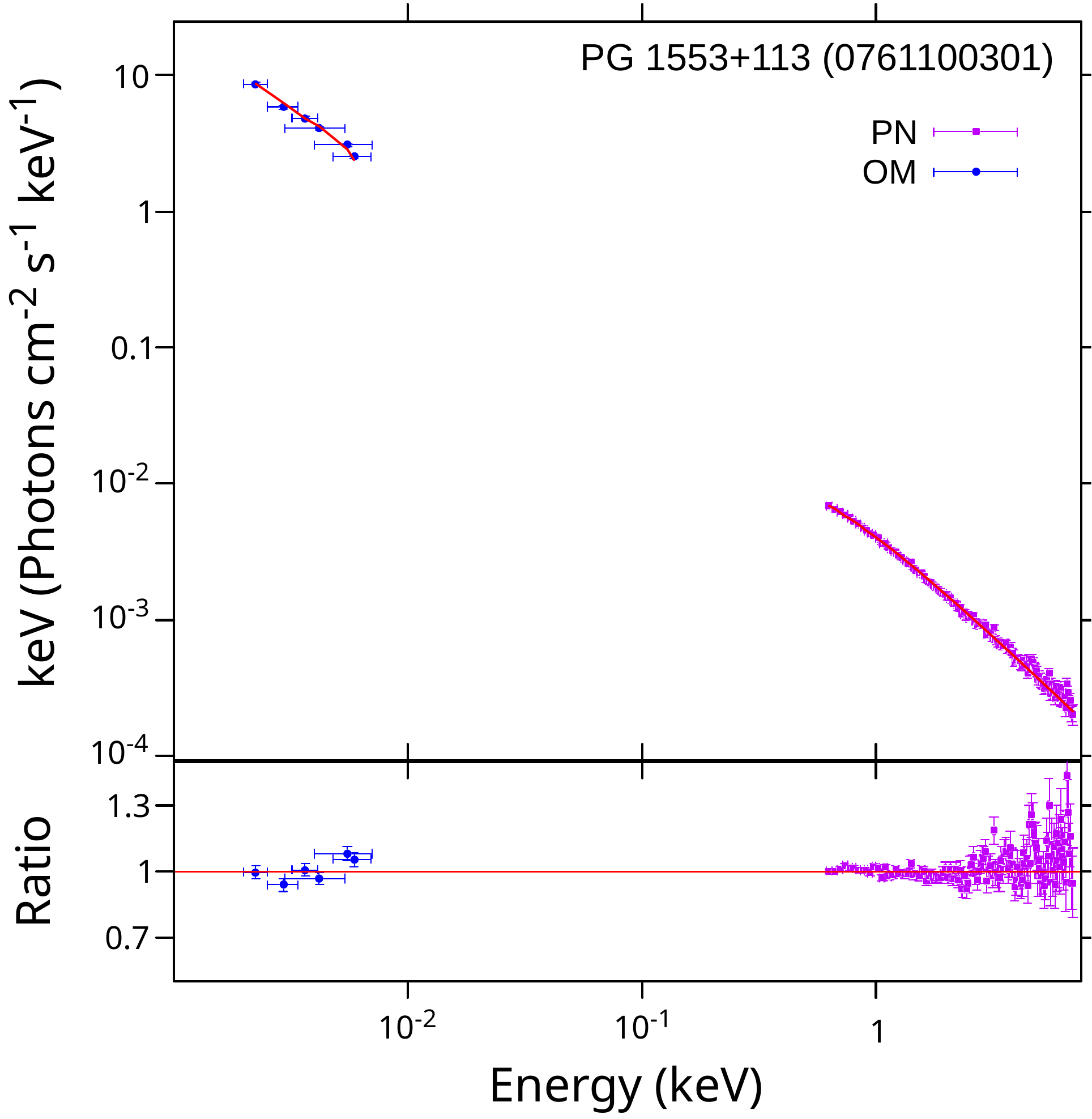}
\vspace{-0.1cm}
\caption{PN+OM joint spectral fitting of the \textit{XMM-Newton EPIC PN} pointed observations of PG 1553+113 with log parabolic  model(\textit{tbabs*redden*logpar}). The upper panel of each plot represents photon flux, while the lower panel represents the data-to-model ratio. For the fitting, reddening E(B-V) is fixed at 0.0446 mag, and galactic absorption n$_H$ is fixed at  3.61$\times 10^20 cm^-2$. PN data in the energy range 0.6-7 keV is represented by dark-magenta filled circles, and the OM data by blue filled circles. Observation IDs are shown in the top right of the plot.\label{A2}}   
\end{figure*}

\clearpage
\setcounter{figure}{1}
\begin{figure*}
\centering
{\vspace{-0.14cm} \includegraphics[width=8.5cm, height=7.5cm]{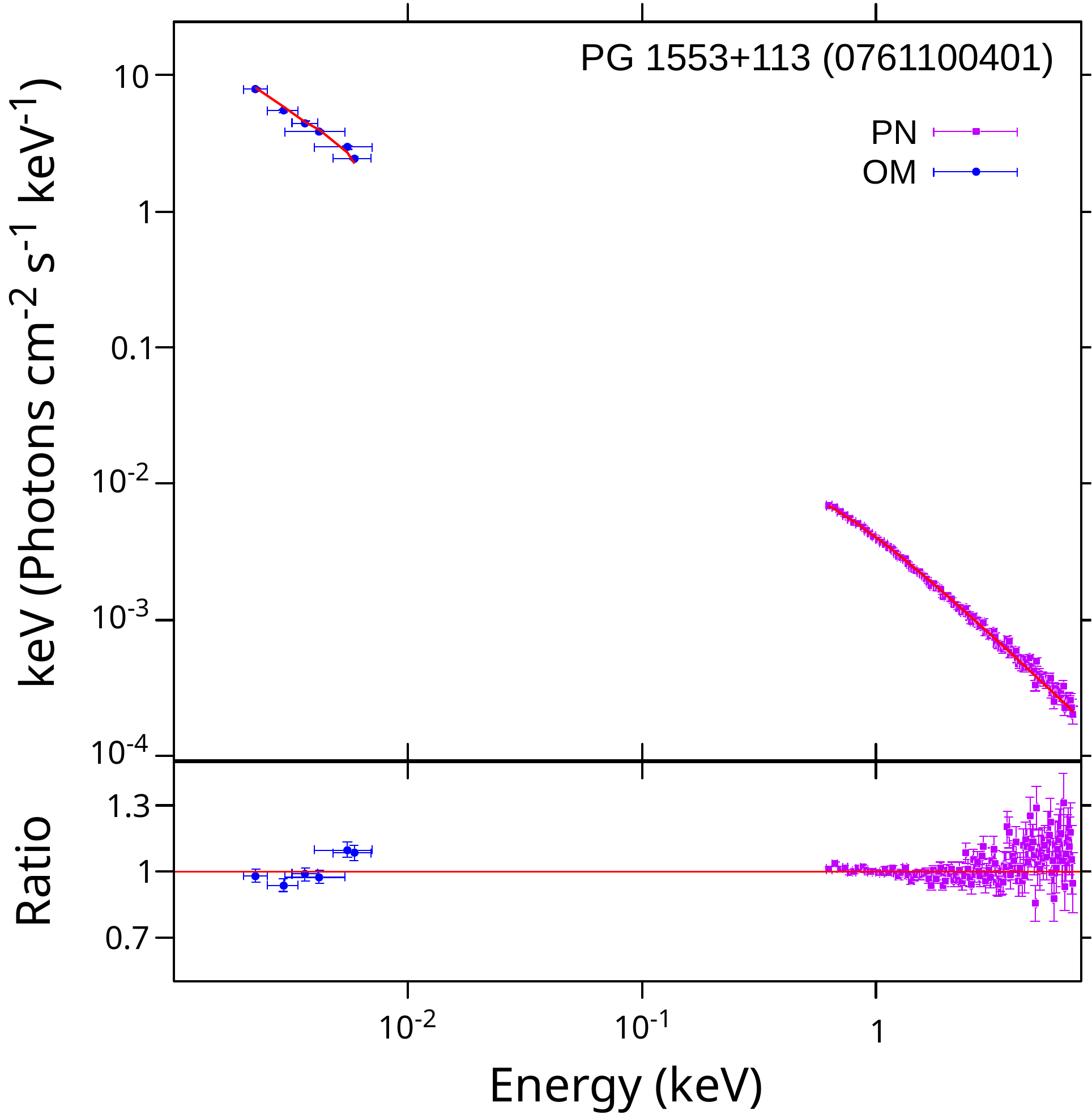}}
\includegraphics[width=8.5cm, height=7.5cm]{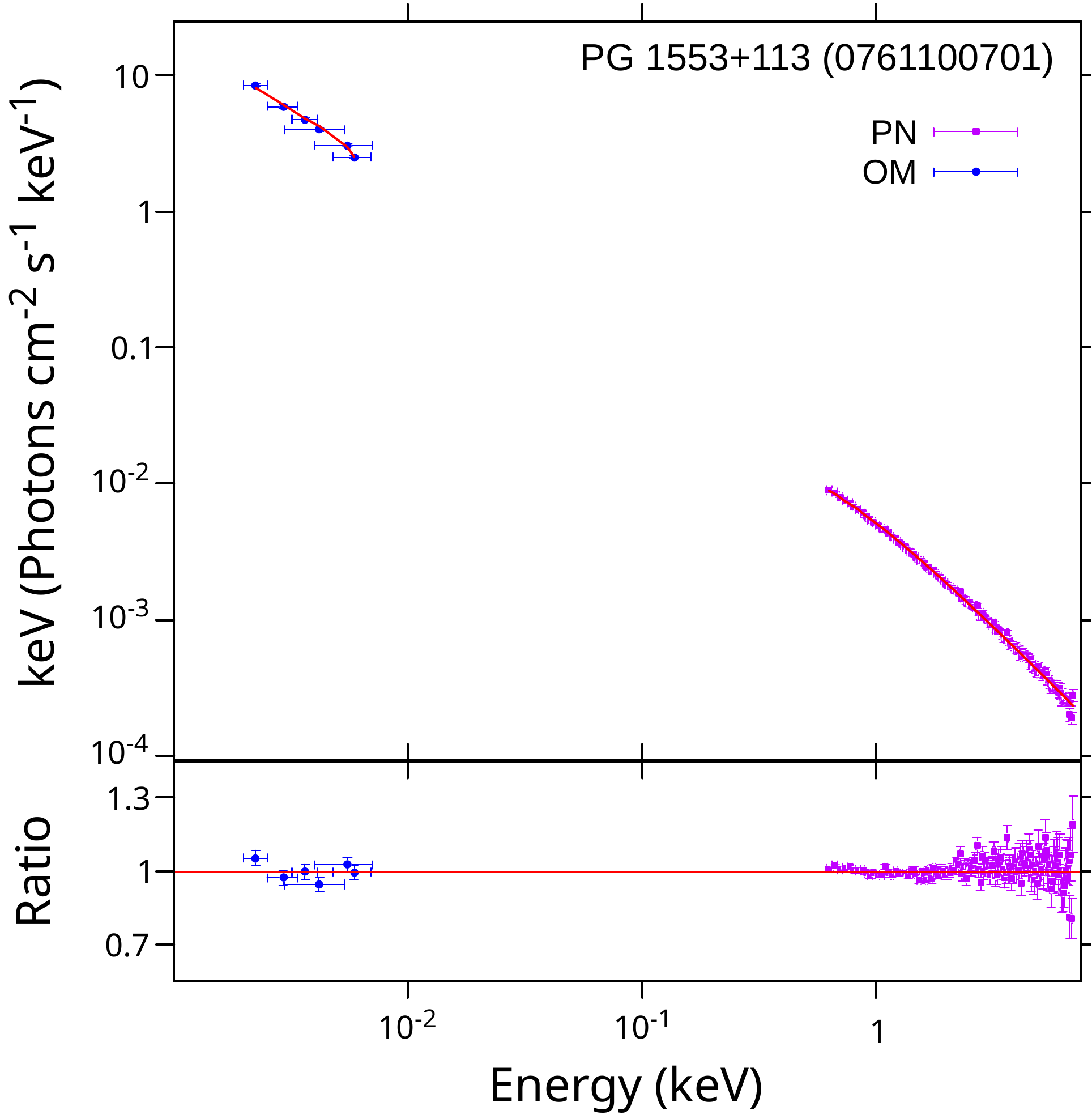}

{\vspace{-0.14cm} \includegraphics[width=8.5cm, height=7.5cm]{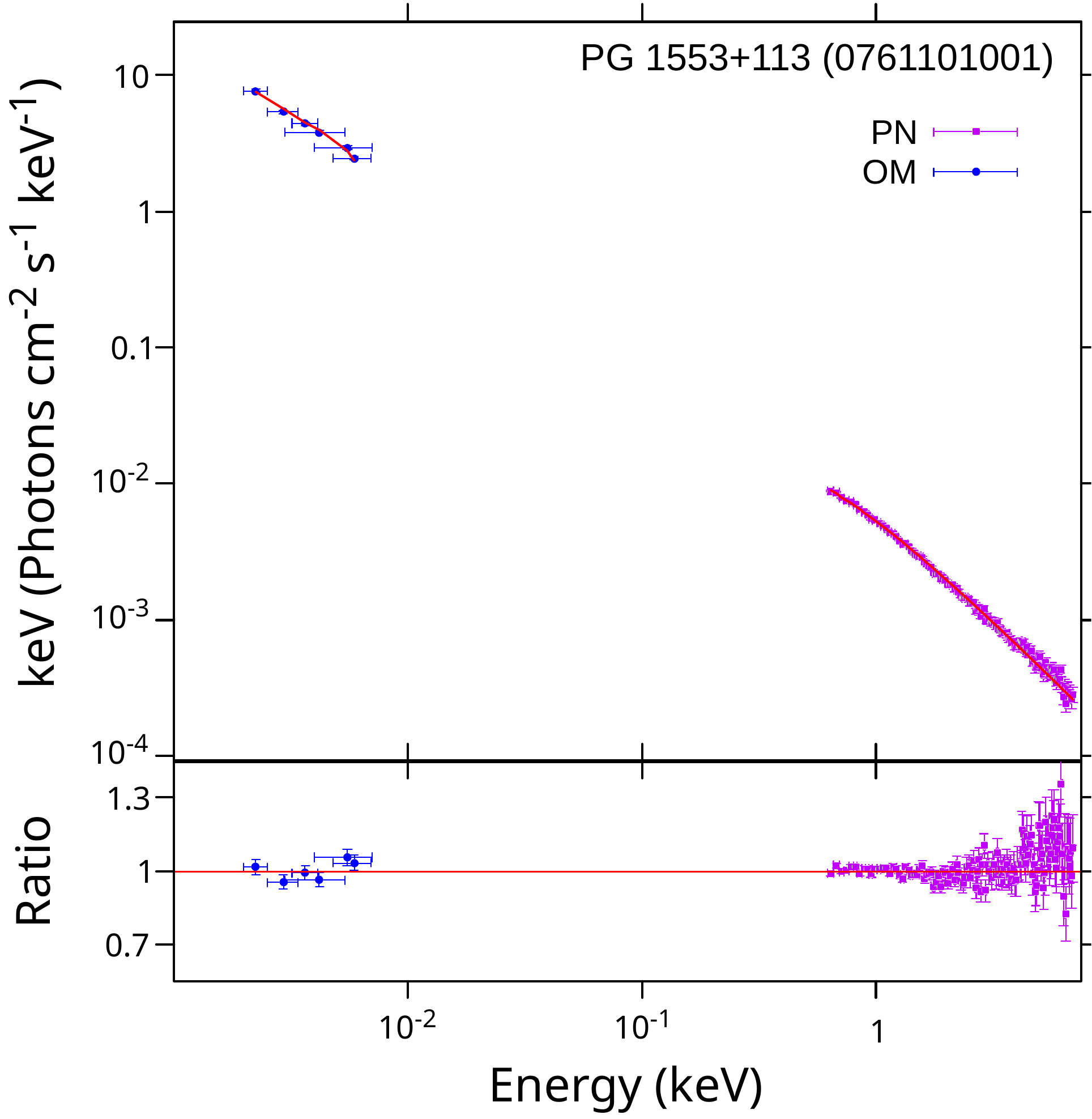}}
\includegraphics[width=8.5cm, height=7.5cm]{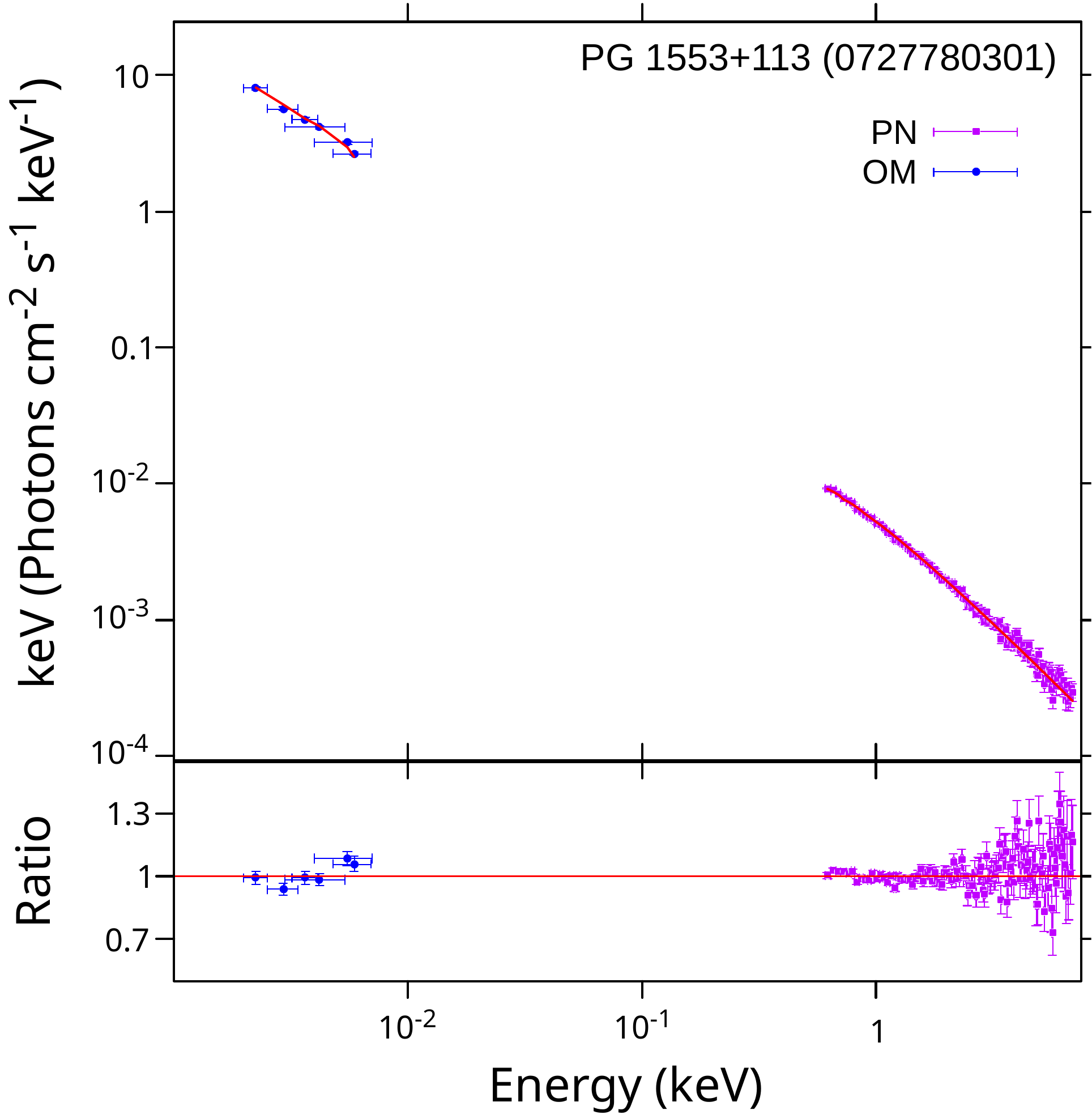}

{\vspace{-0.14cm} \includegraphics[width=8.5cm, height=7.5cm]{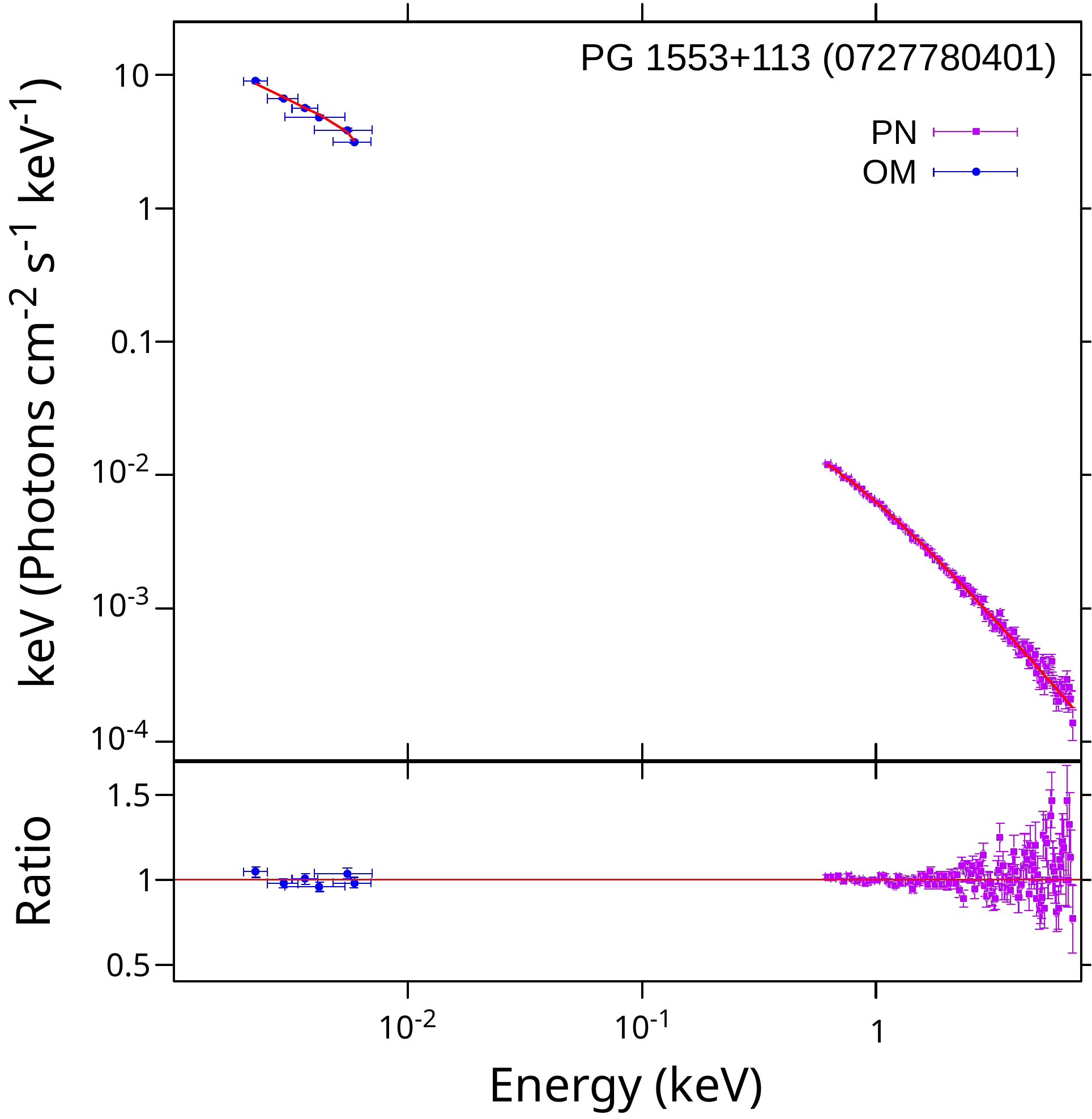}}
\includegraphics[width=8.5cm, height=7.5cm]{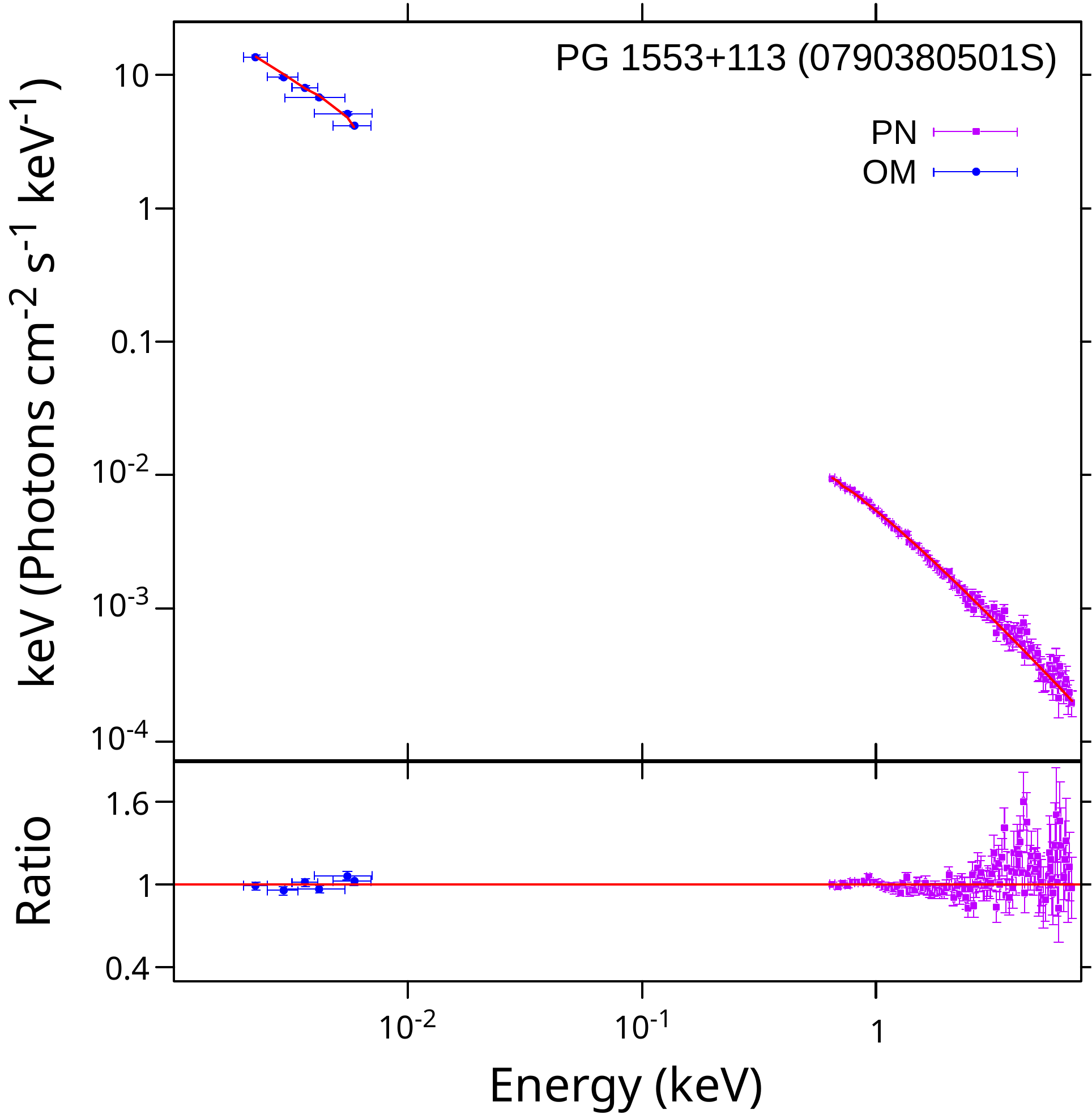}
\caption{Continued} 
\end{figure*}

\clearpage
\setcounter{figure}{1}
\begin{figure*}
\centering
{\vspace{-0.14cm} \includegraphics[width=8.5cm, height=7.5cm]{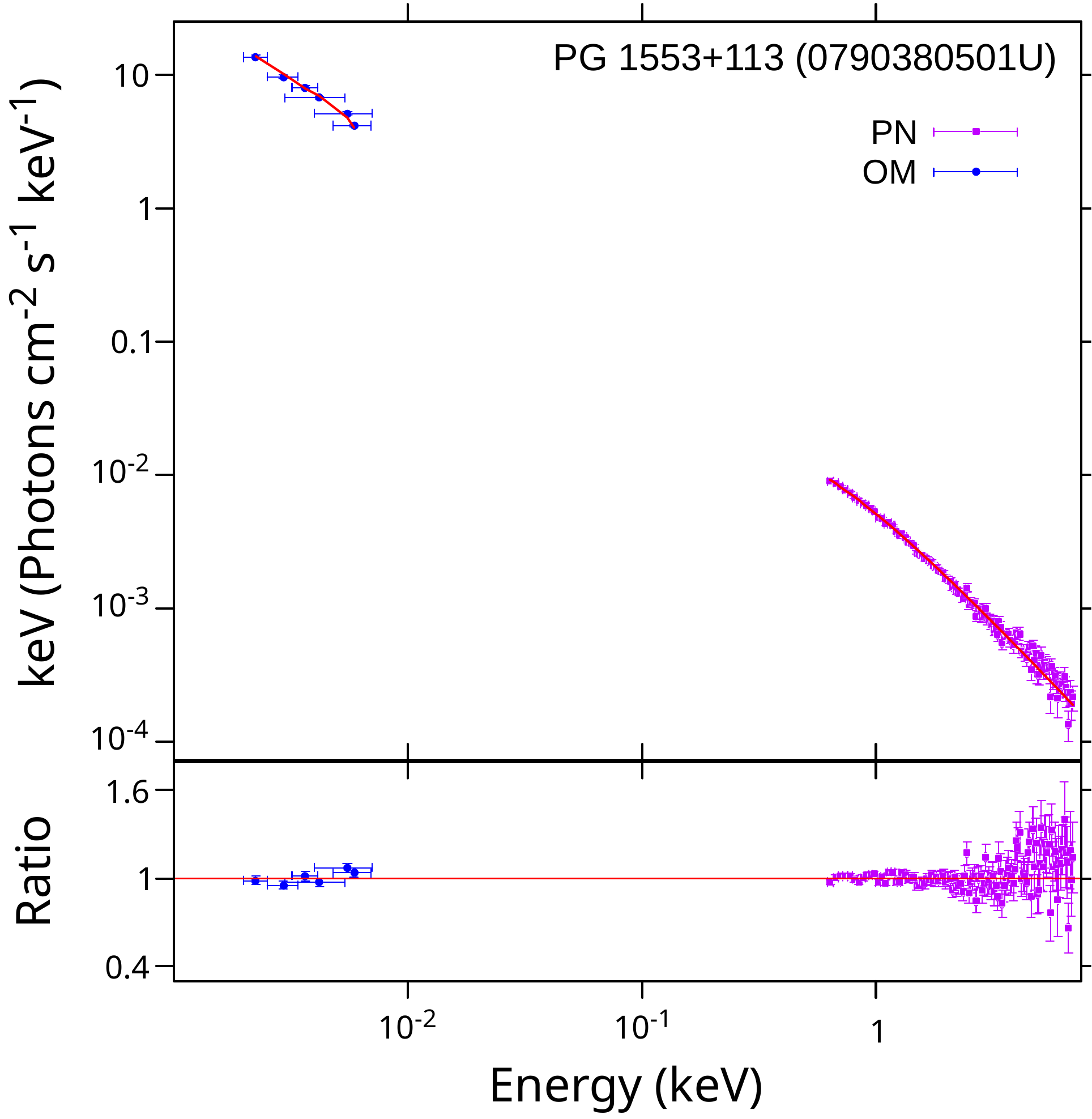}}
\includegraphics[width=8.5cm, height=7.5cm]{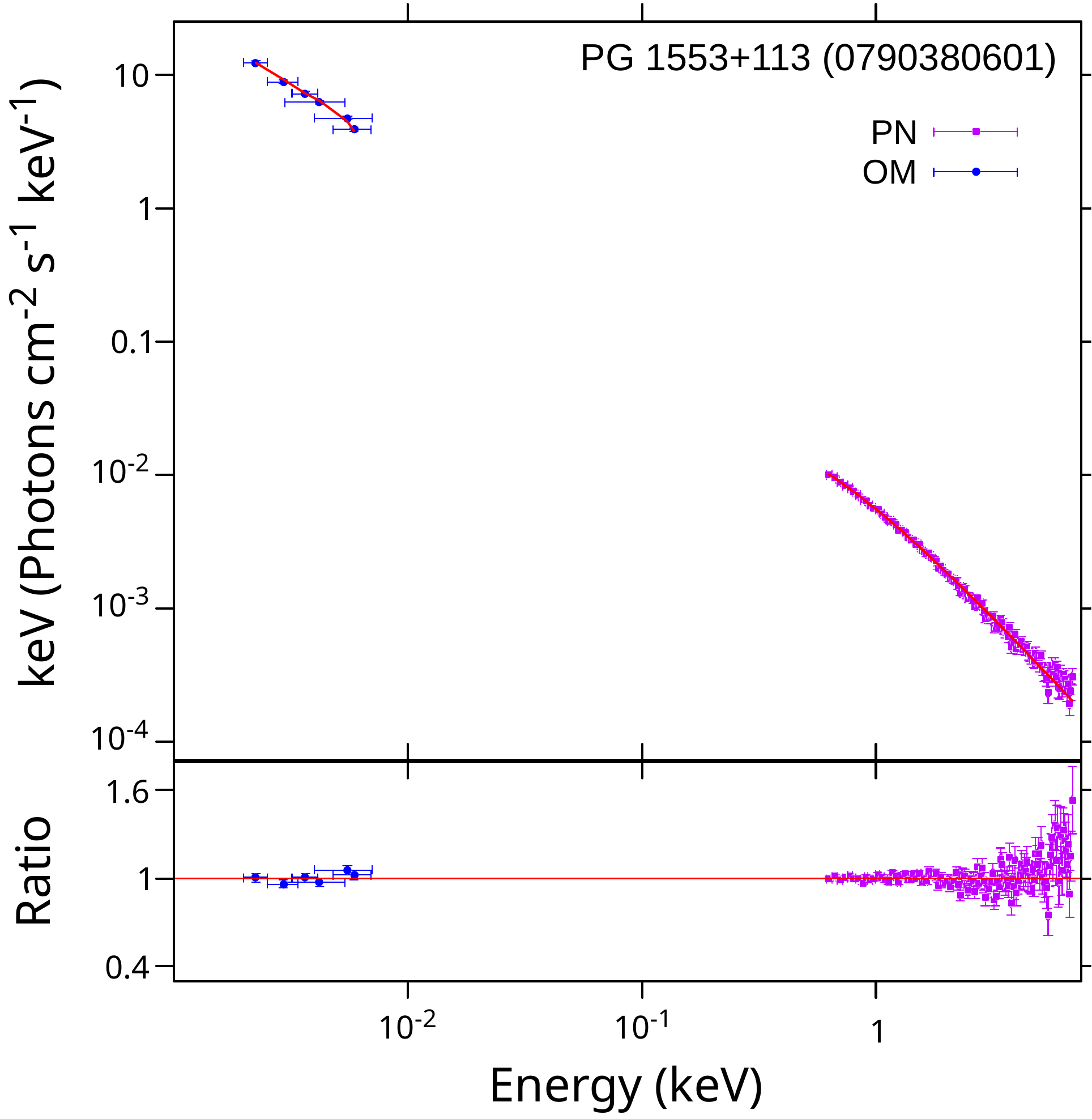}

{\vspace{-0.14cm} \includegraphics[width=8.5cm, height=7.5cm]{fig2_15.pdf}}
\includegraphics[width=8.5cm, height=7.5cm]{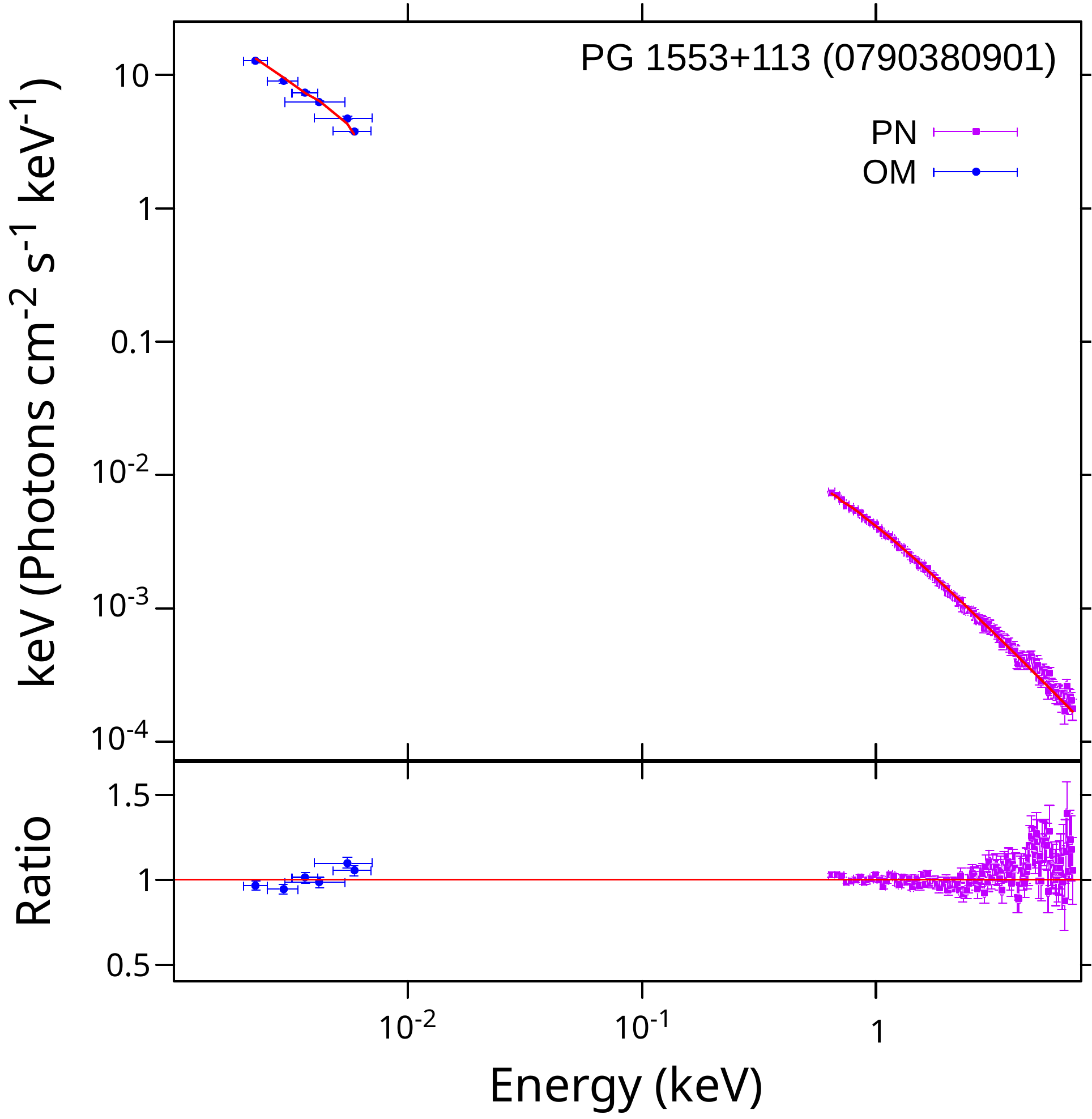}

{\vspace{-0.14cm} \includegraphics[width=8.5cm, height=7.5cm]{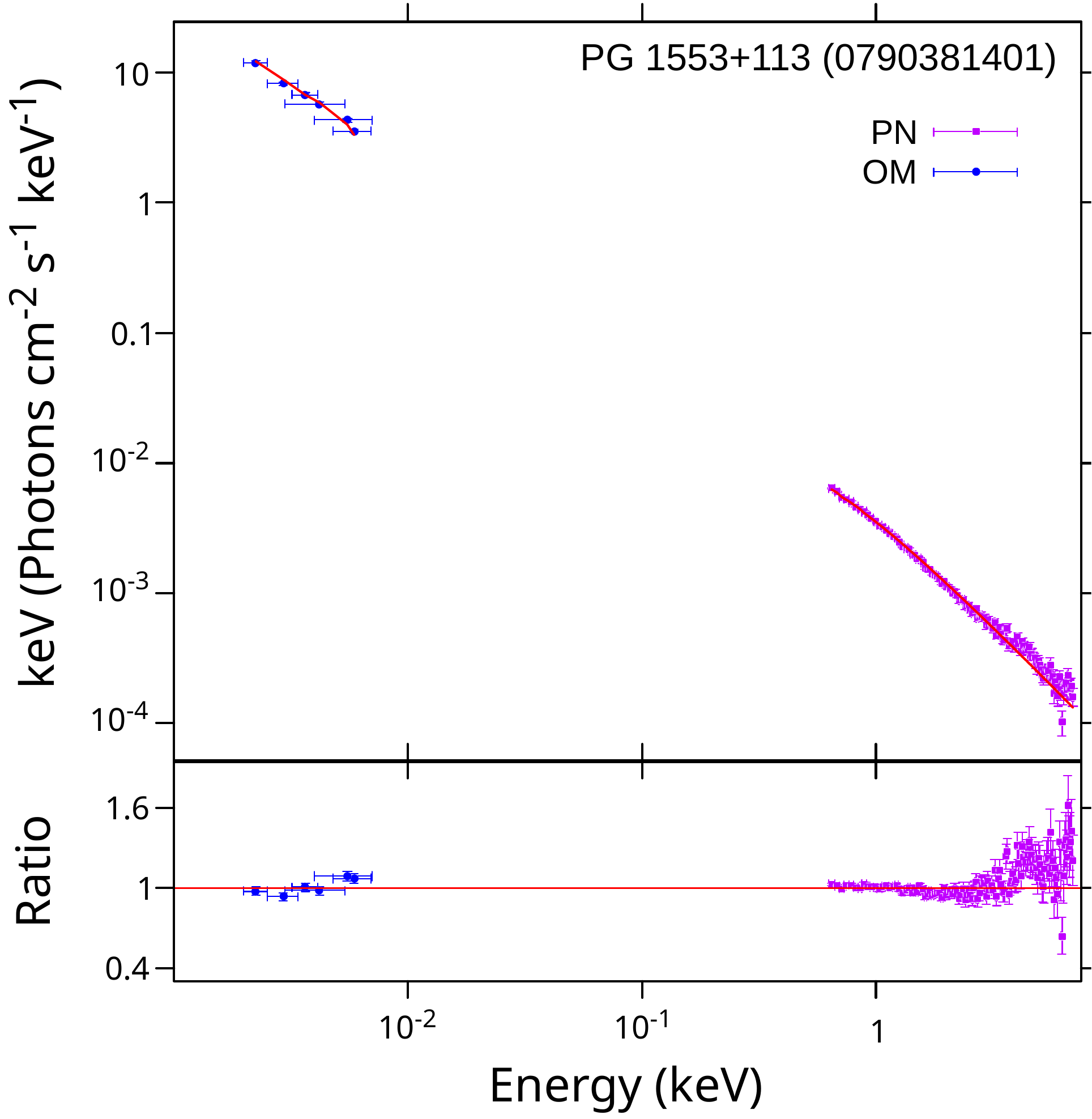}}
\includegraphics[width=8.5cm, height=7.5cm]{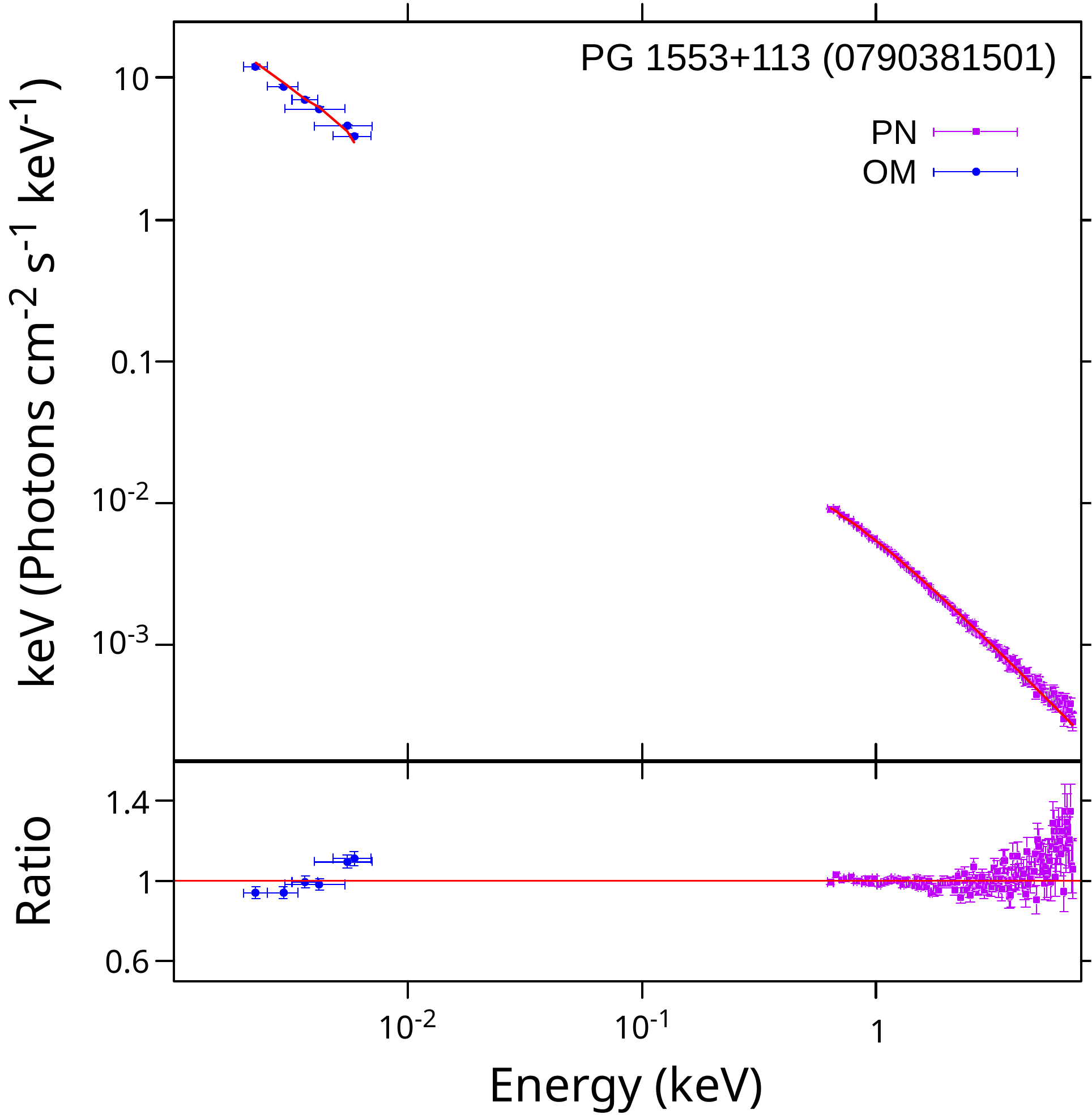}
\caption{Continued} 
\end{figure*}

\clearpage
\setcounter{figure}{1}
\begin{figure*}
\centering
{\vspace{-0.14cm} \includegraphics[width=8.5cm, height=7.5cm]{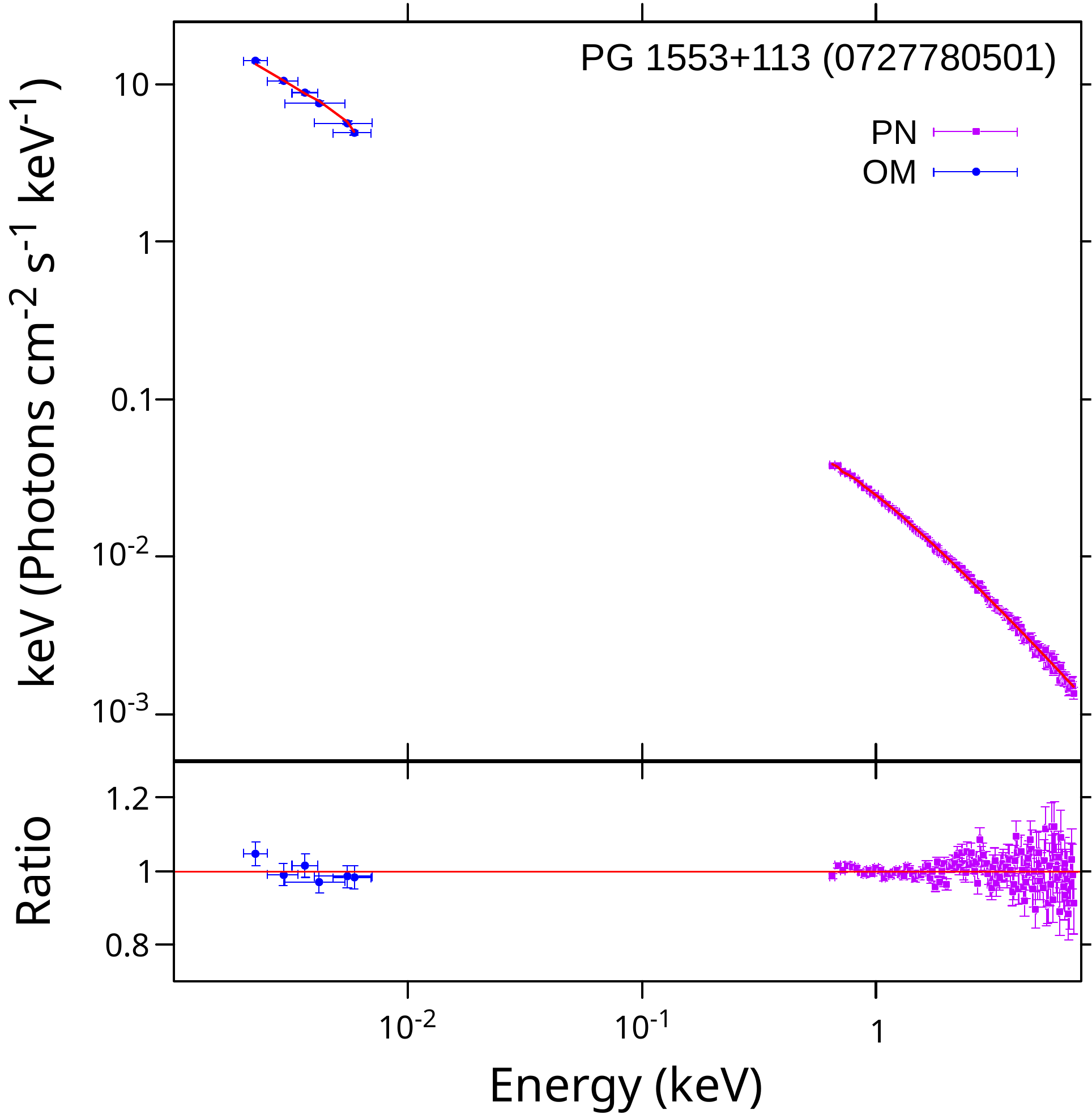}}
\includegraphics[width=8.5cm, height=7.5cm]{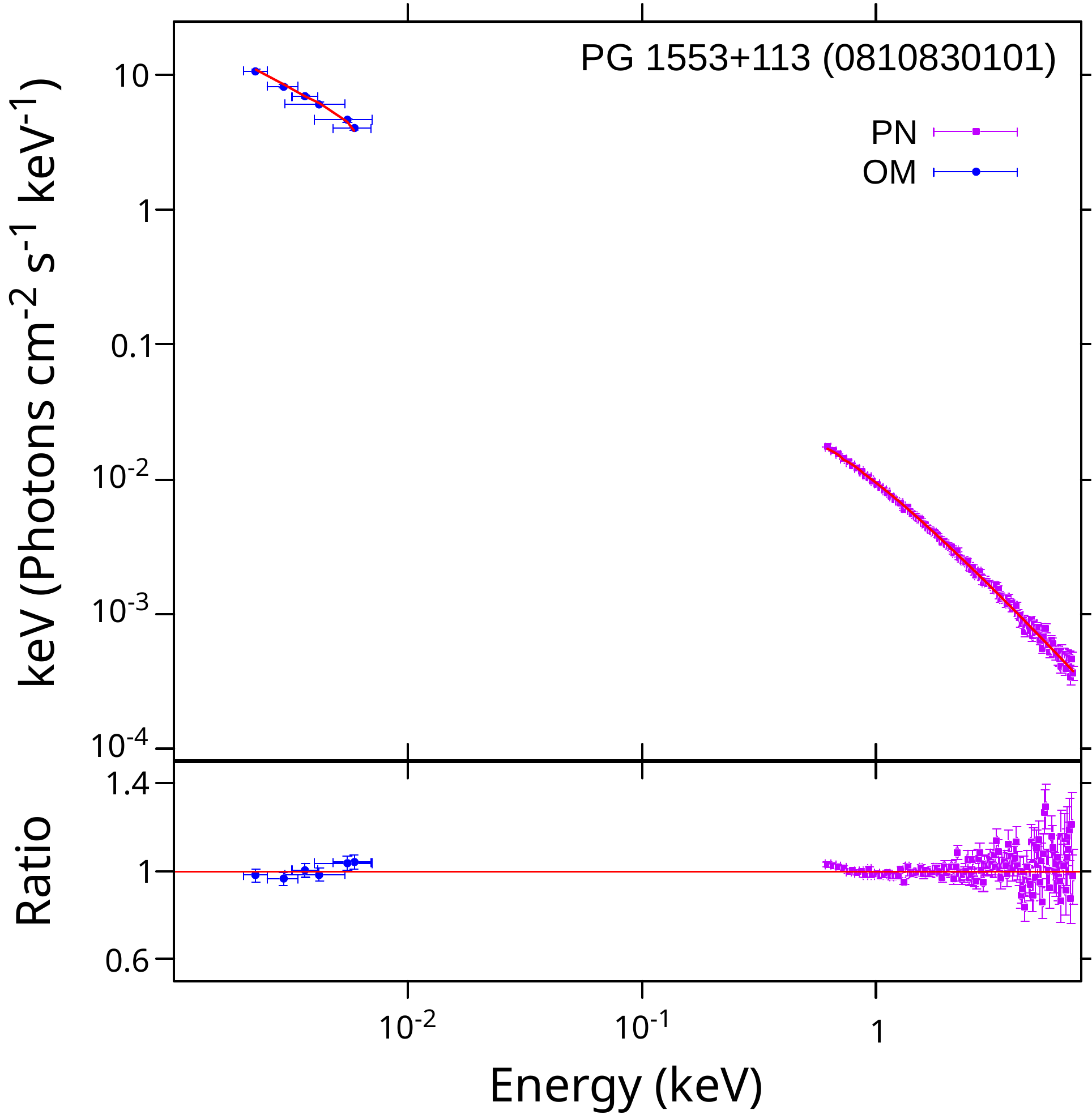}

{\vspace{-0.14cm} \includegraphics[width=8.5cm, height=7.5cm]{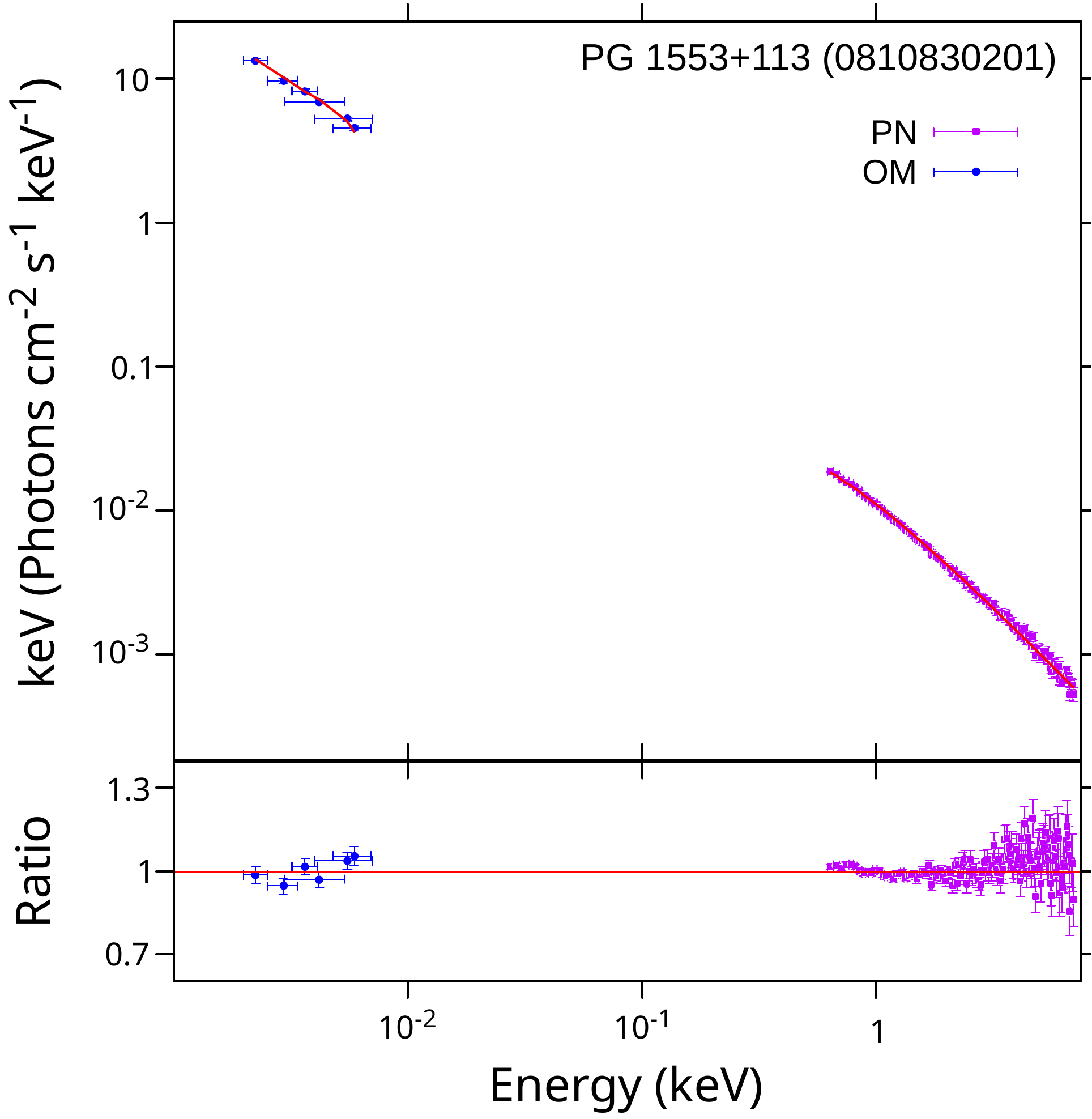}}
\includegraphics[width=8.5cm, height=7.5cm]{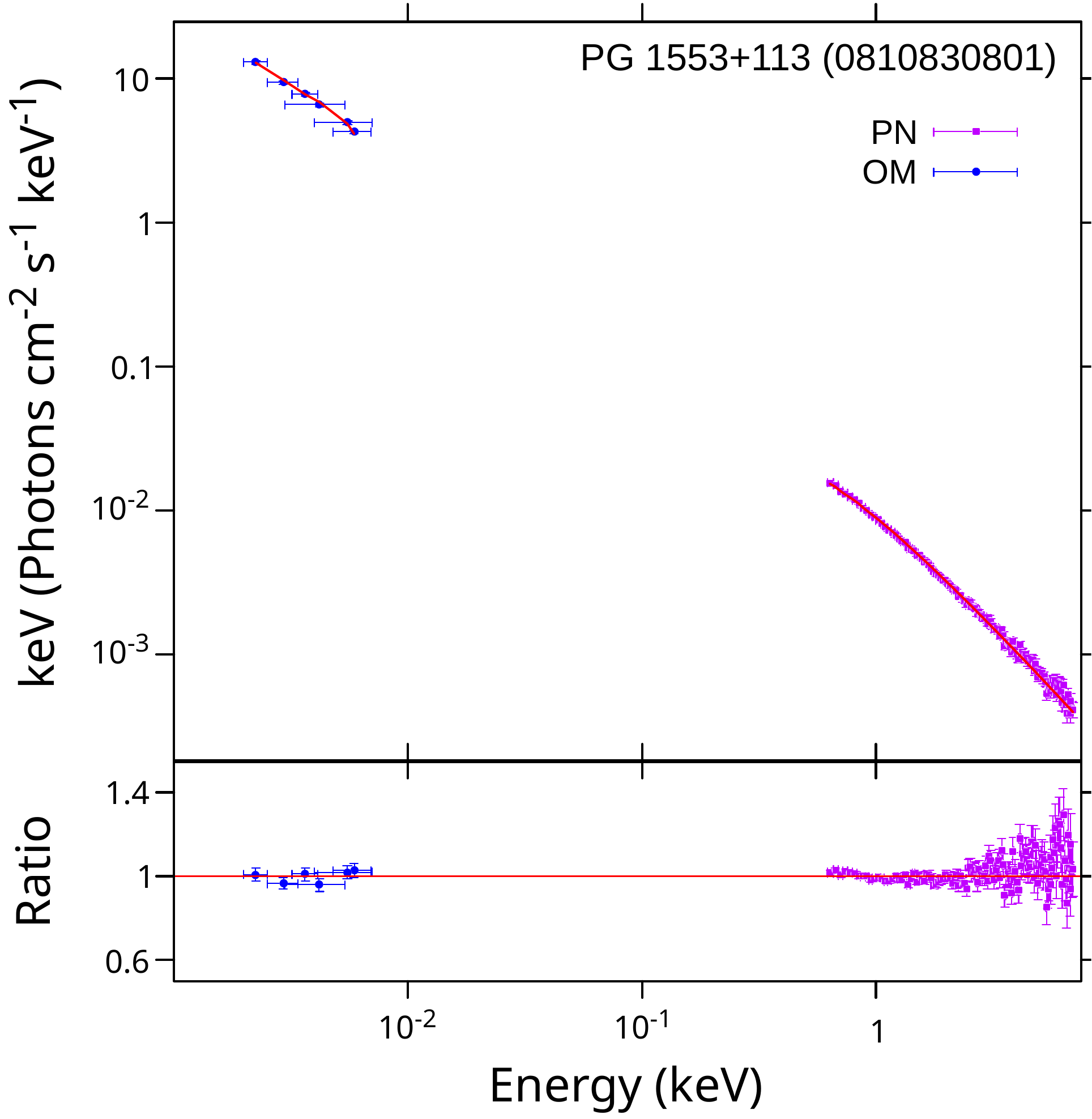}

{\vspace{-0.14cm} \includegraphics[width=8.5cm, height=7.5cm]{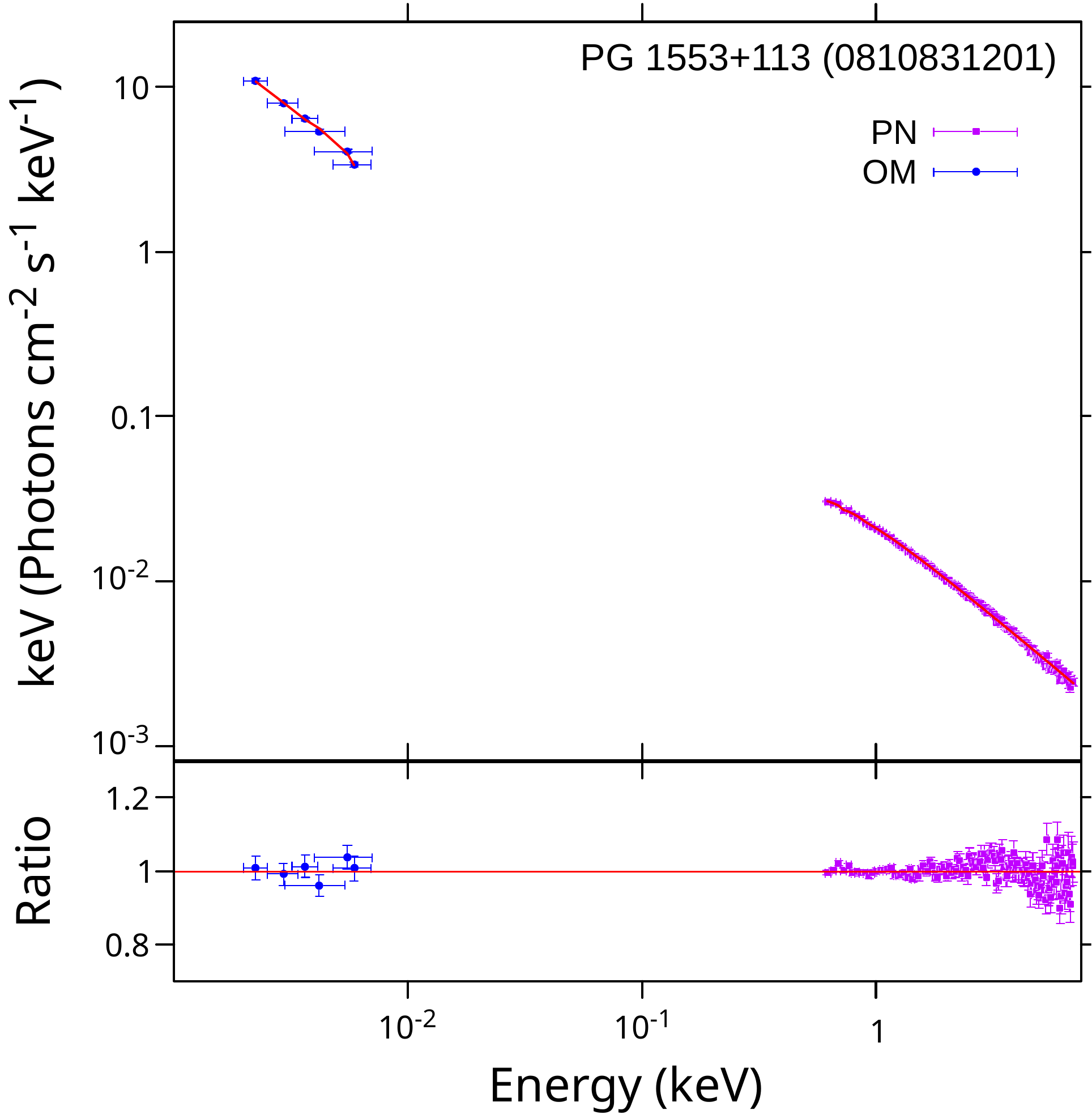}}
\includegraphics[width=8.5cm, height=7.5cm]{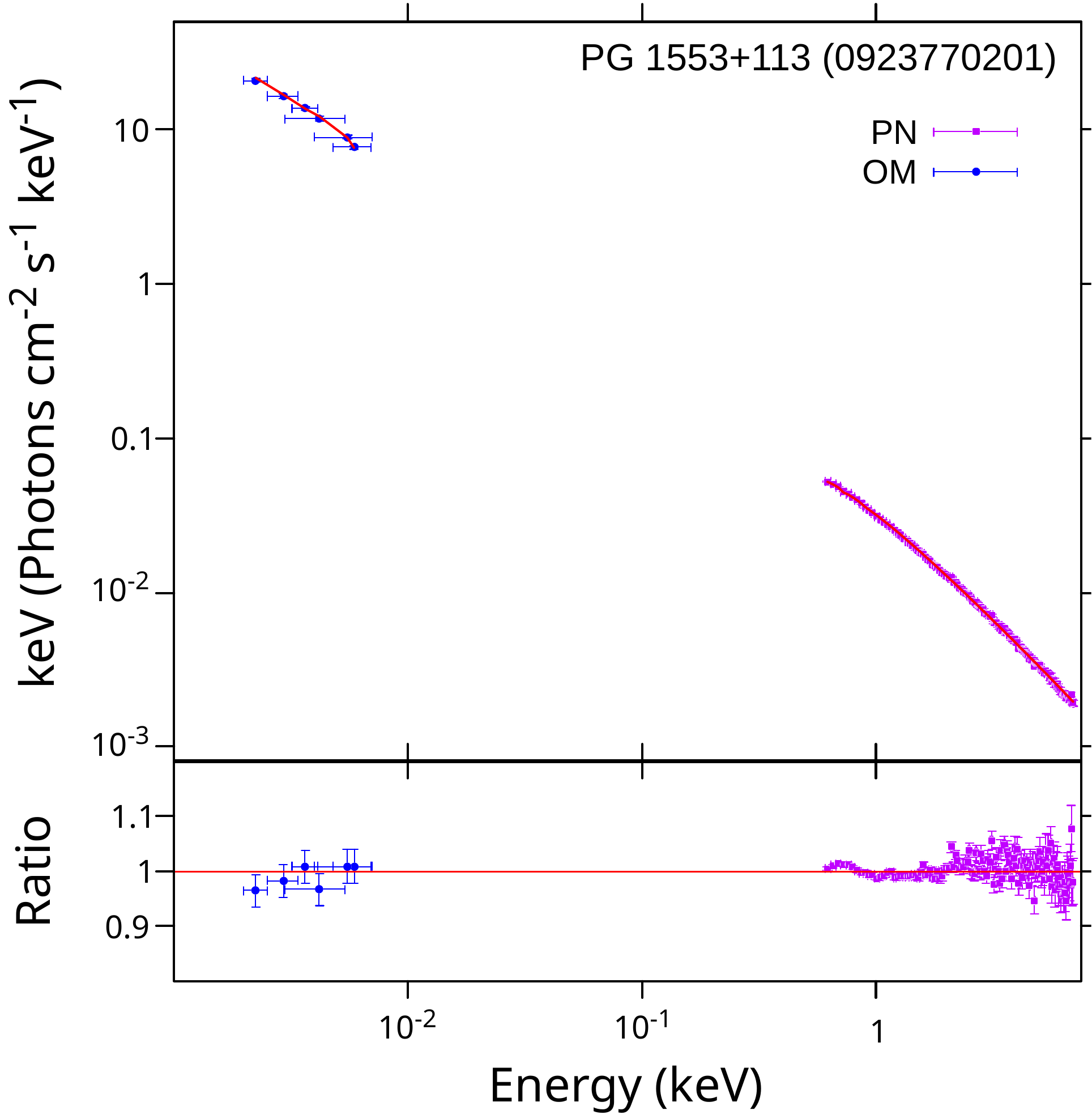}
\caption{Continued} 
\end{figure*}

\clearpage
\setcounter{figure}{1}
\begin{figure*}
\centering
{\vspace{-0.14cm} \includegraphics[width=8.5cm, height=7.5cm]{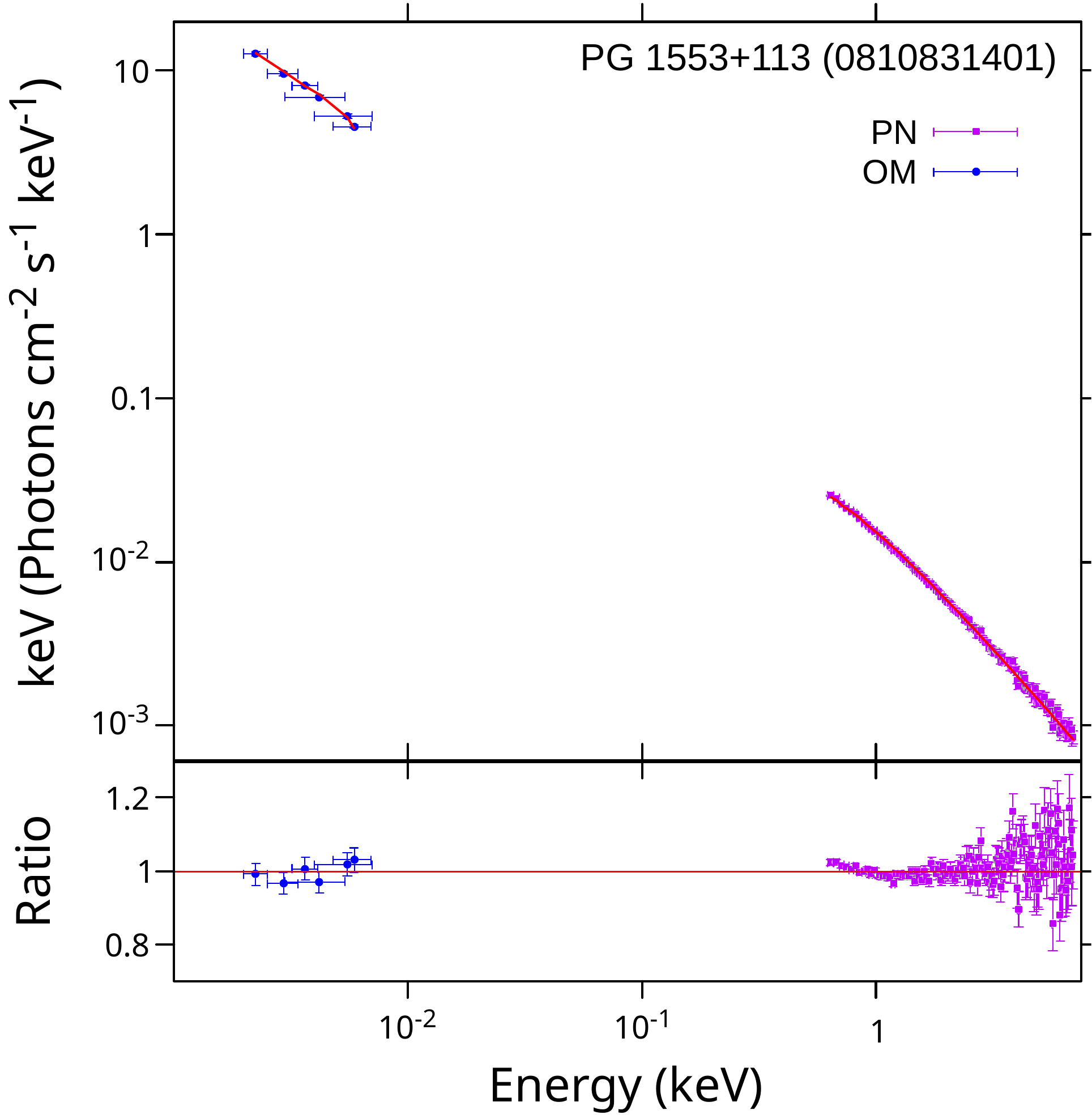}}
\includegraphics[width=8.5cm, height=7.5cm]{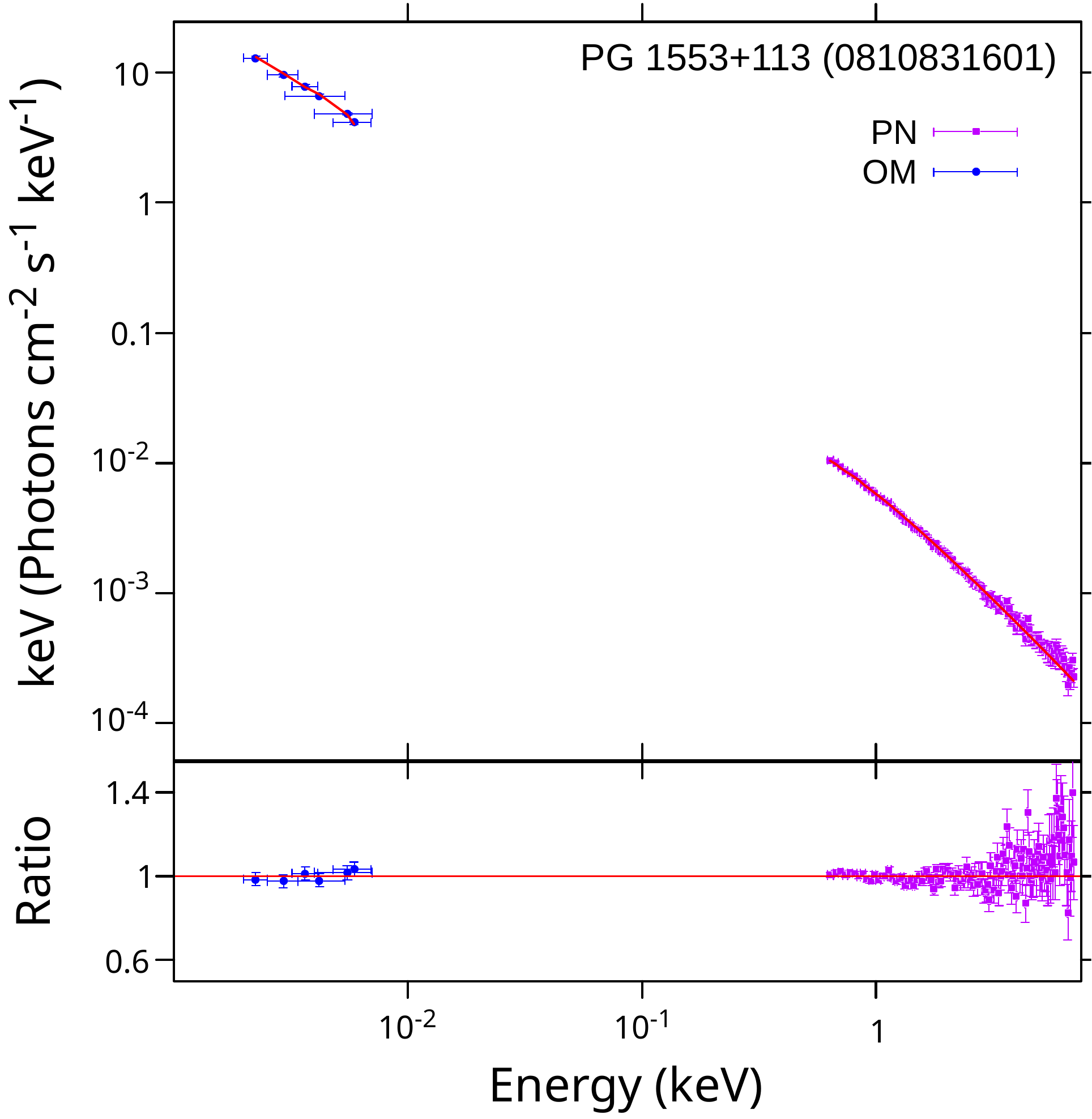}

{\vspace{-0.14cm} \includegraphics[width=8.5cm, height=7.5cm]{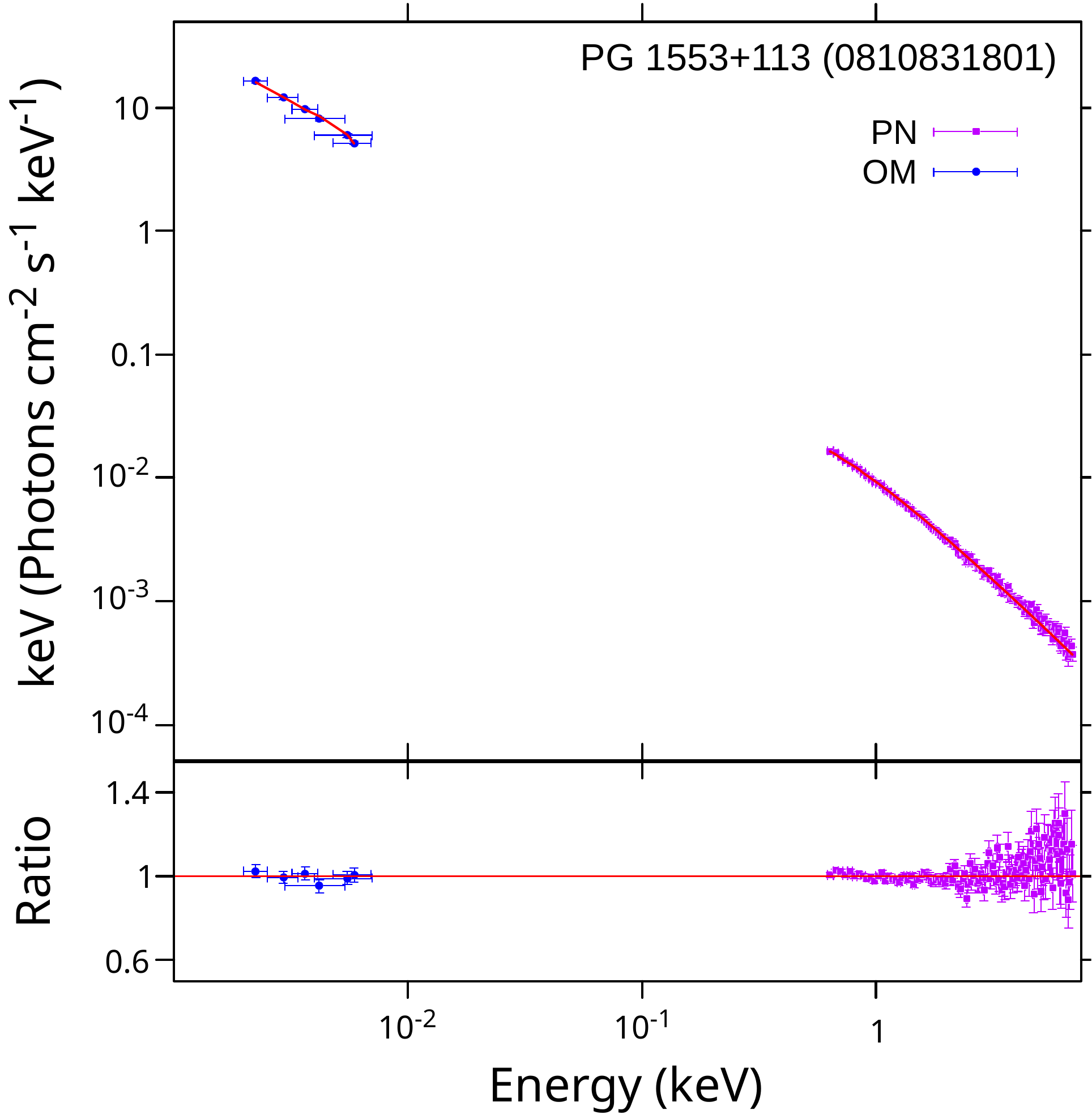}}
\caption{Continued} 
\end{figure*}

\newpage

\setcounter{figure}{2}
\begin{figure*}
\centering
{\includegraphics[width=8.5cm, height=7.5cm]{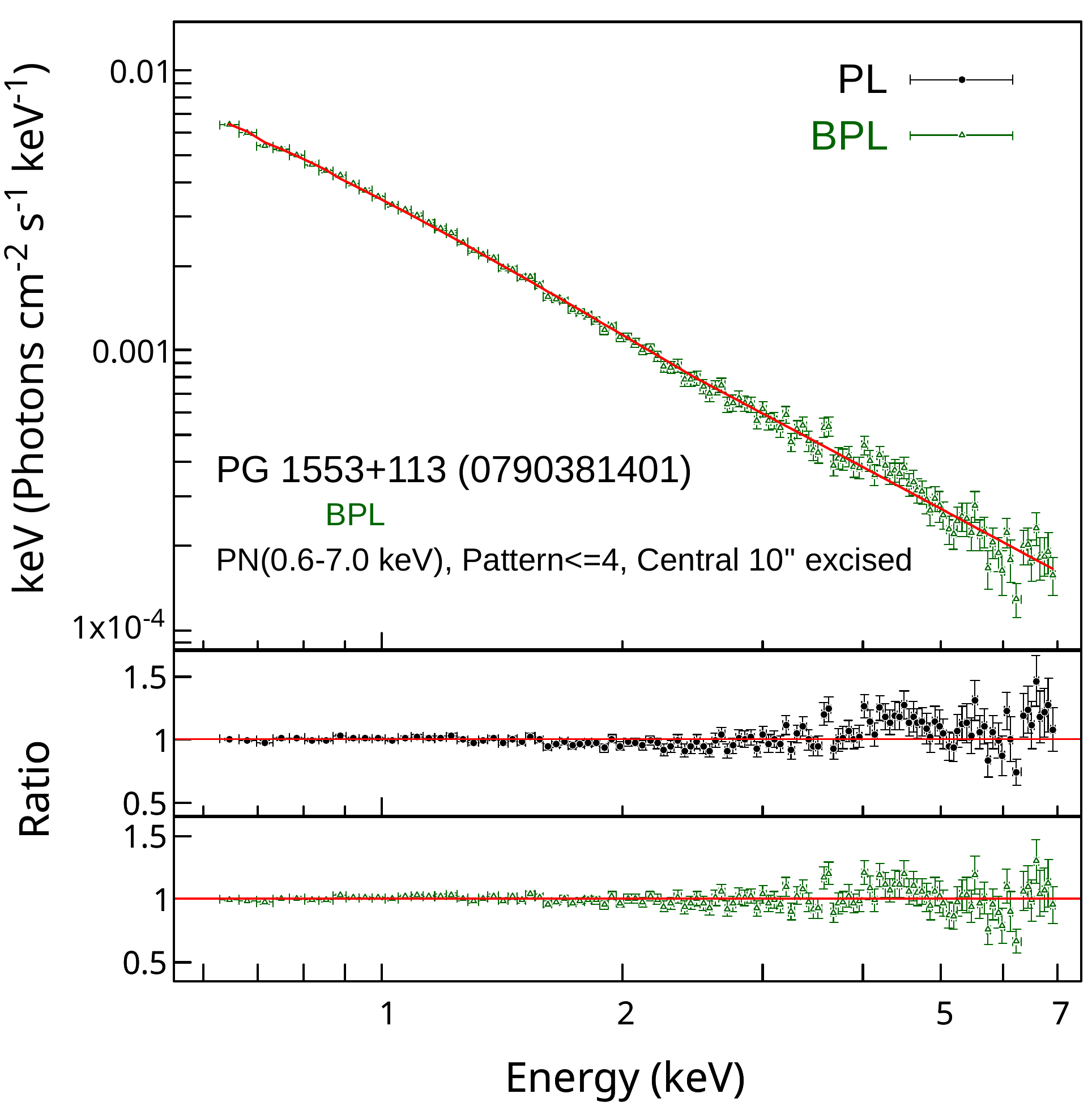}}
\includegraphics[width=8.5cm, height=7.5cm]{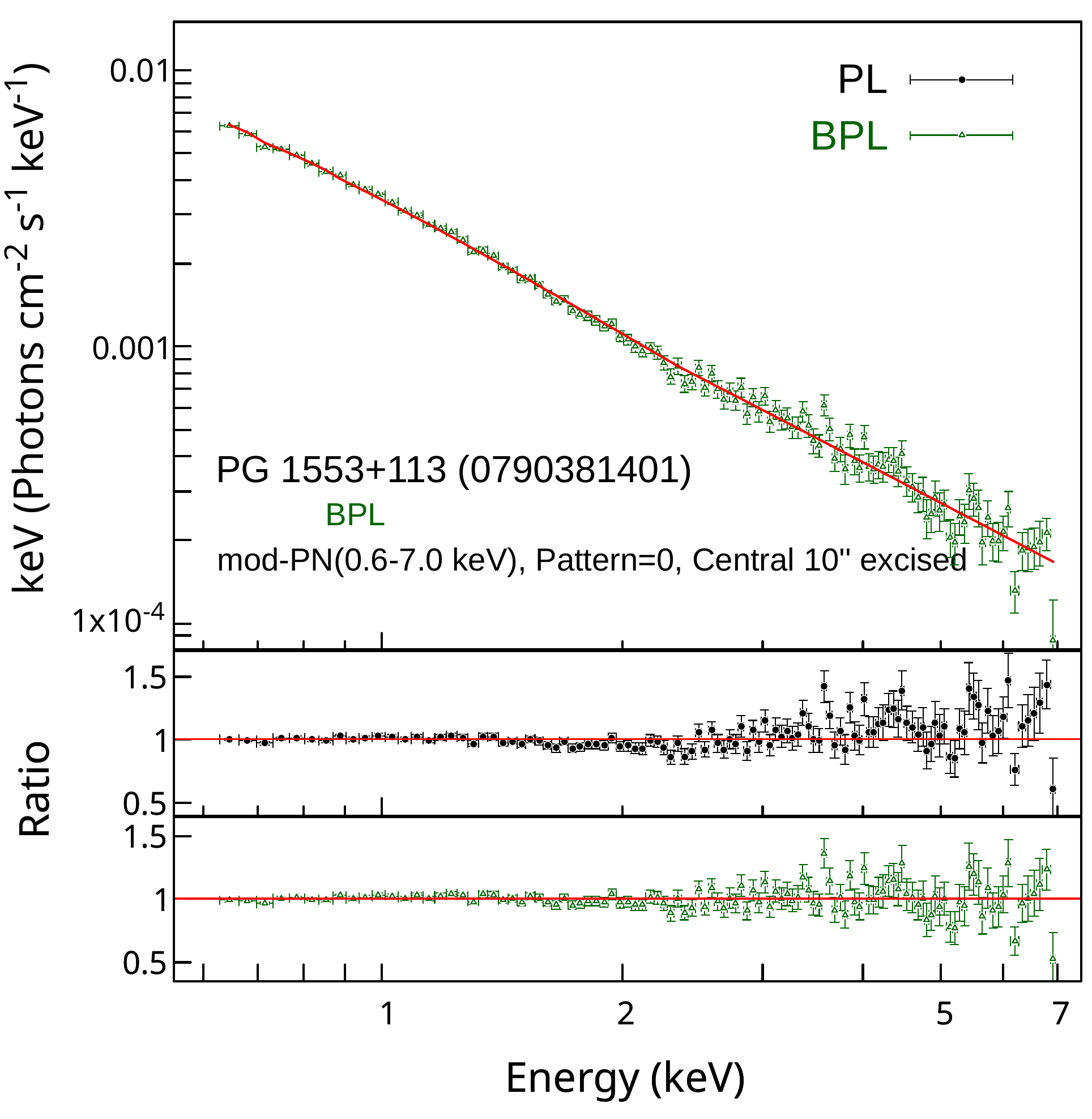}
{\vspace{-0.17cm} \includegraphics[width=8.5cm, height=7.5cm]{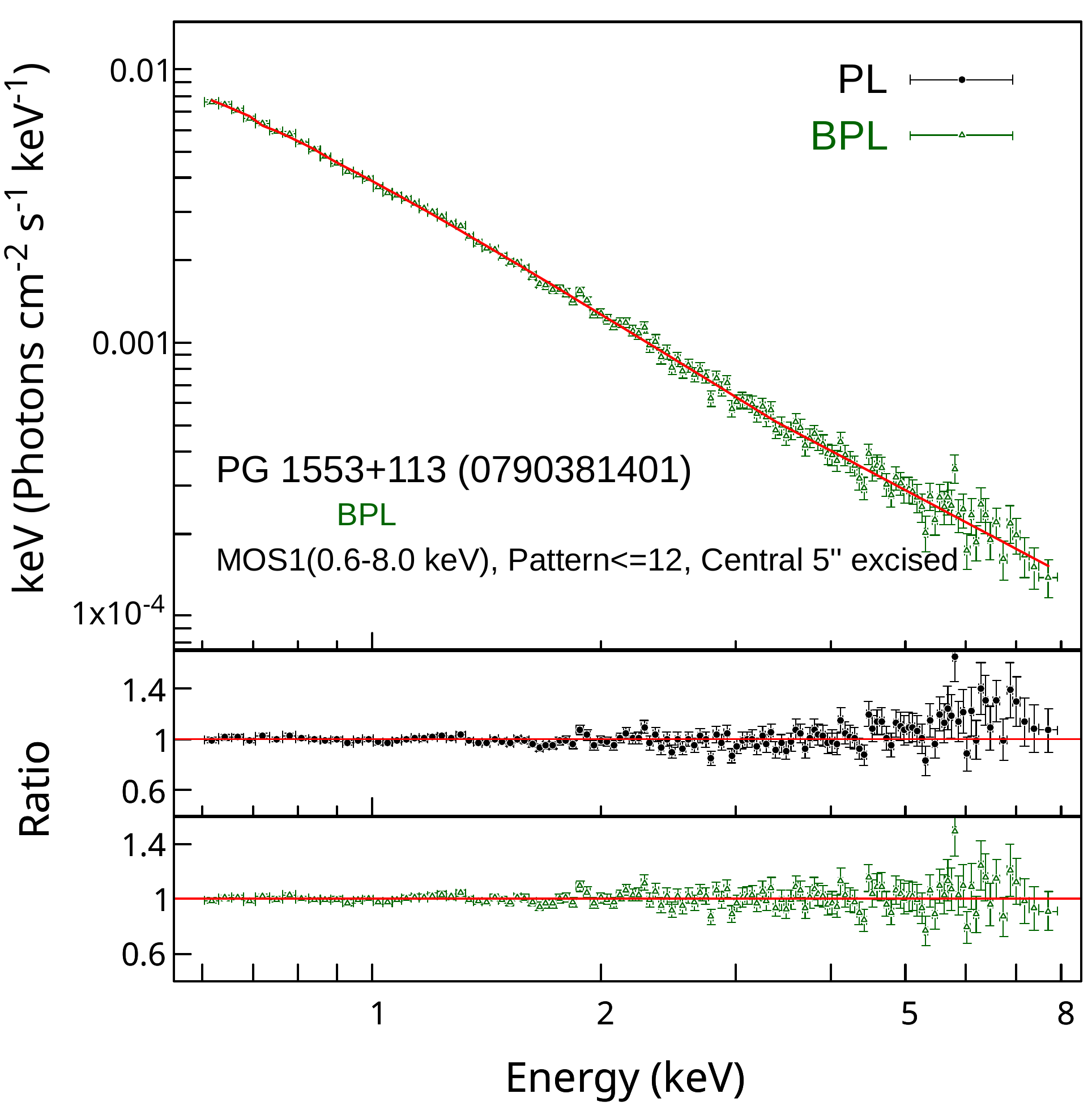}}
\includegraphics[width=8.5cm, height=7.5cm]{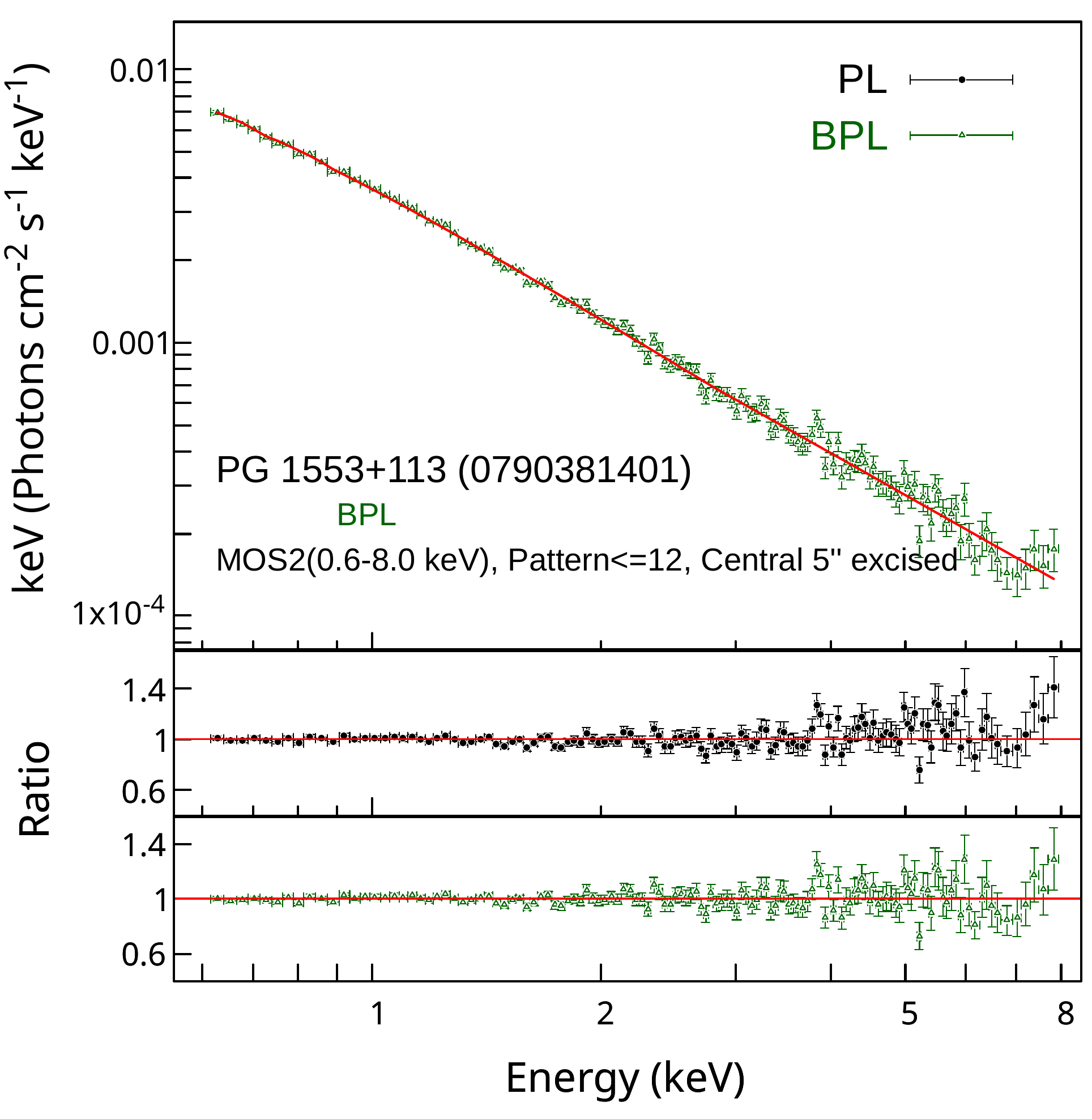}
{\vspace{-0.17cm} 
{\hspace{0.5cm}\includegraphics[width=8.3cm, height=7.4cm]{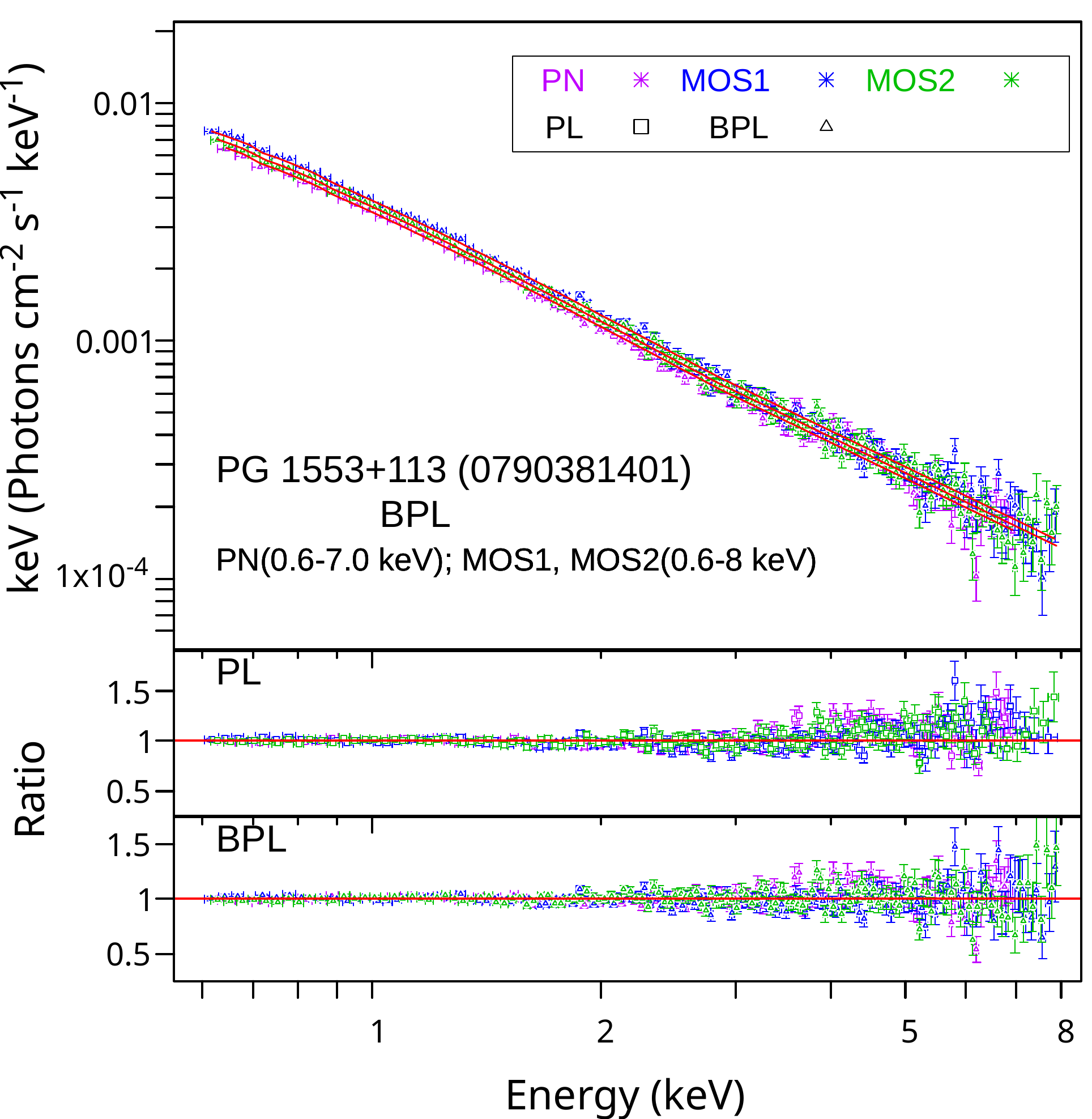}}
{\hspace{0.3cm}\includegraphics[width=8.4cm, height=7.3cm]{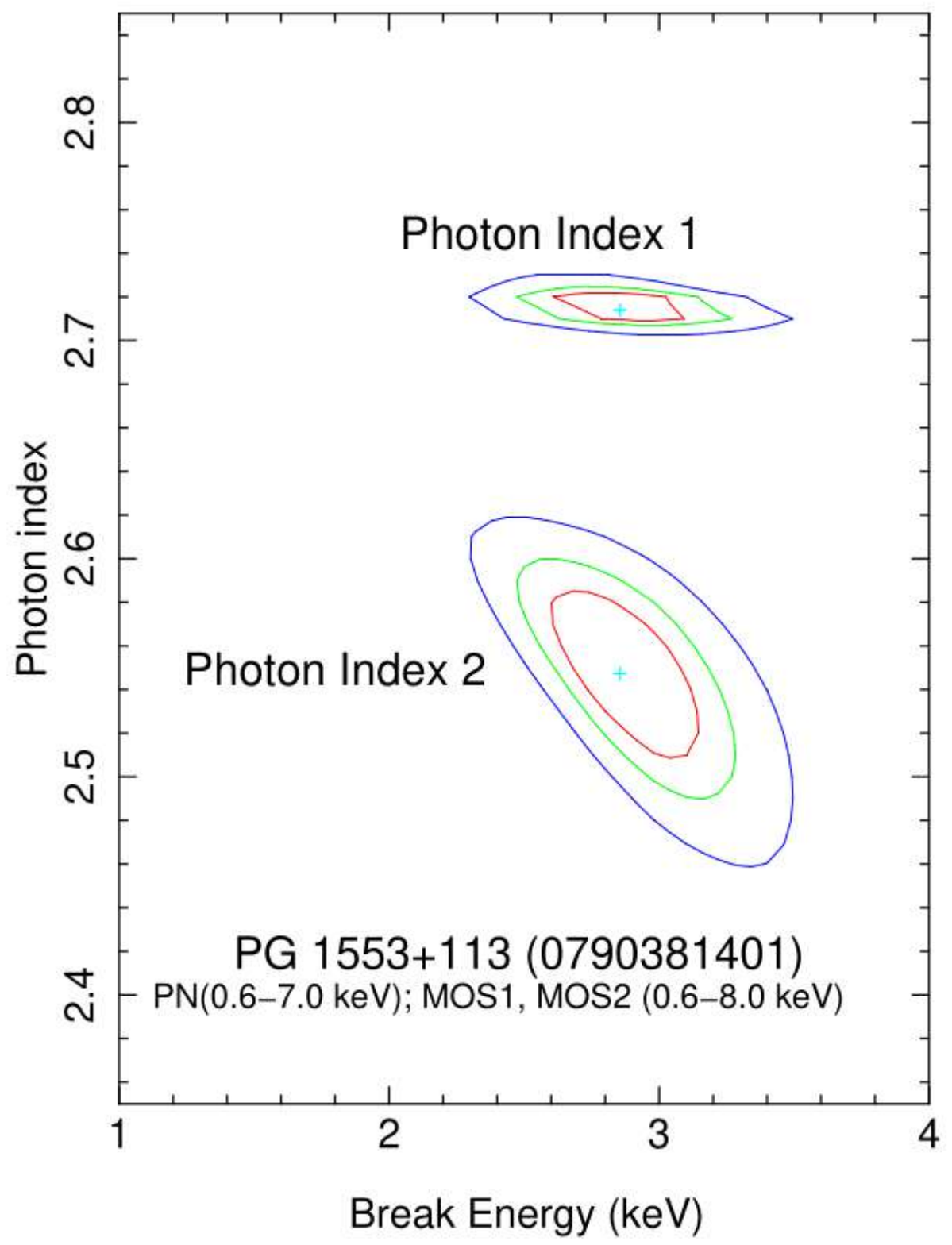}}}
\vspace{-0.1cm}
\caption{(a) X-ray spectral fit plots and contour plots for Obs ID: 0790381401. In the top four plots, PL and BPL models are represented by black filled circles and green-filled triangles, respectively. Observation ID, source name, EPIC camera , energy range, patterns used, and the central region if excised are displayed on each plot.  The bottom left plot is  a joint spectral fit with all  3 EPIC cameras. PL and BPL model fits are shown by squares and triangles, respectively, with \textit{PN}, \textit{MOS1}, \textit{MOS2} data points shown in red, blue, and green colors. 
Broken power law model confidence contour (Chi-squared) plots at 68$\%$(red),  90$\%$(green), and 99$\%$(blue) confidence for photon index 1, photon index 2, and break energy is also shown. \label{A3} }
\end{figure*}

\clearpage
\setcounter{figure}{2}
\begin{figure*}
\centering
{\vspace{-0.14cm} \includegraphics[width=8.5cm, height=7.5cm]{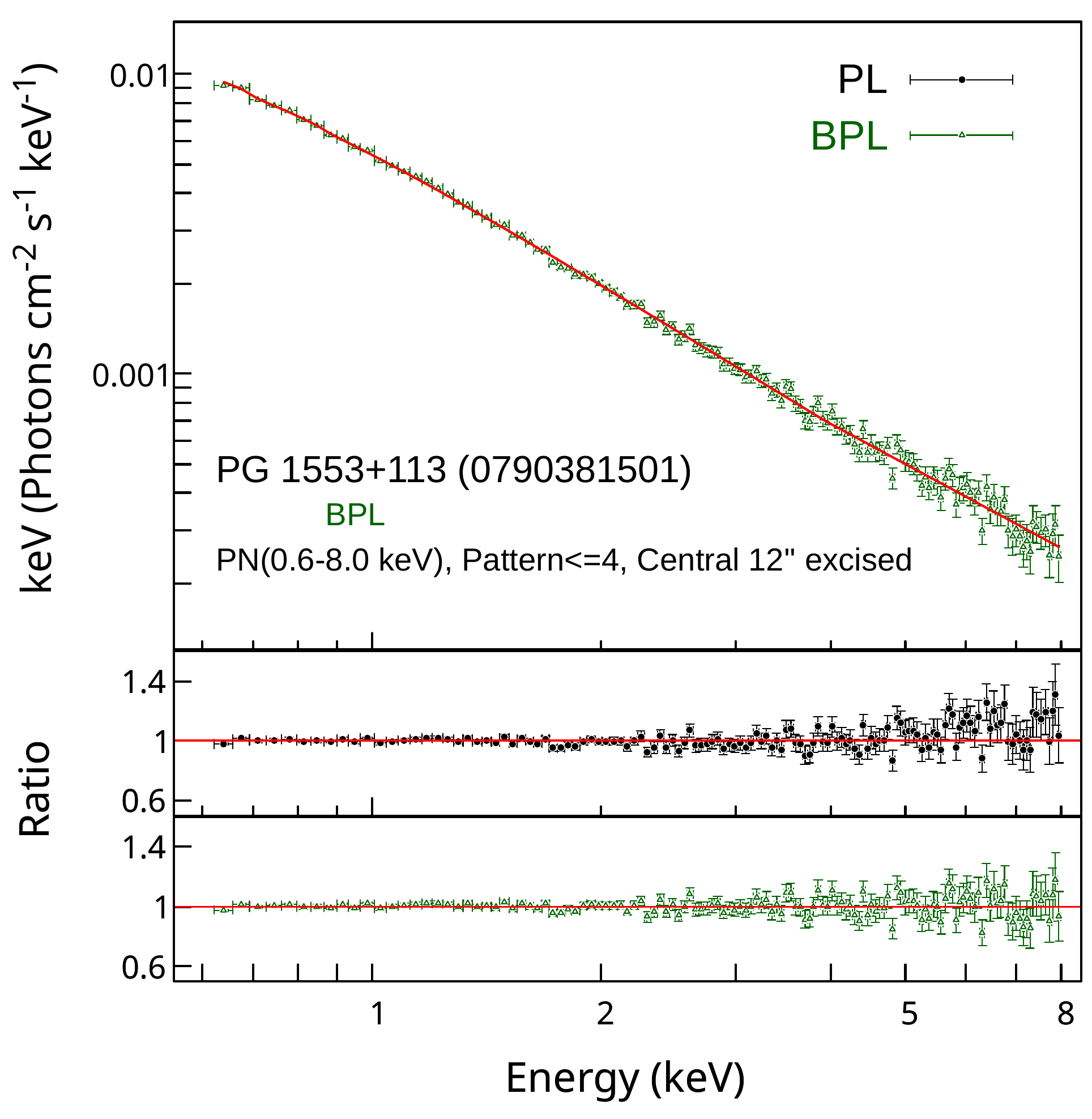}}
\includegraphics[width=8.5cm, height=7.5cm]{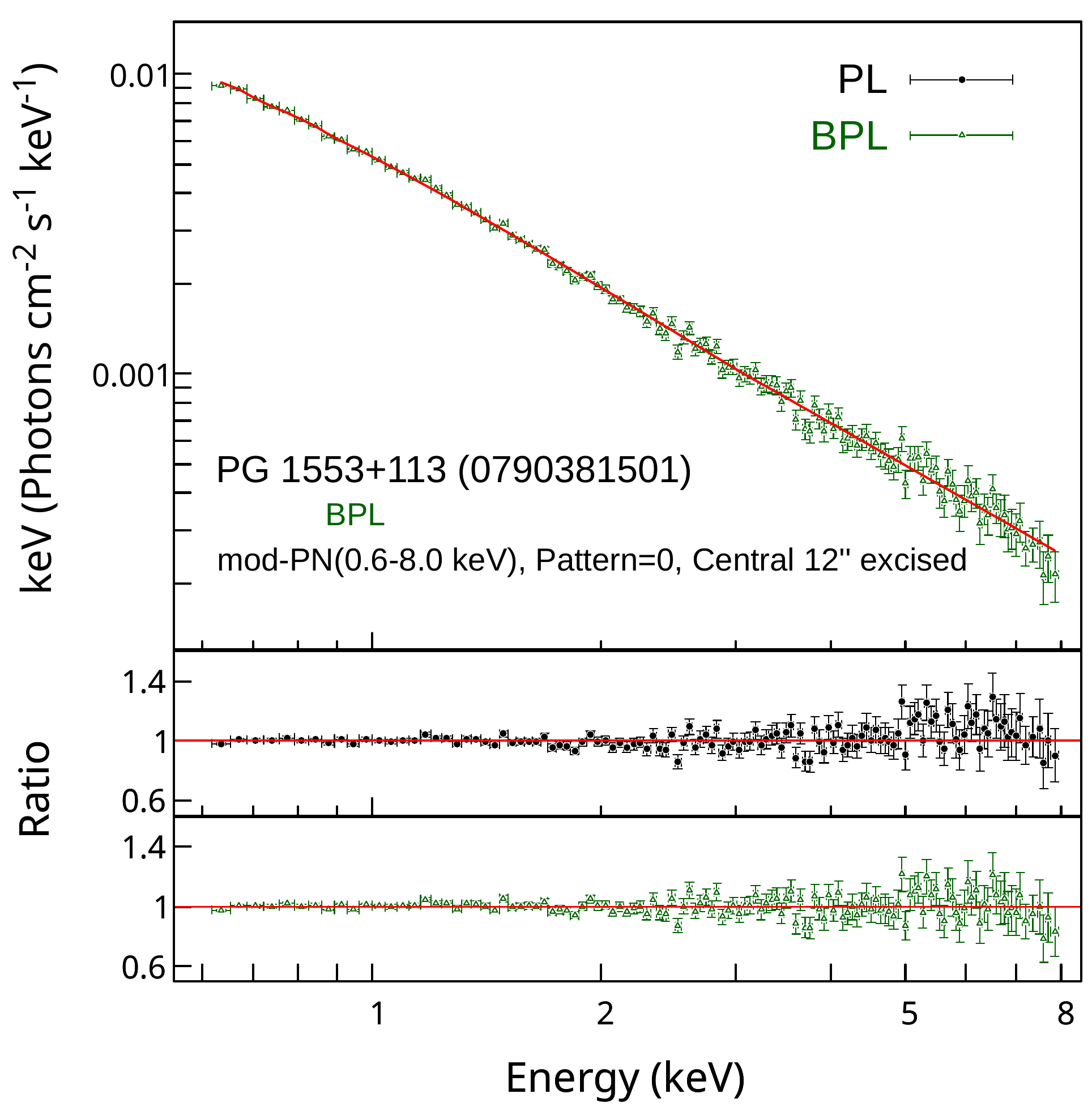}

{\vspace{-0.17cm} \includegraphics[width=8.5cm, height=7.5cm]{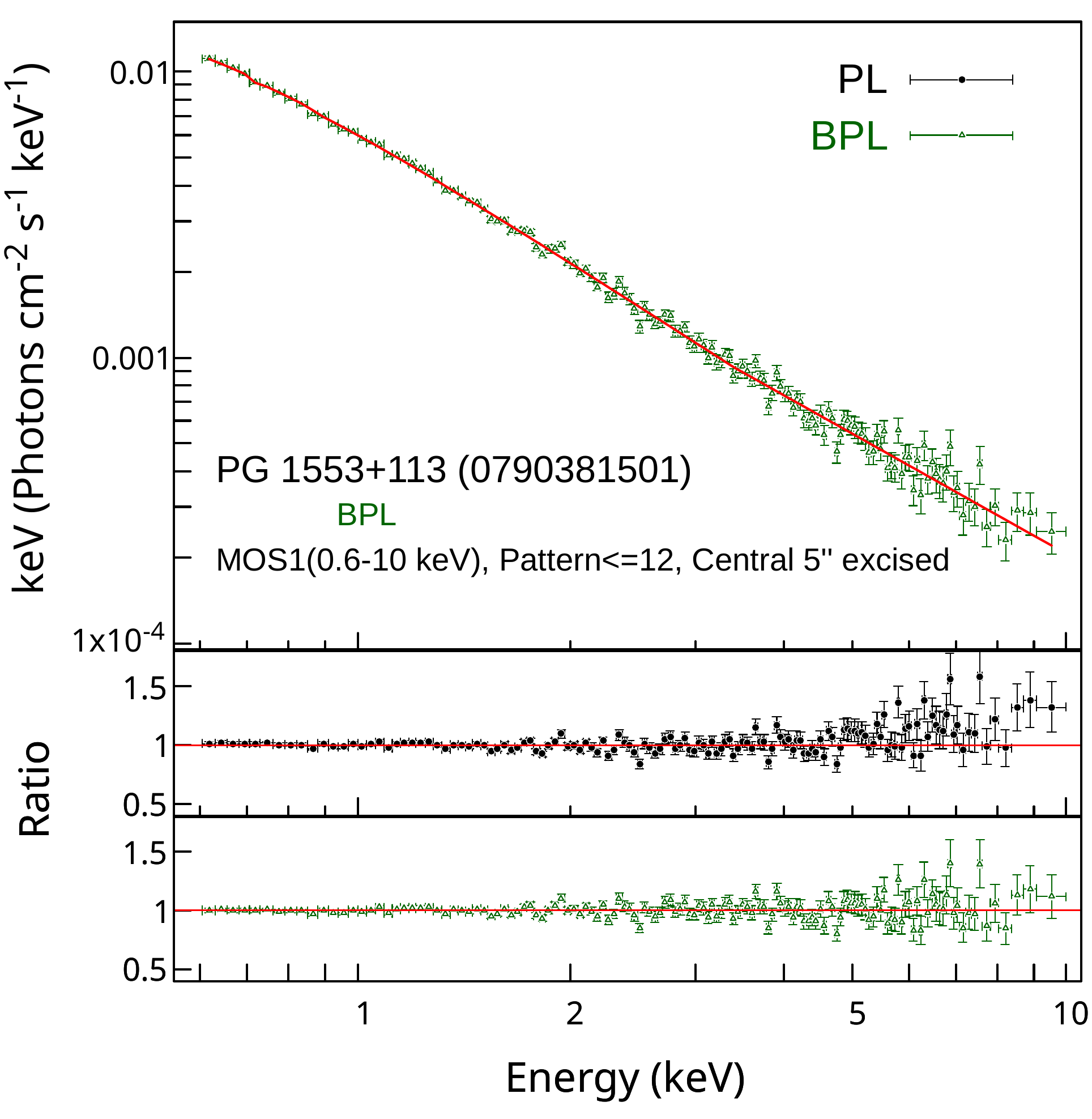}}
\includegraphics[width=8.5cm, height=7.5cm]{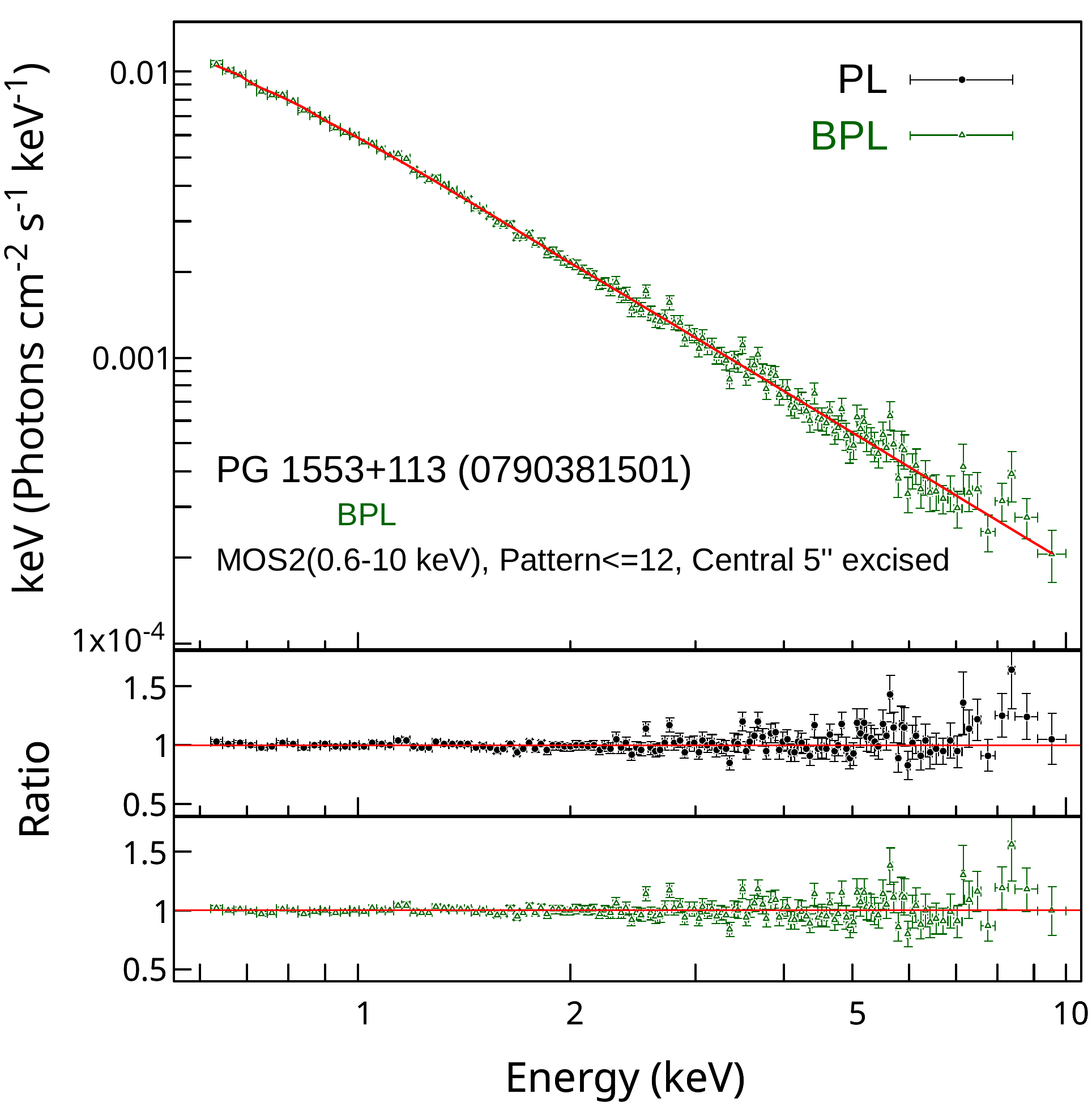}

{\vspace{-0.17cm} {\hspace{0.5cm}\includegraphics[width=8.3cm, height=7.4cm]{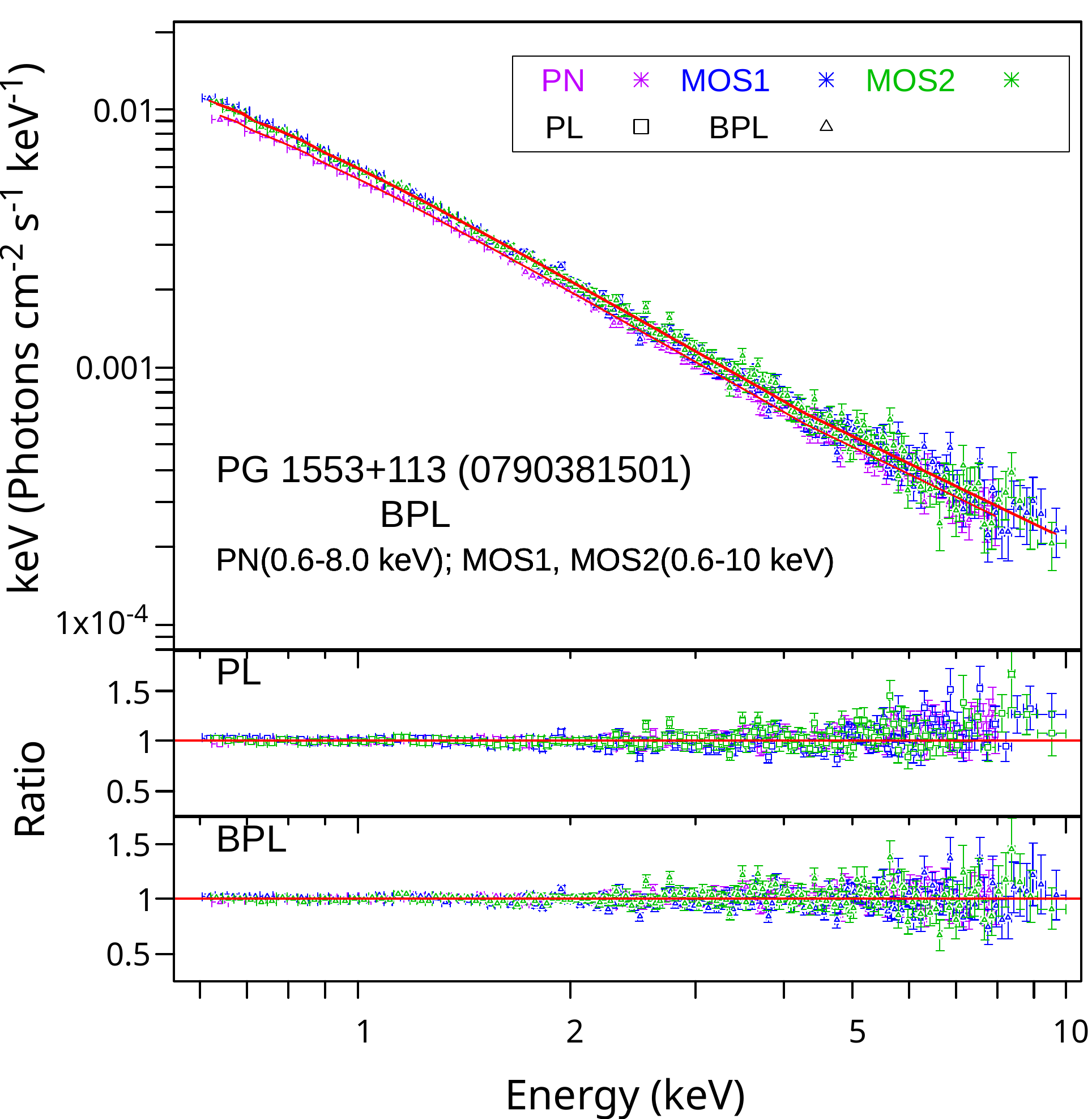}}
{\hspace{0.3cm}\includegraphics[width=8.4cm, height=7.3cm]{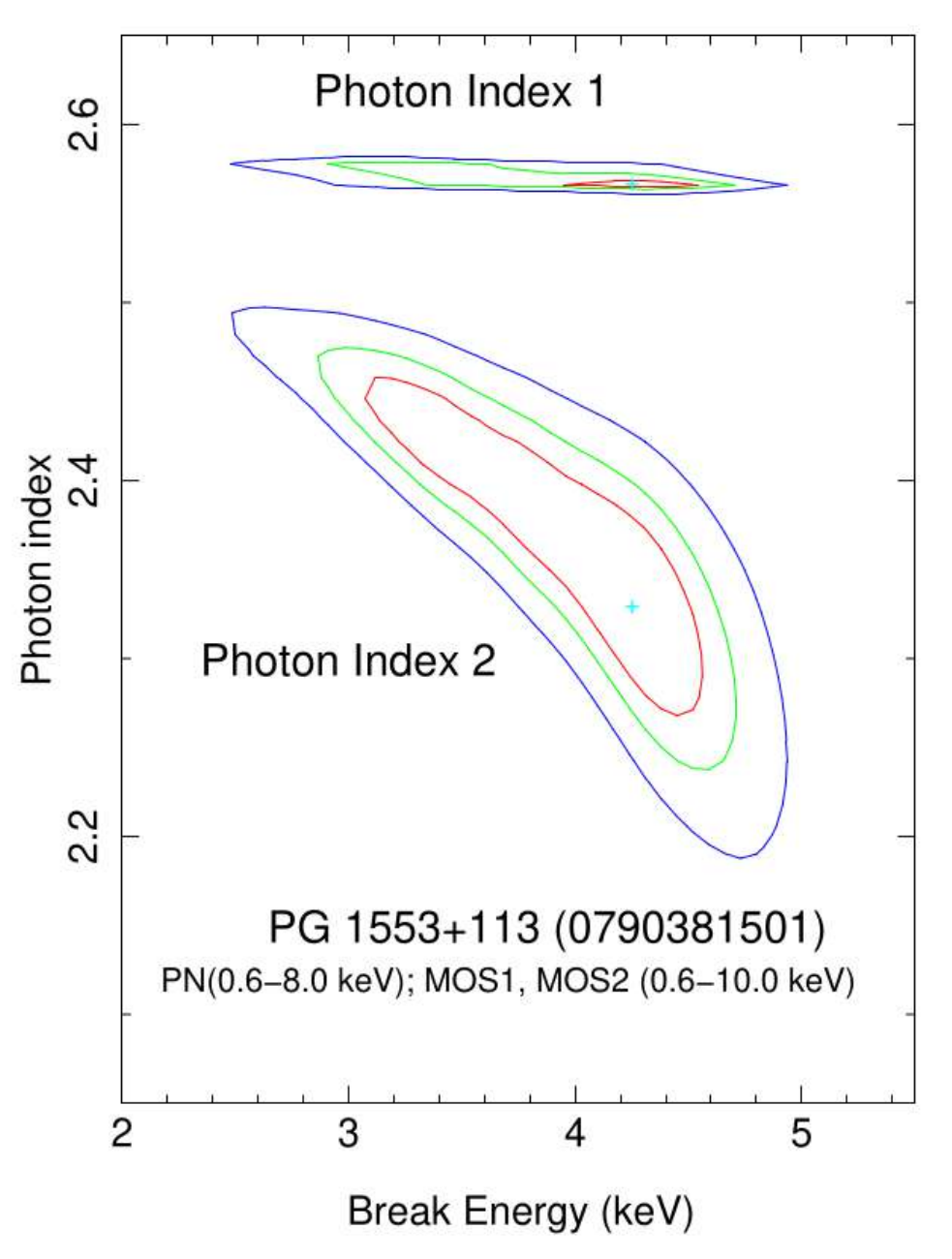}}}
\caption{(b) Similar X-ray spectral fit plots and contour plots for Obs ID: 0790381501} 
\end{figure*}

\clearpage
\setcounter{figure}{2}
\begin{figure*}
\centering
{\vspace{-0.14cm} \includegraphics[width=8.5cm, height=7.5cm]{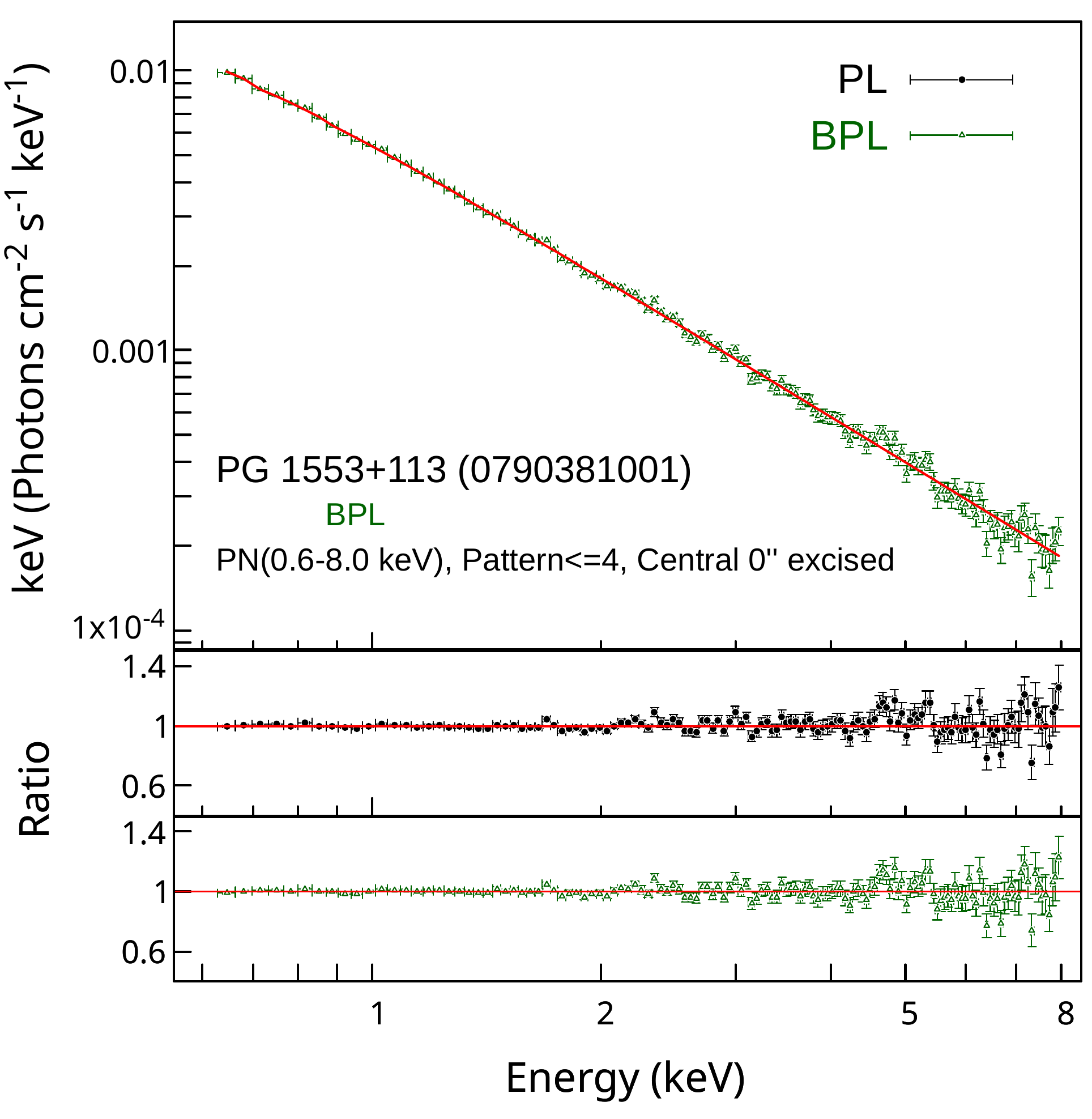}}
\includegraphics[width=8.5cm, height=7.5cm]{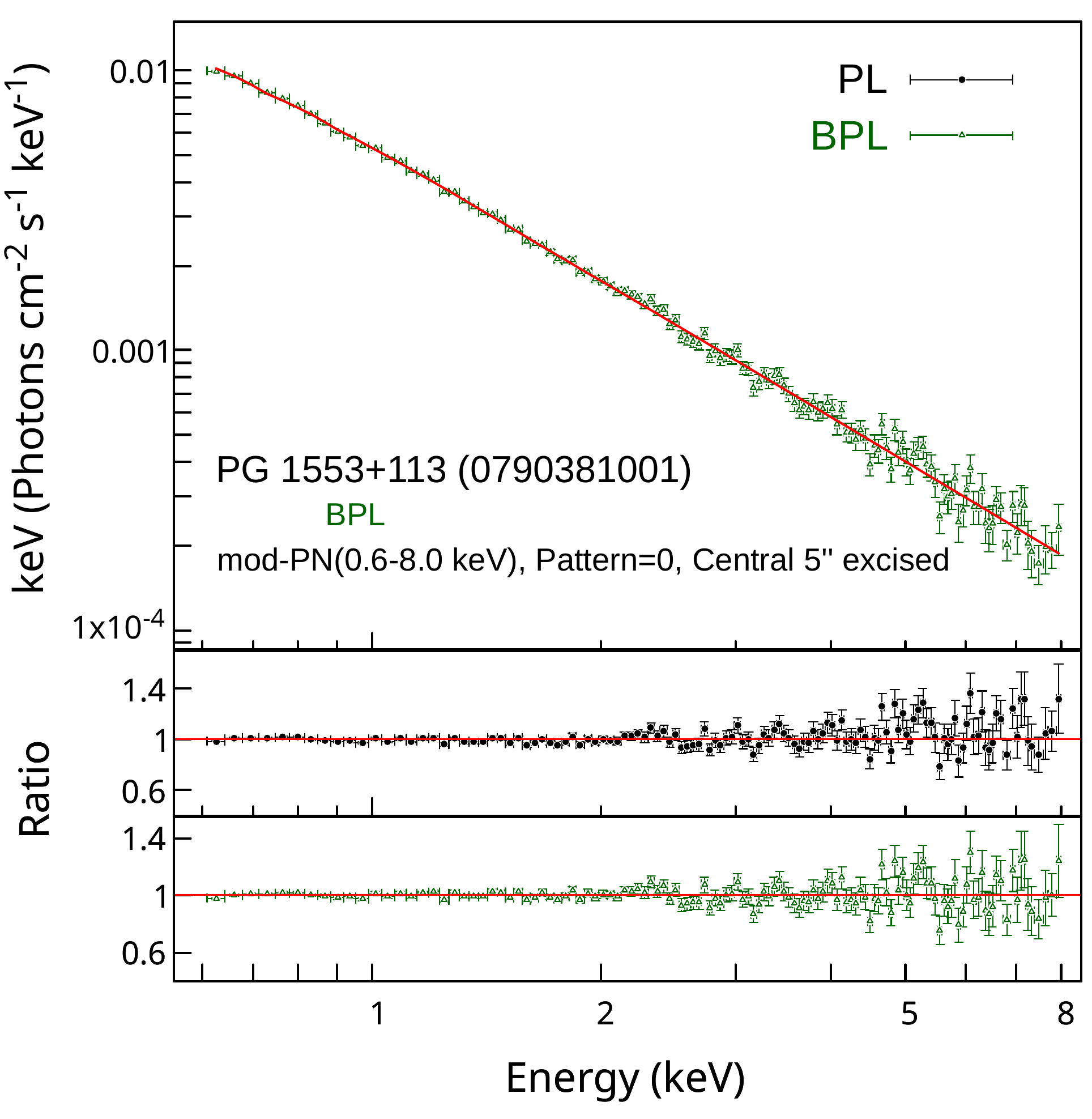}

{\vspace{-0.17cm} \includegraphics[width=8.5cm, height=7.5cm]{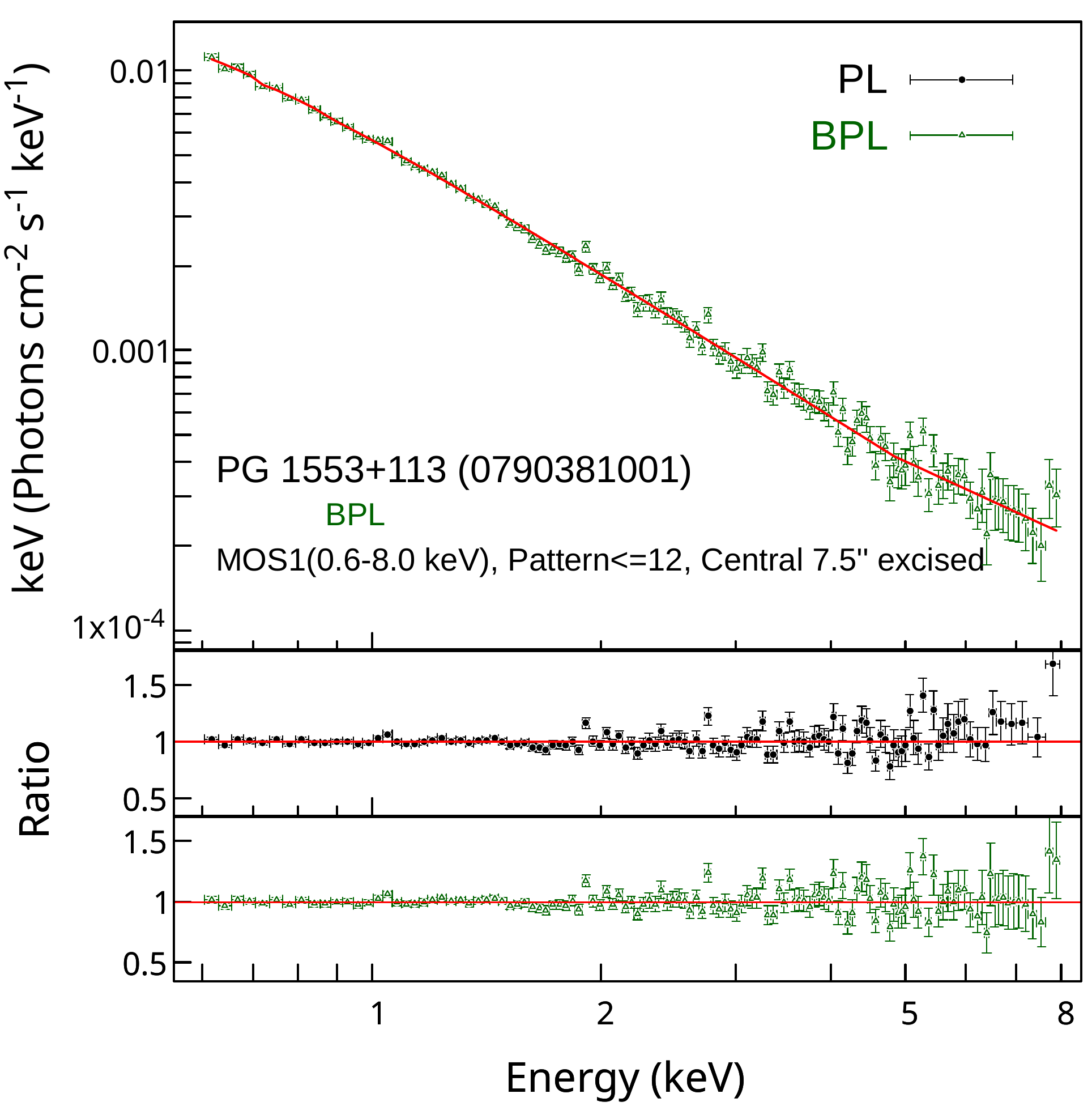}}
\includegraphics[width=8.5cm, height=7.5cm]{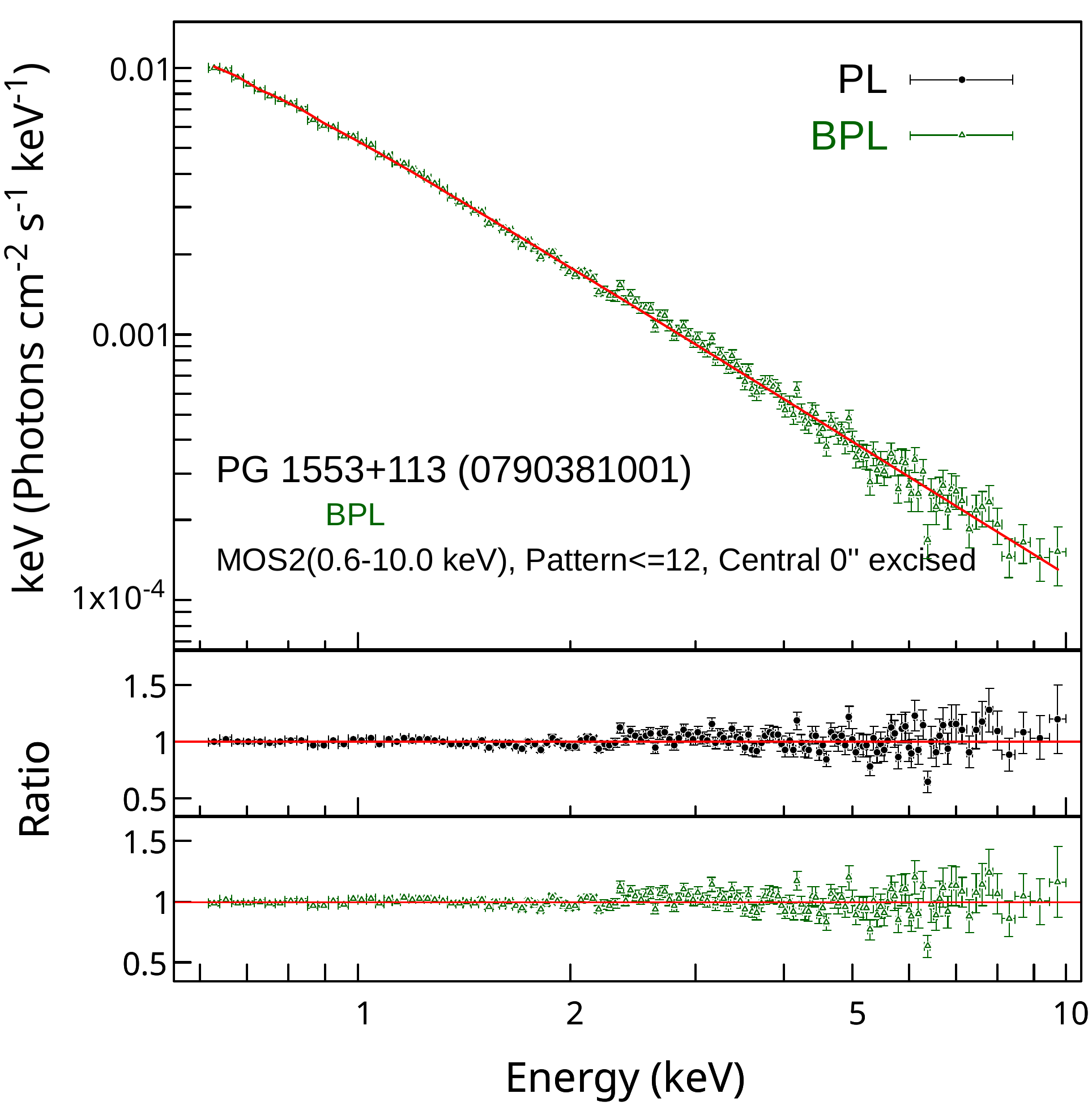}

{\vspace{-0.17cm} {\hspace{0.5cm}\includegraphics[width=8.3cm, height=7.4cm]{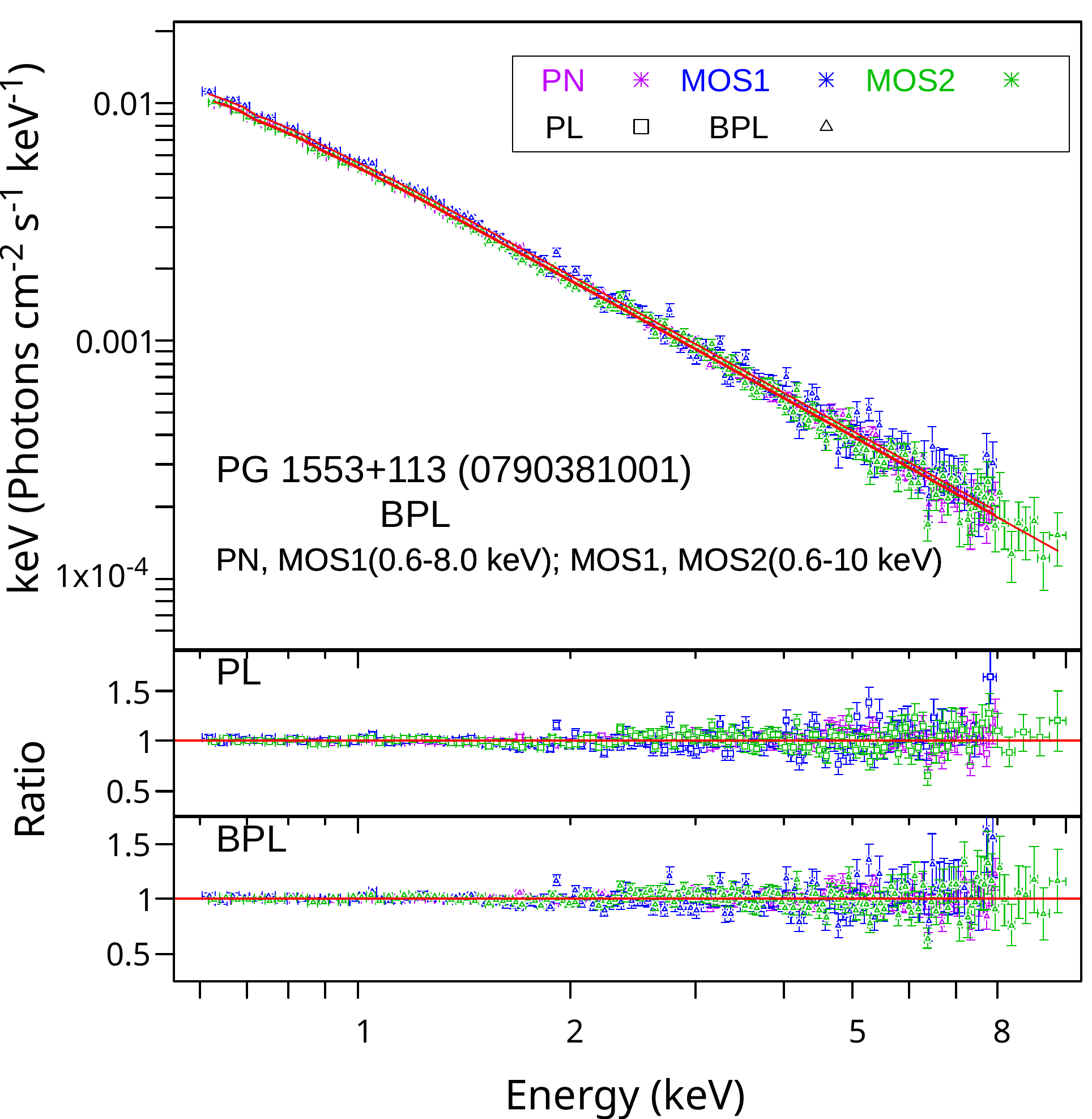}}
{\hspace{0.3cm}\includegraphics[width=8.4cm, height=7.3cm]{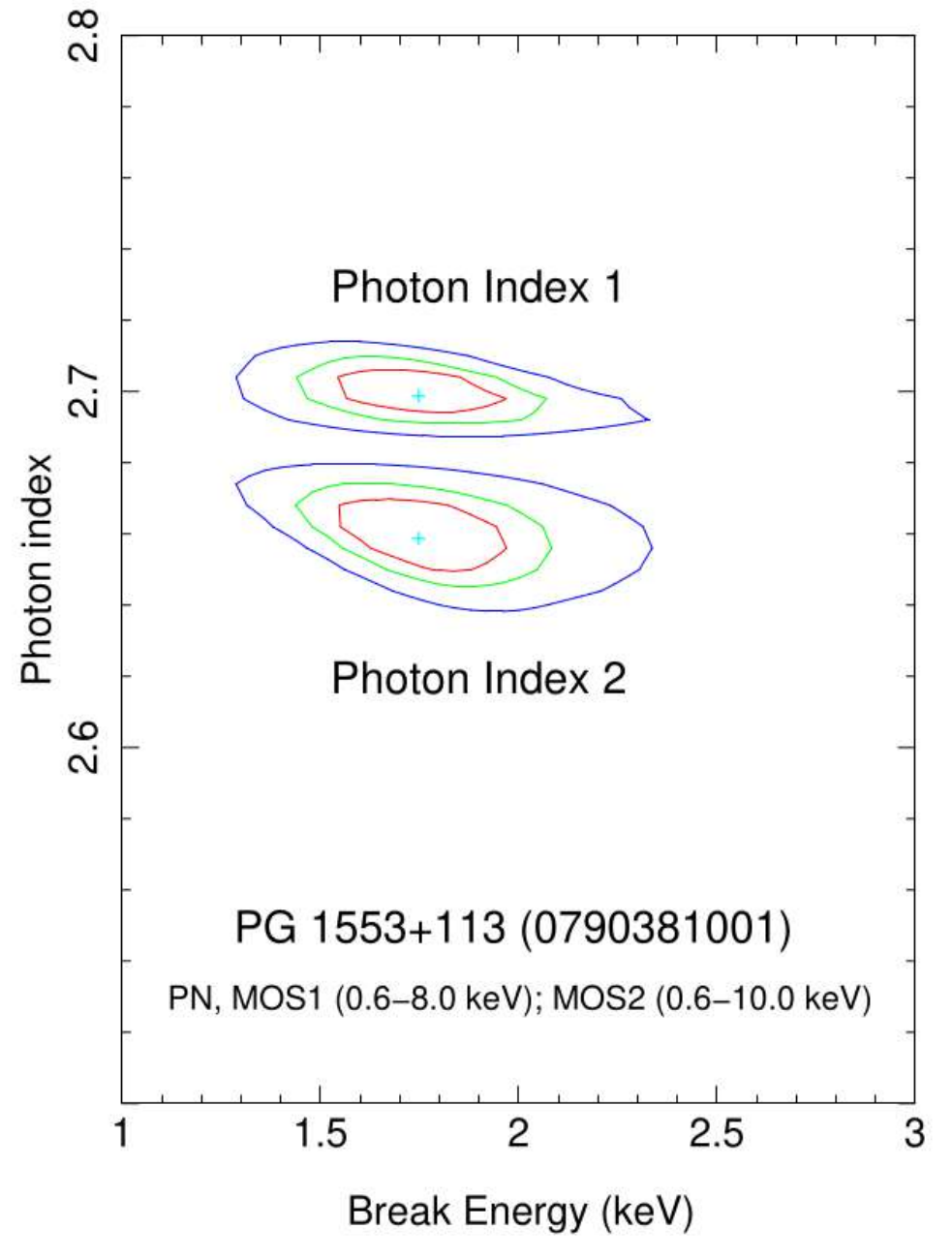}}}
\caption{(c) Similar X-ray spectral fit plots and contour plots for Obs ID: 0790381001} 
\end{figure*}
\clearpage
\newpage
\bibliography{references}{}
\bibliographystyle{aasjournalv7}

\end{document}